\documentclass[10pt,letterpaper,singlecolumn]{article}
\usepackage{ol2}
\usepackage[draft]{hyperref}
\usepackage{amsmath}

\begin{document}

%\twocolumn[ 

\title{Producing slow 
%and stored 
light in warm alkali vapor using electromagnetically induced transparency}

\author{Kenneth DeRose$^{\dag}$, Kefeng Jiang$^{\dag}$, Jianqiao Li, Linzhao Zhuo, Scott Wenner, and Samir Bali*}
\address{
Department of Physics, Miami University, Oxford, Ohio 45056-1866,
USA \\
}

\date{\today}% It is always \today, today,


\begin{abstract}
We present undergraduate-friendly instructions on how to produce light pulses propagating through warm Rubidium vapor with speeds less than 400 m/s, i.e., nearly a million times slower than $c$. We elucidate the role played by electromagnetically induced transparency (EIT) in producing slow 
%\textcolor{black}{and stored} 
light pulses, and discuss how to achieve the required experimental conditions. 
%Various EIT linewidth broadening mechanisms are described. 
The optical setup is presented, and details provided for preparation of pump, probe, and reference pulses of required size, frequency, intensity, temporal width, and polarization purity. 
%We discuss in detail how to adequately magnetically shield the alkali vapor sample. 
%Our instructions include drawings for parts and information on prices and vendors. 
%In the spirit of presenting a stand-alone article for producing slow light, we provide in Supplementary Notes, a detailed semiclassical derivation of EIT and slow light that should be amenable to advanced undergraduates, and 
Further details on
suppressing residual magnetic fields below 0.2 mG over the entire laser-atom interaction region, are provided in the online Supplementary Materials. %Good for grad students and professors who're not experts in EIT/ slow light.
%Further, we offer direct experimental proof of Dicke narrowing in warm alkali vapor by measuring the EIT linewidth as a function of relative pump-probe angle.

\end{abstract}
%FIX keycodes below for aapt AJP
%\ocis{32.30.-r, 32.60.+i, 32.70.-n, 32.70.Fw, 32.70.Jz, 37.10.De, 42.50.Hz, 42.62.Fi}
%\keywords{Raman spectroscopy; pump-probe spectroscopy; light shift; AC Stark shift; optical molasses; magneto-optical trap}

%] %activate for two-column activation

\noindent 
\section{Introduction}
\label{sec:intro}
%%\vspace{-1mm}
It has been two decades since the first demonstration of slow
%and storage~\cite{lukin1,lukin2,lukin3} of 
optical pulses propagating through atomic media.~\cite{hau,kash} Slow 
%and stored 
light was revealed as a striking consequence of a quantum mechanical phenomenon in light-matter interaction known as EIT, i.e., electromagnetically induced transparency. EIT   
%typically 
arises from the interference between probability amplitudes for absorption pathways which are simultaneously excited by two resonant light fields - one strong, referred to as the pump (or coupling) field, and the other weak, referred to as the probe.~\cite{harris1,harris2}
EIT and slow 
%and stored 
light in warm alkali vapor continue 
%continue
to be central topics of research in quantum information and quantum technology, particularly for building robust quantum memories~\cite{irina,ma,gisin}, and stable photon-shot-noise-limited electromagnetic field sensors using Rydberg atoms.~\cite{adams} 
%Application towards the development of a practically implementable quantum memory, for instance, has been widely emphasized~\cite{ma,gisin}. 
In Ref.~\cite{ma}, for instance, the authors provide an overview of current approaches to quantum memory, and state that ``although all of these approaches have been studied and demonstrated, EIT remains the most popular scheme for quantum memory," because, ``in comparison to the other approaches, the EIT approach has a long storage time and is a relatively easy-to-implement and inexpensive solution". 
%Warm vapor experiments, in particular, require only a magnetically shielded sample cell with modest temperature-control and are, therefore, significantly less resource-intensive than laser-cooled experiments for which a full optimized magneto-optical trap is the starting point. 
%\textcolor{blue}{RYDBERG ATOMS AND DAMOP}

An excellent article on EIT-based experiments for undergraduate laboratories was published just over a decade ago.~\cite{ajpirina} Elegant EIT-based experiments carried out at undergraduate institutions have also been reported.~\cite{oleary} However, an undergraduate-friendly \emph{experimental description of slow light} in atomic vapor does not exist. In this paper we endeavor to fill this gap. This is important because an increasing number of physics and engineering majors wish to get involved in the development of cutting-edge quantum technologies.
%We elucidate the role played by EIT in producing slow light pulses and emphasize the importance of optimizing the probe pulse duration and pump intensity. 
There are several pedagogical advantages to introducing the concepts of EIT and slow light in an optics class for undergraduate seniors and first/second-year graduate students. For instance, in the advanced lecture/lab course ``Optics and Laser Physics" that we teach at Miami University, 
students often ask,``We've read about laser applications in imaging, communications and medicine... can you tell us about \emph{quantum} applications we haven't heard of?" 
EIT and slow light can provide a satisfactorily impactful answer to this question.
%, on the theoretical as well as experimental front. On the theory side 
EIT physics is a natural extension to what these students have already been learning about light-matter interaction. Early in the semester, they are introduced to population-rate equations for a laser system, 
%three- and four-level atoms 
%(since a population inversion for lasing cannot be created with just a two-level atom), 
which sets the stage for an EIT-based slow light system, since the simplest model for either system is provided by a three-level atom. 
%The students are therefore primed to take the step up from simple population rates (diagonal elements of the probability density matrix $\hat{\rho}$ for light-atom interaction, where the caret denotes a quantum mechanical operator) to a more nuanced consideration of the coherences (off-diagonal elements of $\hat{\rho}$) in order to understand the quantum interference underlying EIT. Similarly, 
Furthermore, students know about group versus phase velocity, and therefore have all the necessary background to learn about how a steep positive gradient of the refractive index within a narrow EIT spectral window can lead to a dramatically slow group velocity.

%On the experimental side, the class is divided into four groups of two or three students each. The advanced lab component of the course culminates in each student group constructing an external cavity-tunable diode laser system and performing saturated absorption spectroscopy~\cite{wieman,arnold,bali2012}. Interested students may extend their advanced lab experience, in the form of an independent study for a semester or two, and carry out the experiments on EIT and slow light described in this paper.
%\textcolor{black}{refer to Irina's 2009 AJP here}~\cite{ajpirina}

%\textcolor{black}{Good for noob grad students/profs. Wes (Denison), Charles (ODU) showed specific interest in this paper.}
% What about theory for STOPPED light?} 
In this article, we state the relevant theoretical results 
%present basic theory 
for EIT and slow light, and 
%Calculation details, starting with the Schr$\ddot{{\mbox o}}$dinger equation, are provided in Sec. 1 of the Supplementary Notes~\cite{sup}. 
describe our experimental setup for producing slow light pulses with group velocities less than 400 m/s. A price list for parts is included in the online Supplementary Material.
%We present our data and discuss our experimental results. SAY 1 SENTENCE FOR UG'S.
%Details for suppression of stray magnetic fields in the laser-atom interaction region are provided in the Supplementary Notes.
%Secs. 2 and 3 of Ref.~\cite{sup}.
 
\section{Theoretical description}
\label{sec:theory}
The theory behind EIT and slow light has been explained in several high-quality undergraduate-friendly articles (see, for example, Refs.~\cite{ajpirina,ajpolson,ajpslowlight1,ajpslowlight2,opticsinourtime}), \textcolor{black}{including a textbook-treatment in Ref.~\cite{eberly} which is amenable to seniors.} Here, we 
\textcolor{black}{summarize}
%sum up 
the relevant results. A more complete derivation can be found in the online Supplementary Materials.
\subsection{Three-level atom model} 
\label{subsec:3lvl}
\begin{figure}[h]
\centerline{
\includegraphics[width=6cm]{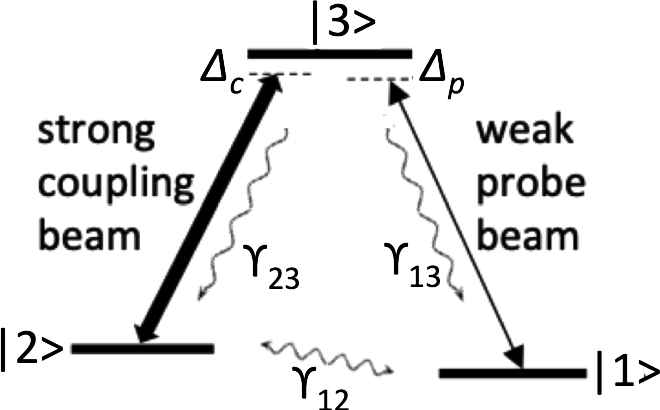}}
%{fig1_rev.pdf}}
\caption{\textsf{Three-level atom in the $\Lambda$-shaped configuration. 
%The energy difference between $|3\rangle$ and $|1\rangle$ is $\hbar \omega_{31}$, and between $|3\rangle$ and $|2\rangle$ is $\hbar \omega_{32}$. 
%A strong coupling (or pump) beam of frequency $\omega_c$ 
%%of frequency $\omega_{c}$ 
%and a weak probe beam $\omega_p$ 
%%of frequency $\omega_{p}$, 
%are tuned to near-resonance with $|3\rangle$, from levels $|2\rangle$ and $|1\rangle$ respectively, with detunings $\Delta_{c} \equiv \omega_c - \omega_{32}$ and $\Delta_{p} \equiv \omega_p - \omega_{31}$. 
%(\textcolor{black}{opposite in sign to Ref.~\cite{eberly}}). 
Excited state decay rates
%3 and 1, 3 and 2, 2 and 1, 
are denoted by $\gamma_{13}$ and $\gamma_{23}$, and the ground state decoherence rate by $\gamma_{12}$. 
%In our experiments, the decoherence from the excited state is dominated by spontaneous emission.
%(see Eqns.~\ref{eq:obedamp} and preceding text). 
} }\label{fig:fig1}
%\vspace{-5.5mm}
\end{figure}

Following the treatment in Ref.~\cite{eberly}, we depict a fictitious three-level atom in Fig.~\ref{fig:fig1}, in which transitions are allowed between states 1 and 3 and between states 2 and 3, but not between states 1 and 2. This is the so-called lambda-system because the lower levels 1 and 2 have close-lying energies, resulting in a $\Lambda$-shaped configuration.
% - is the simplest possible atomic system for understanding the basic concepts of EIT.  
The standard EIT experiment consists of \textcolor{black}{a sample of such atoms illuminated by} a strong coupling (or pump) \textcolor{black}{laser} of frequency $\omega_{c}$ and a weak probe \textcolor{black}{laser} of frequency $\omega_{p}$, both tuned to near-resonance with $|3\rangle$ from levels $|2\rangle$ and $|1\rangle$ respectively. 
The energy difference between levels 3 and 1 is $\hbar \omega_{31}$, and between 3 and 2 is $\hbar \omega_{32}$. The laser detunings are $\Delta_{c} \equiv \omega_c - \omega_{32}$ and $\Delta_{p} \equiv \omega_p - \omega_{31}$.
%~\cite{note_eberly}.
%with detunings $\Delta_{c}$ and $\Delta_{p}$ (note that in Ref.~\cite{eberly}, $\Delta_c$ is set to zero). 
We assume that the pump beam only addresses the coupling transition $|2\rangle \rightarrow |3\rangle$ and the probe beam only addresses the $|1\rangle \rightarrow |3\rangle$ transition. The probe frequency is scanned around the (fixed) pump frequency, and the probe absorption is measured. 
%The pump beam intensity is significantly stronger than the probe, \textcolor{black}{typically by one to two orders of magnitude}. 
%However, the pump intensity is kept below saturation intensity \textcolor{black}{NOT TRUE!} in order to mitigate power broadening which causes unwanted Autler-Townes splitting effects to mix in with EIT~.
%\textcolor{black}{comment on relaxation rates here? ATS vs EIT effects? \cite{2013ATSvsEIT}}
%%\vspace{-2mm}
%%\vspace{-2mm}

In the Supplementary Materials, \textcolor{black}{Sec. S1A - B}, we write the Hamiltonian for an illuminated three-level atom, assuming 
that the incident field-induced electric dipole is the predominant light-atom interaction.
%the predominant light-atom interaction to be that the incident field induces an electric dipole 
%- \textcolor{black}{see Eqns. (S1) - (S9)}. 
We calculate the eigenenergies and eigenstates in \textcolor{black}{Eq. (S9) - (S11)}, taking the pump and probe beams to be on-resonance for simplicity. One of the eigenstates is a ``dark" state which is a \textcolor{black}{coherent} superposition of the two ground states $|1\rangle$ and $|2\rangle$. This dark state has no quantum overlap with the excited state $|3\rangle$. Thus, an atom, once it is transferred to the dark state upon interaction with the light fields, stays there, no longer able to absorb a probe photon. \textcolor{black}{Note that in the absence of the pump, the weak resonant probe would be completely absorbed. Thus the pump laser ``\textcolor{black}{coherently} prepares" the atoms to be transparent to the probe} - this is EIT.

In order to calculate the spectral width of the transparency window, we must take into account the decay of the light-induced dipole moment due to various decohering processes (also called damping, or relaxation, processes). Such processes may arise from atomic motion which may, for example, cause the atom to leave the illuminated region, and the light-atom interaction to cease altogether. Similarly, elastic and inelastic collisions with other atoms and with the container walls, and other processes such as spontaneous emission, affect the phase and amplitude of the light-induced dipole moment. These mechanisms are discussed in more detail in Sec.~\ref{sec:3lvlexpt}\ref{subsec:broaden} and~\ref{subsec:condition} below.  
We denote all decay processes from $3 \rightarrow 1$ and from $3 \rightarrow 2$ by the optical decoherence rates $\gamma_{13}$ and $\gamma_{23}$, respectively. The two ground states are close-lying, so that non-optical processes such as collisions may suffice to cause interchange between the two states - we denote this two-way $1 \leftrightarrow 2$ decay by the ground state decoherence rate $\gamma_{12}$.  
%Some damping mechanisms are mentioned in Sec.~\ref{sec:theory}\ref{subsec:broaden} below. 
%of the induced dipole moment 
In dilute samples where the densities are not so high that non-optical damping processes (e.g., collisions between ground state atoms) begin to dominate over optical damping processes \textcolor{black}{e.g., spontaneous emission or Doppler broadening}, $\gamma_{12}$ is much smaller than $\gamma_{13}$ and $\gamma_{23}$. This is an important requirement for EIT which is satisfied in our experiments, see Sec.~\ref{sec:3lvlexpt}\ref{subsec:condition} below.
%decay vs decoherence - dbl/single arrow
%In our experiments the dominant contribution to $\gamma_{13}$ and $\gamma_{23}$ comes from the excited state spontaneous emission rate $\Gamma$. As explained in Sec. 1D of the Supplementary Notes, $\gamma_{13} = \gamma_{23} = \Gamma/2$.}
%We shall see below that a general requirement for EIT is that $\gamma_{12}$ be very small. 
%In a dilute gas $\gamma_{12}$ is much smaller than $\gamma_{13}$.
% because dissipation due to spontaneous emission is included in $\gamma_{13}$. 
%\textcolor{black}{In our experiments, spontaneous emission is the dominant dephasing mechanism.} 
%(see Fig.~\ref{fig:fig1}(a)). 
%Similar damping processes, but now with spontaneous emission added in, cause the induced dipole moment between the excited state $|3\rangle$ and $|1\rangle$ 
%%(or $|2\rangle$) 
%to relax at a rate which we denote as $\gamma_{13}$. Note that the decoherence rate $\gamma_{12}$ for the non-allowed transition 1 to 2 is much smaller than the decoherence rate $\gamma_{13}$ for the allowed transition 1 to 3.

We show in \textcolor{black}{Sec. S1C - D of the Supplementary Materials}
%\textcolor{black}{Eqns. (S12) - (S23)} 
that the density matrix formalism allows for simple phenomenological inclusion of decay processes into the calculation of the probe-induced atomic dipole moment and polarizability. This allows us to deduce the complex refractive index $n (\omega_p) = n_{r}(\omega_p) + in_{i}(\omega_p)$. 
%\textcolor{black}{as seen by the probe beam.}
%propagating through the coherently prepared medium}. 
The imaginary part $n_{i}$ yields the absorption spectrum $\alpha (\omega_p)$. The complex refractive index is a convenient way to express something well-known: Where there is dispersion $n_r (\omega)$, there must be absorption $\alpha$. This is owing to causality-based Kramers-Kronig relations between the real and imaginary parts of the electric susceptibility $\chi = \chi_r + i\chi_i$.~\cite{saleh} We calculate the absorption $\alpha (\omega_p)$ and real refractive index $n_r (\omega_p)$ seen by the weak probe (see \textcolor{black}{Eq. (S22) - (S23)}) propagating through a sample of $\Lambda$-atoms which is coherently prepared by a strong pump beam in a dark state, as mentioned above. In the following sections we 
%coherently pumped medium, 
discuss some important limiting situations.
%%This is because EIT in atomic media is a resonant phenomenon. 
%%Near-resonance absorption is included in the theory as an imaginary contribution to the refractive index. 

\subsection{Absorption and refractive index with pump field off} 
\label{subsec:controloff}
\begin{figure}[h]
\centerline{
\includegraphics[width=8.5cm]{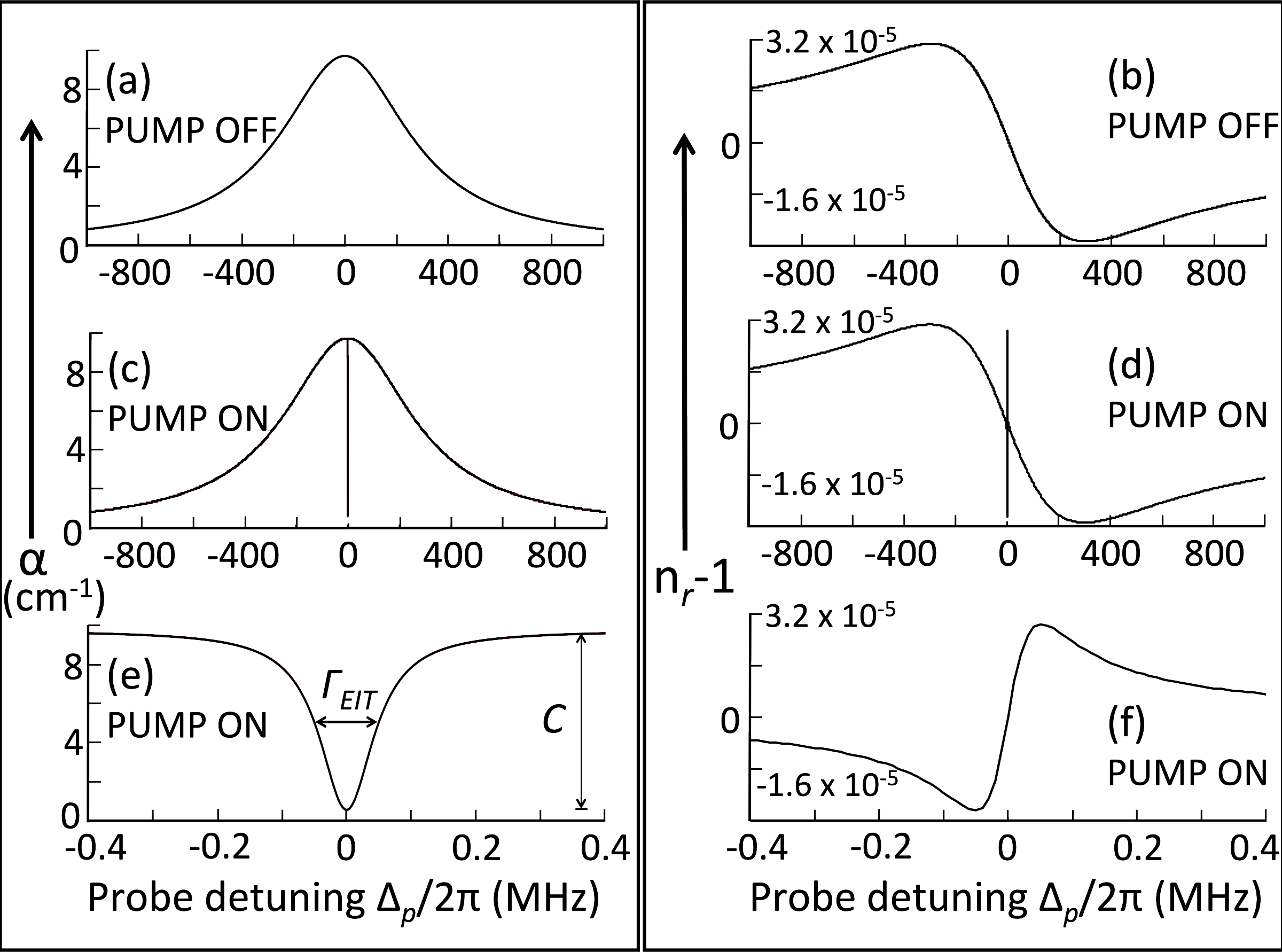}}
%{EITPlots_10_17_21.pdf}}
%%\vspace{-1mm}
\caption{\textsf{
The probe absorption coefficient $\alpha$ (left panel) in cm$^{-1}$ and real refractive index $n_r$ (right panel), versus probe detuning $\Delta_p = \omega_p - \omega_{31}$ as predicted by Eq. (\ref{eq:alpha_chi0}-\ref{eq:n_r}). (a, b) The case of ``no pump" from Eq. (\ref{eq:alpha_chi0}) and Eq. (\ref{eq:nr_chi0}).  (c, d) The ``EIT case" from Eq. (\ref{eq:abs}) and Eq. (\ref{eq:n_r}) with parameter-values from our experiments: $\Delta_c = 0$, $|\Omega_c|/2\pi = 4$ MHz, $\omega_p/2\pi = 3.77 \times 10^{14}$ Hz, $N = 3.36 \times 10^{11}$/cm$^3$, 
%and for the $^{87}$Rb D1 transition: 
$\gamma_{13}/2\pi 
%= \gamma_{23}/2\pi 
= 300$ MHz, $\gamma_{12}/2\pi = 3$ kHz, $|\mu_{13}| = 2.54 \times 10^{-29}$ C.m. (e, f) Magnified view of sharp EIT features in c and d.
%\textcolor{black}{Note that, in practice, the EIT linewidth is narrower by a factor $10^3 - 10^4$ than the prediction of the simple model leading up to Eqns.~\ref{eq:abs} and~\ref{eq:n_r}. Why?} 
}} \label{fig:EIT}
%\vspace{-2mm}
\end{figure}

When the coupling field (or pump) is off, no EIT occurs. We obtain the usual results for probe absorption and real refractive index, which are plotted in Fig.~\ref{fig:EIT} (a) and (b):
%Eqn.~\ref{eq:abs} for the absorption yields
\begin{equation}
\alpha  \xrightarrow{
\Omega_{c} \, = \, 0
%\mbox{\footnotesize{pump off}}
}  \frac{N\omega_{p}}{\epsilon_{0}c} \frac{|\mu_{13}|^{2}}{\hbar} \frac{\gamma_{13}} {{\Delta_{p}}^{2} + {\gamma_{13}}^{2}}
\label{eq:alpha_chi0}
\end{equation}
\begin{equation}
n_{r}  \xrightarrow{
%\mbox{\footnotesize{pump off}}
\Omega_{c} \, = \, 0
}  1 - \frac{N}{2\epsilon_{0}} \frac{|\mu_{13}|^{2}}{\hbar} \frac{\Delta_{p}} {{\Delta_{p}}^{2} + {\gamma_{13}}^{2}}\,\, .
\label{eq:nr_chi0}
\end{equation}
Here 
$\Omega_c$ is the coupling beam {Rabi frequency} defined as $\Omega_{c} \equiv \vec{d}_{32} \cdot \vec{\epsilon}_{c} E_{c} / 2\hbar$, where $\vec{d}_{32}$ is the transition dipole moment between levels $|2\rangle$ and $|3\rangle$, $\vec{\epsilon}_{c}$ is the polarization of the coupling field, and $\hbar$ is Planck's constant divided by $2\pi$.~\cite{factor2} $N$ is the number of atomic dipoles per unit volume, $c$ is the speed of light in vacuum, $\epsilon_0$ is the electric permittivity of vacuum, and
%(\textcolor{black}{see Sec. S1 A in Supplementary Notes}). 
%$\Delta' \equiv$ relative probe-coupling detuning $\Delta_{p}-\Delta_{c}$ (also referred to as the Raman detuning), 
$\mu_{13}$ is defined 
%after \textcolor{black}{Eq. (S19)} 
as the projection of the induced probe dipole moment on the direction of the probe field polarization.
%, i.e., $\mu_{13} = \vec{d}_{13} \boldsymbol{\cdot} \vec{\epsilon}_{p}^{\,\,*}$. 
%to denote the projection of the transition dipole moment on the direction of the field polarization. , 
 
%Note that we define our detunings (see Fig.~\ref{fig:fig1} caption) with opposite sign to that in Ref.~\cite{eberly}. 
Eq. (\ref{eq:alpha_chi0}), plotted in Fig.~\ref{fig:EIT}(a), depicts the expected Lorentzian absorptive line shape,
% for a stationary atom, 
 with half width at half maximum (HWHM) $\gamma_{13}$ and a 
%(\textcolor{black}{is stationary best choice of word? Gaussian for Doppler broadening, Lorentzian for pressure-broadened}), 
maximum at $\Delta_{p} = 0$. 
%for the $^{87}$Rb D1 transition where $2 \, \gamma_{13} = 2\pi(5.75$ MHz)~\cite{steckRb}.
%\noindent For the real refractive index we find from Eqn.~\ref{eq:n_r} that
%For a stationary atom $\gamma_{13}$ may be the spontaneous emission line width. Alternatively, in the case of experiments where buffer gas is used, $\gamma_{13}$ may represent the pressure-broadened line width.   
Eq. (\ref{eq:nr_chi0}), plotted in Fig.~\ref{fig:EIT}(b), depicts the expected dispersive line shape around probe-resonance, again with a HWHM of $\gamma_{13}$. In the immediate vicinity of resonance $\Delta_p \approx 0$, we see that 
%$n_r \rightarrow 1 - \frac{N}{2\epsilon_{0}} \frac{|\mu_{13}|^{2}}{\hbar} \frac{\Delta_{p}} {{\gamma_{13}}^{2}}$, 
$n_r \rightarrow 1 - {N}{|\mu_{13}|^{2}}{\Delta_{p}} / (2\epsilon_{0}{\hbar}{\gamma_{13}}^{2})$,
i.e., $n_r (\omega_p)$ has a negative slope. 
%Eqn.~\ref{eq:nr_chi0} , clearly depicting that $n_r$ indeed decreases with frequency near resonance. 
This is the well-known ``anomalous dispersion" effect that occurs close to an atomic resonance and in which $n_r$ decreases with increasing optical frequency.
By contrast, as we move further away from resonance we see that $n_{r}$ increases with frequency on either side. This is normal dispersion, consistent with the classic experiment of propagating white light through a glass prism and observing that red light deviates the least and blue the most.

\subsection{Absorption and refractive index with pump field on} 
\label{subsec:controlon}
When the strong coupling field is on, and assumed to be on-resonance for simplicity ($\Delta_c = 0$), we obtain for the probe absorption $\alpha (\omega_p)$ and real refractive index $n_r (\omega_p)$ in the ``weak-probe approximation" ($\Omega_p \ll \Omega_c$; here $\Omega_p$ is the probe Rabi frequency and is defined as $\Omega_{p} \equiv \vec{d}_{31} \boldsymbol{\cdot} \vec{\epsilon}_{p} E_{p} / 2\hbar$, where $\vec{d}_{31}$ is the transition dipole moment between levels $|1\rangle$ and $|3\rangle$, and $\vec{\epsilon}_{p}$ is the polarization of the probe field):~\cite{factor2}
%\textcolor{black}{see Sec. S1 A in Supplementary Notes)}: 
\begin{eqnarray}
\alpha (\omega_p) & = & \frac{N\omega_{p}}{\epsilon_{0}c} \frac{|\mu_{13}|^{2}}{\hbar}  \label{eq:abs} \\
& \times& \frac{\gamma_{13}{\Delta_p}^{2} + \gamma_{12} \, (\gamma_{12}\gamma_{13} + |\Omega_{c}|^{2})}{{\left[ {\Delta_{p}}^2 - \gamma_{12}\gamma_{13} - |\Omega_{c}|^{2} \right]^{2} + {\Delta_p}^2\left[ \gamma_{12} + \gamma_{13} \right]^{2}}} \nonumber
\end{eqnarray}
%(\Delta'^{2} + \gamma_{12}^{2})(\Delta_{p}^{2} + \gamma_{13}^{2})  - 2 |\Omega_{c}^{2}| (\Delta'\Delta_{p} - \gamma_{12}\gamma_{13}) + |\Omega_{c}|^{4}} 
%and for the real part of the refractive index $n_{r}$
\begin{eqnarray}
n_{r} (\omega_p) & = & 1- \frac{N}{2\epsilon_{0}} \frac{|\mu_{13}|^{2}}{\hbar}  \label{eq:n_r} \\
& \times & \frac{\Delta_p \, ({\Delta_p}^2 - |\Omega_{c}|^2 + \gamma_{12}^2)}{\left[ {\Delta_p}^2 - \gamma_{12}\gamma_{13} - |\Omega_{c}|^{2} \right]^{2} + {\Delta_p}^2\left[ \gamma_{12} + \gamma_{13} \right]^2}. \nonumber 
\end{eqnarray}
Eq. (\ref{eq:abs}-\ref{eq:n_r}) are plotted in Figs.~\ref{fig:EIT}(c) - (f). \textcolor{black}{Eq. (S22) and Eq. (S23)} are generalized forms of Eq. (\ref{eq:abs}) and Eq. (\ref{eq:n_r}) for $\Delta_c \neq 0$. 
%Here $\Omega_c$ is the coupling beam {Rabi frequency} defined as $\Omega_{c} \equiv \vec{d}_{32} \cdot \vec{\epsilon}_{c} E_{c} / 2\hbar$, where $\vec{d}_{32}$ is the transition dipole moment between levels $|2\rangle$ and $|3\rangle$, and $\vec{\epsilon}_{c}$ is the polarization of the coupling field~\cite{factor2}. 
%\textcolor{black}{The Rabi frequency $\Omega_c$ and coupling beam intensity $I$ have a well-known relation~\cite{metcalf}: 
%\begin{equation}
%\frac{{(2 |\Omega_c|})^2}{{\Gamma}^2} \equiv \frac{I}{2I_{sat}}\,, \,\, \,\,\, \mbox{where} \,\, I_{sat} \equiv \frac{ \pi h c }{ 3{\lambda}^3 } {\Gamma}
%\label{eq:chi2_I}
%\end{equation}
%Here $\lambda$ is the optical wavelength for the $2 \leftrightarrow 3$ transition and the saturation intensity $I_{sat}$ for the transition is defined as the excitation intensity at which the stimulated emission rate is half the spontaneous emission rate $\Gamma$. }

The absorption $\alpha$ 
and refractive index $n_r$ in the lower panels of Fig.~\ref{fig:EIT} are strikingly different from the top panel where the coupling field is off. It is instructive to verify that Eq. (\ref{eq:alpha_chi0}-\ref{eq:n_r}) satisfy the Kramers-Kronig relations between $\chi_r = 2 \, (n_r - 1)$ and $\chi_i = c \, \alpha/\omega_p$, namely, $\chi_r (\omega_p) = (2/\pi) \, \int_0^{\infty} d\omega \frac{\omega \chi_i(\omega)}{ {\omega_2}^2 - {\omega_p}^2}$ and $\chi_i (\omega_p) = (2/\pi) \, \int_0^{\infty} d\omega \frac{\omega_p \chi_r(\omega)}{ {\omega_p}^2 - {\omega}^2}$.

\subsection{The EIT ``window"} 
\label{subsec:window}
EIT manifests itself as a dramatic drop in absorption $\alpha \, (\omega_p)$ in Figs.~\ref{fig:EIT} (c, e) when the coupling and probe beams are on-resonance. 
%Claire suggestion:
In our experiments, $\gamma_{12}/2\pi \sim$ few kHz, $|\Omega_c|/2\pi \sim$ few MHz, $\gamma_{13}/2\pi \sim$ few hundred MHz, and $|\Delta_p|/2\pi \leq 100$ kHz, 
%(see Fig.~\ref{fig:eitdata} below) 
so that we verify the ``strong coupling field approximation":
\begin{equation}
|\Omega_c| >> {|\Delta_p|}, {\gamma_{12}}
\label{eq:cond1}
\end{equation} 
which is one of two conditions (see Eq. (\ref{eq:cond2}) below) that $\Omega_c$ must satisfy in order to produce slow light. Using $\gamma_{12}\ll\gamma_{13}$, we find from Eq. (\ref{eq:abs}), in the $\Delta_p\rightarrow 0$ limit :
%yielding a narrow transparency window near resonance. 
%This is readily seen by considering Eqn.~\ref{eq:abs} at $\Delta_p \approx 0$ in the ``strong coupling field approximation":
%\begin{equation}
% |\Omega_c| >> {|\Delta_p|}, {\gamma_{12}}
%\label{eq:cond1}
%\end{equation} 
%In our experiments we see below that $\gamma_{12} \sim$ few kHz, $|\Omega_c| \sim$ few MHz, $\gamma_{13} \sim$ few hundred MHz, and from Fig.~\ref{fig:eitdata} we see that $|\Delta_p| \leq 100$ kHz. 
%%$\gamma_{13} > |\Omega_c|$. 
%Using $\gamma_{12} << \gamma_{13}$, we find from Eqn.~\ref{eq:abs}: 
\begin{equation}
\alpha \rightarrow \frac{N\omega_p}{\epsilon_{0} c} \frac{|\mu_{13}|^{2}}{\hbar} \frac{\gamma_{12}}{|\Omega_c|^2} \approx 0
\label{eq:alpha_res}
\end{equation}
which yields transparency in a narrow ``EIT window".

In order to simply obtain an approximate expression for the {EIT linewidth} $\Gamma_{EIT}$ (see Fig.~\ref{fig:EIT}e)
%and contrast $\Delta a$ (or amplitude of the transparency window) 
it is convenient to set $\Delta_p = 0$ in Eq. (S22) 
%for the absorption $\alpha (\omega_p)$ 
while allowing $\Delta_c$ to vary around the resonance condition $\Delta_p - \Delta_c = 0$ (also known as the Raman, or two-photon, resonance condition):
%, see text just below Eq. (S23) in Supplementary notes):
\begin{equation}
%\frac{ |\Omega_c|^2 }{ \gamma_{13} } = 1
\alpha  \xrightarrow{\Delta_{p} \, = \, 0}  
\alpha_{\Omega_{c}= 0} 
\left[ 1 -   
\,\,\frac{|\Omega_c|^2 }{\gamma_{13}} \,\,\frac { 
{\gamma_{12} +  \frac{|\Omega_c|^2 }{\gamma_{13}}}  }{ {\Delta_{c}}^{2} + \left( \gamma_{12} + \frac{|\Omega_c|^2 }{\gamma_{13}} \right)^2 } 
\right],
\label{eq:alpha_eit}
\end{equation} 
where $\alpha_{\Omega_{c}= 0} = N\omega_p|\mu_{13}|^2 / (\epsilon_0c\hbar\gamma_{13})$ is obtained by setting $\Delta_p = 0$ in Eq. (\ref{eq:alpha_chi0}). The first term inside the parentheses in Eq. (\ref{eq:alpha_eit}) represents the baseline absorption without EIT. The second term is the EIT window depicted in Fig.~\ref{fig:EIT}(e), which is  
%- the dip in absorption depicted in Fig.~\ref{fig:EIT}(b). This dip in absorption is 
a Lorentzian with full-width-half-maximum $\Gamma_{EIT}$ given by:
\begin{equation}
\Gamma_{EIT}  =  2 \left( \gamma_{12} + \frac{|\Omega_c|^2 }{\gamma_{13}} \right).
\label{eq:eitlinewidth}
\end{equation}
%In Eqn.~\ref{eq:eitlinewidth} the term dependent on the pump intensity, through Eqn.~\ref{eq:chi2_I}, is the power-broadened contribution. 
In Eq. (\ref{eq:eitlinewidth}), $\Omega_c$ depends on the pump intensity $I$ (see Eq. (\ref{eq:chi2_I}) below), so that the EIT window broadens when the intensity of the pump beam increases.
\textcolor{black}{Upon substituting typical experimental parameter-values (see caption under Fig.~\ref{fig:EIT}) into the approximate expression in Eq. (\ref{eq:eitlinewidth}), we predict $\Gamma_{EIT} \approx 100$ kHz for the EIT window. This is in close agreement with 
%the rigorous expression in 
Eq. (\ref{eq:abs}) which is plotted in Fig.~\ref{fig:EIT}(e), as revealed by direct inspection of the figure.} 
The probe absorption $\alpha$ is minimum when $\Delta_c$ is minimum, so that an expression for the {EIT contrast}, or amplitude of the transparency window, denoted by $C$ (see Fig.~\ref{fig:EIT}e)
may be simply obtained by setting $\Delta_c = 0$ in the second term of Eq. (\ref{eq:alpha_eit}):
%\begin{equation}
$C = { |\Omega_c|^2 }/{ (\gamma_{13}\gamma_{12} + |\Omega_c|^2 )}$
%\label{eq:contrast}
%\end{equation}
which, in the strong coupling approximation (Eq. (\ref{eq:cond1})), predicts $\approx 100$\% transparency. 
%in Fig.~\ref{fig:EIT}(a) by Eqn.~\ref{eq:abs}, in agreement with the approximate expression in 
%However, the theoretical predictions in Eqns.~\ref{eq:abs} -~\ref{eq:contrast} for the EIT linewidth and contrast yield unrealistic numbers. In order to understand why and also gain a physical understanding of the origin of the transparency window, we ask the question: What are the new eigenstates describing the three-level atomic system  under illumination by the coupling and probe fields in Fig.~\ref{fig:fig1}? In Sec. 1 of the Supplementary Notes~\cite{sup} we show that the atom is transferred to a ``dark" state which is a quantum superposition of the two ground states $|1\rangle$ and $|2\rangle$. This dark state has no quantum overlap with the excited state $|3\rangle$, thus an atom once transferred to the dark state stays there.
%because the atom is unable to absorb a photon from the coupling or probe fields and be excited to state $|3\rangle$. 

In practice, the {observed} 
%EIT linewidth is far narrower and the 
contrast and linewidth {are much lower} (see Sec.~\ref{sec:protocol}). This is 
%as we shall see in . 
because, in steady-state the fraction of atoms settling in the dark state is significantly reduced for reasons discussed in Sec.~\ref{sec:3lvlexpt}\ref{subsec:broaden}. In short, our theory is based on a closed three-level model, 
%of \textcolor{black}{stationary} atoms 
whereas, in real atoms, many energy levels are addressed by the laser fields. This provides
%optically dense samples of 
%\textcolor{black}{moving} atoms with multi-levels \textcolor{black}{cross out!}. 
extraneous energy-levels to which the population can leak, severely reducing contrast (in our experiments the maximum contrast is about 25\%). The lowered contrast, in turn, leads to a significant reduction in the observed EIT linewidth: Only the probe spectral components which lie at the center of the EIT window can propagate through and are detectable in transmission even though considerably reduced in intensity, while frequency components near the edge of the window are too heavily attenuated to be detected.~\cite{lukin,lukin2} In Ref.~\cite{lukin} it is shown that the power-broadened component of the linewidth in Eq. (\ref{eq:eitlinewidth}) is estimated to 
reduce by a factor 
%$\sqrt{\eta kL}$ where $\eta = 3N{\lambda^3}/4{{\pi}^2}, k = 2\pi/\lambda$, and $L$ is the illuminated length of atom sample by laser beams of wavelength $\lambda$~\cite{od,lukin}.
$\sqrt{OD}$, where $OD$ is the on-resonance optical depth seen by the probe as it traverses an atom sample of length $L$ (with the coupling beam turned off; see Supplementary Notes, Sec. S1 D). The $OD$ is 
defined by the product of $L$ with the on-resonance probe absorption $\alpha_{\Omega_c=0}$ (see Eq. (\ref{eq:alpha_eit}) above): 
%$OD \equiv \frac{N\omega_{p}}{\epsilon_{0}c} \frac{|\mu_{13}|^{2}}{\hbar} \frac{L}{\gamma_{13}}$. 
$OD =  L\alpha_{\Omega_c = 0} = {N\omega_{p}} {|\mu_{13}|^{2}}{L} / ({\epsilon_{0}c}{\hbar}{\gamma_{13}})$. 
%at  = 3N{\lambda^3}/4{{\pi}^2}, k = 2\pi/\lambda$, and $L$ is 
%the illuminated length of the atomic sample 
%(see Supplementary Notes).
%by the laser beams of wavelength $\lambda$~\cite{od,lukin}. 
%Thus $OD = \frac{N\omega_{p}}{\epsilon_{0}c} \frac{|\mu_{13}|^{2}}{\hbar} \frac{L}{\gamma_{13}}$. 
To take this attenuation into account, we amend Eq. (\ref{eq:eitlinewidth}) to
\begin{equation}
%\Gamma_{EIT}  =  2 \left( \gamma_{12} + \frac{|\Omega_c|^2 }{\gamma_{13} \sqrt{OD}} \right) 
\Gamma_{EIT}  =  2 \left( \gamma_{12} + \frac{|\Omega_c|^2 }{\gamma_{13} \sqrt{OD}} \right),
%\xrightarrow{\text{Eqn. 6}} 
%\, \mbox{where} 
%\, \, OD = \frac{N\omega_{p}}{\epsilon_{0}c} \frac{|\mu_{13}|^{2}}{\hbar} \frac{L}{\gamma_{13}} 
%\propto \frac{I}{\sqrt{N L \gamma_{13}}} 
%\, \, \mbox{from Eqns.~\ref{eq:chi2_I}-\ref{eq:cond1}}
\label{eq:eitlinewidth2}
\end{equation}
where the power-broadened term is identical to Eq. (5) in Ref.~\cite{irina} and Eq. (2) in Ref.~\cite{lukin}. 

\subsection{The refractive index and slow light} 
\label{subsec:slow}
We shift focus now to the real refractive index $n_r$ (Eq. (\ref{eq:n_r})), plotted in Fig.~\ref{fig:EIT} (d, f). 
%Slow light arises from the dramatic change in $n_r$ within the narrow EIT transparency window. 
In particular, using Eq. (\ref{eq:cond1}), we find that a strong coupling field induces a \emph{positive} slope in $n_r$ within the narrow EIT window,
\begin{equation}
n_r \rightarrow 1 + \frac{N}{2\epsilon_{0}} \frac{|\mu_{13}|^{2}}{\hbar} \frac{\Delta_p}{|\Omega_c|^2},
\label{eq:n_res}
\end{equation}
as clearly depicted in Fig.~\ref{fig:EIT} (f).
We see from Eq. (\ref{eq:n_res}) that, if we ensure that the coupling field satisfies the condition
\begin{equation}
 |\Omega_c|^2 <<  \frac{ N |\mu_{13}|^2 }{ 2\epsilon_{0} \hbar } \omega_p,
\label{eq:cond2}
\end{equation}
we may produce a {steep positive gradient} about $\Delta_p \sim 0$. 
\newline The group velocity of light $v_g$ is defined by the relation $v_g/c \equiv { (n_r + \omega_p dn_r/d\omega_p)^{-1} }$. A steep positive gradient therefore leads to a significantly reduced group velocity for the probe. Applying Eq. (\ref{eq:n_res} -~\ref{eq:cond2}), we have
\begin{equation}
%\frac{v_g}{c} \approx \,\,\frac{ |\Omega_c|^2 }{ \frac{ N |\mu_{13}|^2 }{ 2\epsilon_{0} \hbar } \omega_p} < < 1
\frac{v_g}{c} \approx \,\, { |\Omega_c|^2 } \bigg/ \left( { \frac{ N |\mu_{13}|^2 }{ 2\epsilon_{0} \hbar } \omega_p} \right) < < 1.
\label{eq:v_g}
\end{equation}
From the definition of the on-resonance optical depth $OD$ above, and assuming that the power-broadened term in Eq. (\ref{eq:eitlinewidth2}) is dominant, we may re-cast Eq. (\ref{eq:v_g}) as
\begin{equation}
\frac{v_g}{c} = \frac{\Gamma_{EIT}}{\sqrt{OD}}\, \frac{L}{c} \, \, \propto \frac{I}{N},
%= \frac{ \gamma_{13} }{ \frac{ N |\mu_{13}|^2 }{ 2\epsilon_{0} \hbar } \omega_p } \, \,  
% \propto  \frac {I}{N} \, \, \mbox{from Eqn.~\ref{eq:chi2_I}}.
\label{eq:v_g1}
\end{equation}
where we have used 
%$\Omega_c$ and $I$ are related as in 
Eq.~\ref{eq:chi2_I}. 
%If the probe is in the form of a pulse of duration $\tau_p$, it exits the sample after a delay $\tau_d$ relative to a reference pulse traveling through free space. Our sample length $\sim$ cm, and $\tau_d \sim \mu$s, so $\tau_d$ is simply $L/v_g$. We find
This means that the probe light is delayed by a time $\tau_d$, given by:~\cite{lukin2}
\begin{equation}
%\tau_d = \frac{\sqrt{OD}}{\Gamma_{EIT}} \, \propto \,  \frac {NL}{I}
\tau_d=\frac{L}{v_g}\sim\frac{\sqrt{OD}}{\Gamma_{EIT}}\propto\frac{NL}{I}.
\label{eq:delay}
\end{equation}
Thus,
%reducing the pump intensity and increasing the atom density is expected to lower the group velocity and raise the pulse-delay $\tau$. The 
the narrower the EIT window
%, the tighter the pinch in the $n_r$-curve 
%in Figs.~\ref{fig:EIT} (c, e) 
%yielding larger $dn_{r}/d\omega_p$, 
%and 
the larger the pulse-delay $\tau_d$.~\cite{identical1} 
%\textcolor{black}{For the parameters in Fig.~\ref{fig:EIT}'s caption and $L = 2.5$ cm, Eqn.~\ref{eq:delay} predicts $\tau_d \approx \sqrt{25}/ \left[ 2\pi (20 \, kHz) \right] \sim 40 \, \mu$s, which is comparable to the observed delay of 68 $\mu$s in Fig.~\ref{fig:slow}(b).} 
%Eqn.~\ref{eq:delay} is identical to Eqn. 6 in Ref.~\cite{irina} and Eqn. 38 in Ref.~\cite{lukin2}.
See Ref.~\cite{klein} and Sec. 4.1 of Ref.~\cite{irina} for more accurate predictions of the maximum achievable $\tau_d$, that are achieved by making detailed measurements of the EIT lineshape and contrast, and incorporating some of these empirical values into the theory.  

We wish to clarify a couple issues that often confuse students. First, the dramatic changes in $n_r$ near resonance 
%in Figs.~\ref{fig:EIT}(d, f) 
do not much impact the \emph{phase velocity} of light ($c/n_r$), because {$n_r$ never significantly departs from unity} for a dilute vapor - this is clearly visible in Fig.~\ref{fig:EIT} (d, f). 
%One may visualize there being a ``kink" or ``pinch" near zero pump-probe detuning in the $n_r$-curve which is otherwise flat. 
No physical significance is ascribed to the phase velocity, so that the fact that $n_r$ dips below unity, causing the phase velocity to exceed $c$, is no cause for alarm. 
%Recall that information and energy carried by a light pulse propagates at $v_g$, not the phase velocity~\cite{griffiths}. 
Second, the definition of $v_g$ above suggests that when $dn_r/d\omega$ is negative, as is the case near-resonance just outside the EIT window, 
%- see Figs.~\ref{fig:EIT} (b, d) - 
the group velocity may, in principle, exceed the phase velocity and even $c$. However, just outside the EIT window also happens to be where the absorption is high, thus no signal is actually able to propagate faster than $c$.~\cite{opticsinourtime} In the context of our experiments, recall the discussion in the previous sub-section of EIT linewidth narrowing owing to low EIT contrast, where we pointed out that even probe spectral components that lie within the EIT window but not near the center are too heavily attenuated to be detectable in transmission, let alone components outside the transparency window.       

\subsection{Stored light and EIT-based quantum memory} 
\label{subsec:store}
In effect, {a pulse of probe light is slowed and stored in the sample for a time $\tau_d$}~\cite{lukin2}, and the medium then acts as a quantum memory where information can be stored. A variety of schemes, some not based on EIT, have reported values of $\tau_d$ ranging from $\mu$s to ms, and even higher~\cite{ma}. Notice that the pulse undergoes a dramatic spatial compression inside the sample: The front end slows down upon entry into the sample and propagates at $v_g$, even as the pulse-rear, which is still outside, propagates at $c$. The compression factor is given by $L_p/L_0 = v_g/c$ where $L_p$ is the length of the compressed pulse inside the medium and $L_0$ is the free-space pulse length.~\cite{lukin2}

However, from Sec.~\ref{sec:theory}\ref{subsec:window}
%the discussion preceding Eqn.~\ref{eq:eitlinewidth2}, 
it is clear that Eq. (\ref{eq:delay}) and the subsequent discussion apply {only to a probe pulse for which the frequency bandwidth ${\tau_p}^{-1}$ fits within the EIT linewidth $\Gamma_{EIT}$}, i.e. for a pulse such that ${\tau_p}^{-1} \leq \Gamma_{EIT}$. In other words, if $\tau_p$ is so short that the probe bandwidth is broader than $\Gamma_{EIT}$, then the probe frequency components outside the transparency window suffer significant absorption, causing the pulse to become distorted. Substituting this condition in Eq. (\ref{eq:delay}) we find:~\cite{identical2}
\begin{equation}
{\tau_p}^{-1} \tau_d \leq \sqrt{OD} \, .
\label{eq:product}
\end{equation}
%Eqn.~\ref{eq:product} is identical to Eqn. 39 in Ref.~\cite{lukin2}. 
This relation between the bandwidth of the probe pulse and the pulse delay is known as the {``delay-bandwidth product"}, which is a figure of merit for a storage/memory device, as it is a measure of how many probe pulses fit within the sample (without being absorbed to the extent that they undergo distortion).~\cite{irina,lukin2} A drawback of EIT-based quantum memory is that, owing to the narrow bandwidth requirements of EIT, this product remains low, e.g., for our longest observed $\tau_d$ of 68 $\mu$s (see Fig.~\ref{fig:slow}b) and $\tau_p = 170 \, \mu$s \textcolor{black}{(see Sec.~\ref{sec:optsetup}\ref{subsec:pulses})}, the delay-bandwidth product is 0.4. In our experiments, $\sqrt{OD} \approx 5$ (see Sec.~\ref{sec:delay}\ref{subsec:compare}), so Eq. (\ref{eq:product}) is satisfied. Our $OD$ and delay-bandwidth product are comparable to previous values reported for EIT-based quantum memory in warm $^{87}$Rb vapor.~\cite{irina,2008Phillips} On the other hand, non EIT-based techniques in the same medium have yielded a delay-bandwidth product that is two orders of magnitude higher.~\cite{irina,2011natcom}

\section{Three-level atoms: Implementation in the lab}
\label{sec:3lvlexpt}
%As indicated above, the narrower the EIT window the slower the group velocity of light propagating through the medium. To combat broadening of the EIT line width and reduction in contrast, unwanted contributions from energy-levels that are extraneous to the three-level scheme in Fig.~\ref{fig:fig1}(a) must be avoided as far as possible. 
%Further, we remind ourselves that Eqns.~\ref{eq:eitlinewidth} and~\ref{eq:contrast} tell us that optimizing both EIT contrast and linewidth requires finding just the right coupling beam strength.
%\textcolor{black}{When you know your pump strength, say a few words here}

\begin{figure}[h]
\centerline{
\includegraphics[width=8cm]{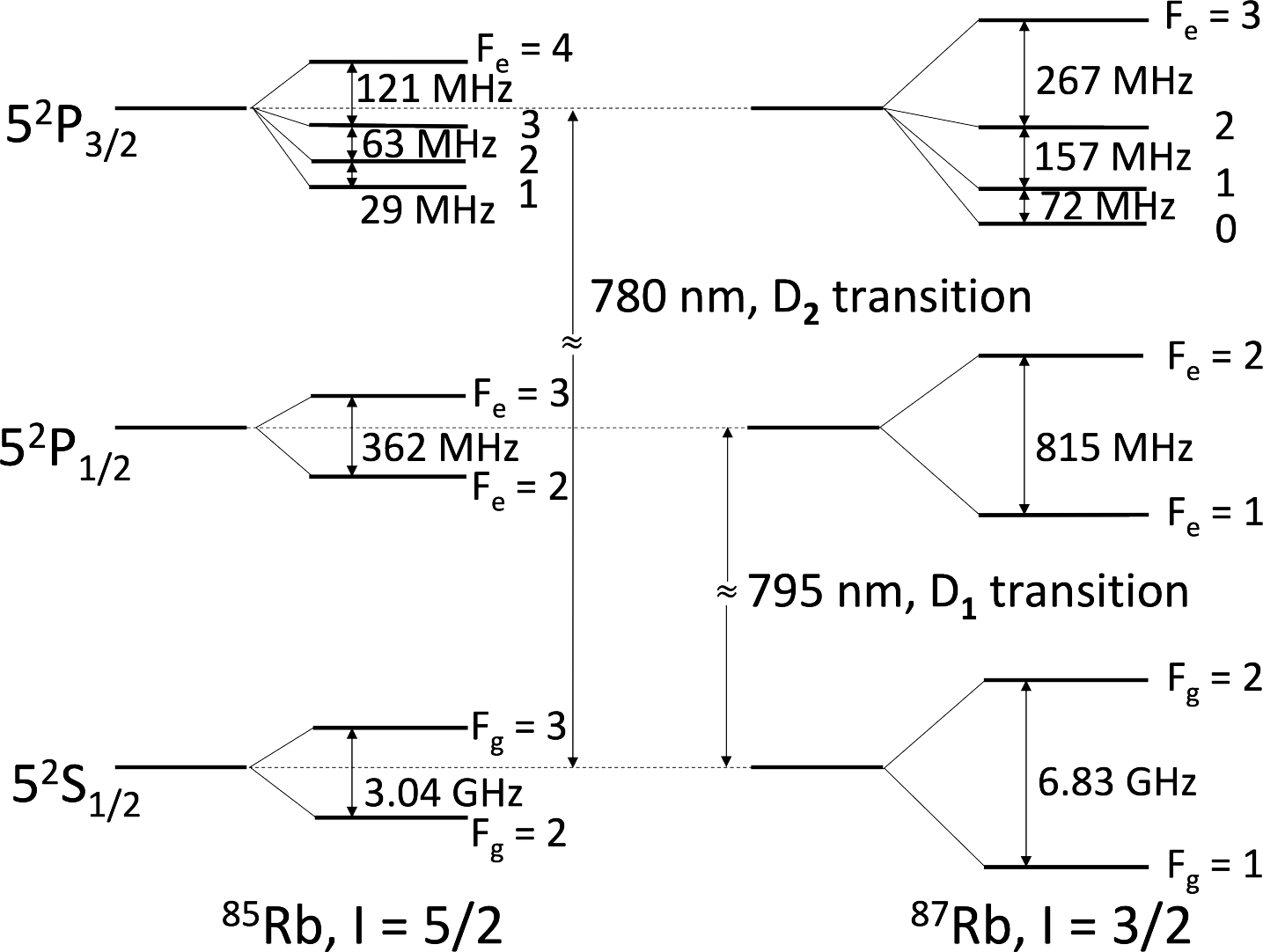}}
%{FigRbenergy_levels.pdf}}
\caption{\textsf{Rb energy-levels with hyperfine structure. 
}} \label{fig:Rb_energylevels}
\end{figure}

In the lab, Rubidium is a popular choice of alkali, owing to the ready availability of inexpensive single mode diode lasers at the resonance wavelength for transitions from the $5^{2}S_{1/2}$ ground state to the $5^{2}P_{1/2}$ excited state (D1-transition; 795 nm) or the $5^{2}P_{3/2}$ excited state (D2-transition; 780 nm). Fig.~\ref{fig:Rb_energylevels} shows the hyperfine energy-level structure for both stable isotopes $^{85}$Rb (nuclear spin I $ = 5/2$) and $^{87}$Rb (I $ = 7/2$).~\cite{steckRb}
%As shown in Fig.~\ref{fig:Rb_energylevels}, %When we account for electron spin-orbit interaction, the hydrogenic states $5S$ and $5P$ are converted to $5S_{1/2}$ and $5P_{1/2}, 5P_{3/2}$ energy-levels - this is the \emph{fine} structure. The interaction between electron angular momentum and nuclear spin yields the \emph{hyperfine} structure, denoted by the $F$-quantum numbers. A weak magnetic field causes Zeeman splitting of each of these $F$-levels into $2F+1$ magnetic sub-levels with quantum numbers $-m_F, -m_F+1,...m_F-1, m_F$ (not shown).

Our experiments are performed on the D1 transition $5^{2}S_{1/2}, F_{g} = 2 \rightarrow 5^{2}P_{1/2}, F_{e} = 1$ in $^{87}$Rb atoms. The hyperfine D1 transition for $^{87}$Rb involves the $5^{2}S_{1/2}$ ground state and the $5^{2}P_{1/2}$ excited state, which are split by the interaction of the valence electron's total angular momentum (orbital + spin) with the nuclear spin, creating $F_{g} = 1, \, 2$ ground energy states (separated by 6.8
%6.8347 
GHz), and $F_{e} = 1, \, 2$ excited energy states (separated by 815 MHz). Each hyperfine state $F$ has $2F+1$ degenerate substates with quantum number $m_F$ ranging from $m_F = -F$ to $F$ in increments of 1. %where the subscripts $g$ and $e$ are used to denote Zeeman sub-levels in the ground and excited states, respectively. In total, there are 8 ground state levels and 8 excited state levels.
Measurement will find the atom with an angular momentum component $m\hbar$ along a chosen quantization axis - we choose $z$ as our axis of quantization. For the $F_g = 2$ ground state, for example, five magnetic substates are revealed in the presence of a weak external magnetic field $B_z$ (along $z$) 
which lifts the degeneracy by imparting a Zeeman shift ${\Delta} E = g_{f}\mu_{B}m_{F_g} B_z$ to each magnetic sub-level. Here $g_{f}$ is the Lande g-factor for the hyperfine state ($g_f=$1/2 for $F_g = 2$), $\mu_{B}$ is the Bohr magneton ($9.274 \times 10^{-28}$ Joules/Gauss), 
%$B_z$ is the external magnetic field magnitude in Gauss 
and $m_{F_g}$ is the magnetic sub-level number. The Zeeman shifts between the magnetic sub-levels of the $F_g = 2$ ground state are 0.7 kHz/mG.~\cite{steckRb}

A pump or probe laser tuned near-resonance to a particular $F_g \rightarrow F_e$ transition pumps other close-lying $F_e$ states off-resonantly, creating extraneous channels for atomic population leakage, which diminishes the steady-state fraction of atoms in the dark state. 
%For this reason, the 87 isotope is chosen over 85 because the hyperfine levels in $^{87}$Rb are farther spaced, see Fig.~\ref{fig:Rb_energylevels}. 
%The D1 transition is preferred to D2 for the same reason with regard to the excited state splittings. 
{The $^{87}$Rb D1 transition is advantageous in that respect because the $F_e$ level-separation of 815 MHz is largest}, compared to that of the D1 transition in $^{85}$Rb (362 MHz), and of the D2 transitions (Fig.~\ref{fig:Rb_energylevels}). \textcolor{black}{See Sec.~\ref{sec:optsetup}\ref{subsec:tune} for further details.}

\subsection{Zeeman EIT: Spin polarization via optical pumping}
\label{subsec:zeemaneit} 
\begin{figure}[h]
\centerline{
\includegraphics[width=8cm]{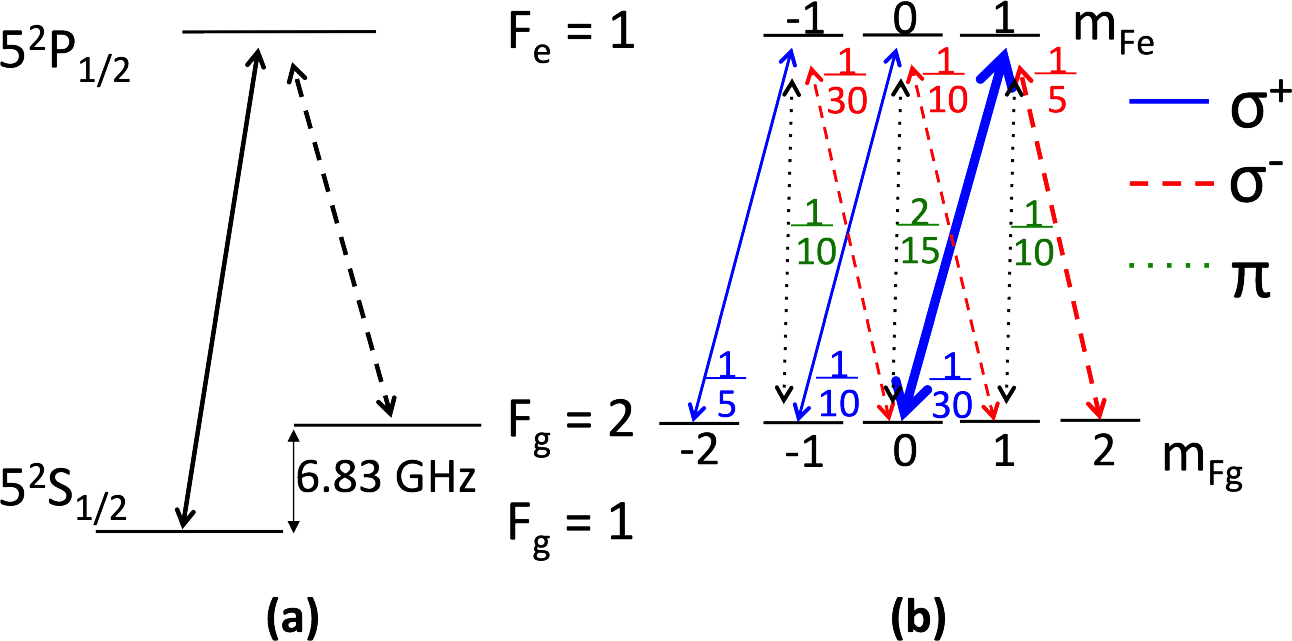}}
%{fig5_hypzeeEIT_rev.pdf}}
%%\vspace{-4mm}
\caption{\textsf{Three-level lambda schemes for a) hyperfine EIT and b) Zeeman EIT. 
In (b) the transition strength between a pair of magnetic sub-states depends on both the light polarization and the magnetic quantum numbers of the two levels involved. The relative strengths are indicated 
%as numerical fractions
%~\cite{steckRb,budker}. 
%In (b), relative transition strengths are indicated 
%by Clebsch-Gordan (CG) coefficients 
for $\sigma^+$ (solid lines), $\sigma^{-}$ (dashed lines), and linear ($\pi$; dotted lines) polarizations of light.~\cite{steckRb,budker1} The ground states $m_{F_{g}} = 2, 0$ and the excited state $m_{F_{e}} = 1$ approximate the three levels $|1\rangle, |2\rangle, |3\rangle$ in Fig.~\ref{fig:fig1}(a), respectively. }} 
\label{fig:hypzeeEIT}
%%\vspace{-4mm}
\end{figure}

Three-level schemes with alkali atoms
%as depicted in Fig.~\ref{fig:fig1}(a) 
for EIT-based slow light experiments fall into two categories:
% involving optical transitions of the D lines of alkali metal atoms such as Rb or Cs: 
 Hyperfine EIT and Zeeman EIT.  
% As shown in Fig.~\ref{fig:hypzeeEIT}(a), the hyperfine $5^{2}S_{1/2}$ ground state for the D1 transition is split by the interaction of the valence electron's total angular momentum (orbital + spin) with the nuclear spin, creating $F_{g} = 1, \, 2$ ground energy states separated by 6.8
%%6.8347 
%GHz. 
Hyperfine EIT with $^{87}$Rb is depicted in Fig.~\ref{fig:hypzeeEIT} (a). A strong pump beam (coupling laser; solid line) is tuned to one $F_g \rightarrow F_e$ transition, while a weak beam (probe; dashed line) is tuned to the other transition, 6.83 GHz apart. The probe is created by splitting off a small amount  from the coupling beam, using an expensive electro-optic modulator. 
% see Mason Klein thesis pgs 26 and 39
%to split off a weak sideband frequency (probe; dashed line), 6.83 GHz apart from the main frequency (coupling beam; solid line), each beam tuned to an $F_g \rightarrow F_e$ transition. 
Alternatively, one may employ two phase-locked lasers, each tuned to either transition. 
Zeeman EIT, depicted in Fig.~\ref{fig:hypzeeEIT} (b), is less resource-intensive, requiring a single passively-stable laser system and less expensive acousto-optical modulators. Hence, we 
%do not discuss hyperfine EIT further, choosing instead to 
confine our attention to this method. 
%In Zeeman EIT, each hyperfine level $F$ is split into $2F + 1$ magnetic sub-states by a small magnetic field $B_z$ created from a solenoid along its longitudinal $z$-axis (see Sec.~\ref{sec:chamber} below). 
%We refer to $z$ as the axial, or longitudinal, direction. %To the best of our knowledge, slow light IN ZEEMAN EIT has only ever been demonstrated in D1 transition of 87Rb, not in D2 transitions, or in the D1 transition of $^{85}$Rb(?) 
Zeeman EIT relies upon the optical pumping of the magnetic Zeeman sub-levels created by applying $B_z$, using a strong coupling beam and a weak probe beam of mutually orthogonal circular-polarization (one beam $\sigma^{+}$, the other $\sigma^{-}$). 
%(linear polarization is denoted as $\pi$), 
Both beams propagate collinearly with $B_z$.
%, as described below. 

%I think this paragraph should be reorganized for clarity. I suggest the following outline:
%1. There are dipole selection rules valid for specific polarizations (give a ref + mention section S4)
%2. Describe the pumping
%3. Say a word on the relative transition rates, and how they are determined (or give a ref)
%4. Insert the sentence "Eventueally...sub-states"
%5. Describe fig 4b
%6. Verify the assumptions of Sec 2
%7. Discuss the lifetimes of the excited states vs the pumping times for Rb and for other alkali
%8. Introduce the discussion on the spin relaxation

Fig.~\ref{fig:hypzeeEIT}(b) depicts a strong $\sigma^{+}$ coupling field (solid arrows) and weak $\sigma^{-}$ probe field (dashed arrows), tuned to near-resonance with the $5^{2}S_{1/2}, F_g = 2 \rightarrow 5^{2}P_{1/2}, F_e = 1$ transition.~\cite{steckRb,budker1} 
%The transition strength between a pair of magnetic sub-states depends on both the light polarization and the magnetic quantum numbers of the two levels involved (the relative strengths are indicated as numerical fractions)~\cite{steckRb,budker}. 
The dipole selection rules for transitions induced between magnetic substates by $\sigma^+ (\sigma^-)$ light are $\Delta m = +1 (-1)$ for absorption and $\Delta m = -1 (+1)$ for stimulated emission, in addition to the $\Delta m = 0, \pm 1$ selection rule for spontaneous emission transitions.~\cite{steckRb} 
This means that after many absorption-emission cycles have occurred, the strong $\sigma^{+}$ field optically pumps the atoms into the $m_{F_{g}} = 0$ sub-state. From here, the pump excites the atoms to the $m_{F_{e}} = 1$ sub-state, which is followed either by spontaneous decay to the $m_{F_{g}} = 1, 2$ spin sub-states (where they stay) or by emission back into the $m_{F_{g}} = 0$ state (in which case they are re-excited). \textcolor{black}{Eventually, after a time that is long compared to the excited state lifetime of 27.7 ns for the Rb D1 line, the atoms accumulate into the $m_{F_{g}} = 1, 2$ spin sub-states (typically, after pumping on a time-scale of ms).} The action of the weak $\sigma^{-}$ probe is to optically pump some of these atoms 
%that collect in the $m_{F_{g}} = 1, 2$ sub-statesafter a time that is long compared to the excited state lifetime (27.7 ns for the Rb D1 line)
%(thick dashed arrow in Fig.~\ref{fig:hypzeeEIT}(b)) 
%pumps atoms in the $m_{F_{g}} = 1$ state are optically pumped 
%toward states with 
toward lower $m_{F_{g}}$-states.
%, thus preventing any atoms from piling up in the $m_{F_{g}} = +1$ sub-state.
Eventually, in the steady-state, the atomic population is concentrated in the $m_{F_{g}} = 0, 2$ spin sub-states. Thus the atom is 
%said to be 
``spin-polarized" via optical pumping.~\cite{budker}  
%\textcolor{black}{Optical pumping times on a time-scale of milliseconds are typically used.}
% for spin-polarizing alkali atoms such as Na, Rb, and Cs (with excited state lifetimes of 16 ns, 27 ns, and 31 ns, respectively). 
%, as shown in Fig.~\ref{fig:hypzeeEIT}(b). 

Indeed, the ideal three-level lambda system (states $|1\rangle$, $|2\rangle$, and $|3\rangle$ 
%illuminated by strong pump and weak probe 
in Fig.~\ref{fig:fig1}) is well-approximated by the ground states $m_{F_{g}} = 2, 0$ and the excited state $m_{F_{e}} = 1$. In Fig.~\ref{fig:hypzeeEIT}(b) the pump (thick solid arrow) and probe (rightmost dashed arrow) are highlighted. The method for fine-tuning the pump and probe laser frequencies to resonance with these particular Zeeman sub-levels is described in Sec.~\ref{sec:optsetup}\ref{subsec:tune}. These approximate three-level atoms then evolve into the dark state, enabling EIT and slow light. Factors causing spin relaxation that reduce the dark state population are discussed in Sec.~\ref{sec:3lvlexpt}B below.
% and brought back to the right hand side of the diagram. 
%where they can participate again in the three-level system transitions. 

Note that the assumption made at the beginning of Sec.~\ref{sec:theory}, that the coupling beam only addresses levels $|2\rangle$ and $|3\rangle$, and the probe beam only addresses levels $|1\rangle$ and $|3\rangle$, is automatically satisfied in Zeeman EIT because $\sigma^{+}$ pump photons cannot be absorbed by the $|m_{F_{g}} = 2\rangle \rightarrow |m_{F_e} = 1\rangle$ transition and $\sigma^{-}$ probe photons cannot be absorbed by the 
$|m_{F_{g}} = 0\rangle \rightarrow |m_{F_e} = 1\rangle$ transition. 
%$|0\rangle \rightarrow |1\rangle$ transition. 

%Optical pumping times of $\sim$ 0.5 - 1 $\mu$s are adequate for ``spin-polarizing" in this manner alkali atoms such as Na, Rb, and Cs (with excited state lifetimes 16 ns, 27 ns, and 31 ns, respectively). 
%The dipole matrix element $|\mu_{13}| = CG_{13} \, |\mu_0|$, where $|\mu_0| = 2.54 \times 10^{-29}$ Cm and for the spin-polarized atom $CG_{13} \approx 1$. 

%Once the atoms are spin-polarized by optical pumping to approximate a three-level system they 

%We introduced the theory behind the dark state in Sec.~\ref{sec:theory}\ref{subsec:dark} and remarked toward the end of that subsection that, practically speaking, EIT contrast and linewidth and hence slow light effects are limited by the fraction of atoms that actually settle in the dark state. 
%One of the limiting factors is spontaneous emission, which we neglected in the theoretical discussion of Sec.~\ref{sec:theory}\ref{subsec:dark}. Note that spontaneous emission from the excited $m_{F_{e}} = 1$ state down to the $m_{F_{g}} = 1$ state in Fig.~\ref{fig:hypzeeEIT}(b) causes  atomic population to ``fall out" of the three-level system which is comprised of the $m_{F_g} = 0, 2$ and the $m_{F_e} = 1$ states. However, as depicted in Fig.~\ref{fig:hypzeeEIT}(b), this problem is mitigated by the process of optical pumping:    
%\textcolor{black}{Is there an easy way to see why conditions for slow light are more stringent than those for EIT?.}
%%\vspace{-5mm}
\subsection{Spin relaxation mechanisms: Role of buffer gas}
\label{subsec:broaden} 
There are several spin depolarization mechanisms that reduce the EIT contrast and affect its linewidth.  The spin-polarized Rb atoms prepared in the previous subsection exist within the sample volume that is jointly illuminated by the pump and probe beams. In a glass cell filled with warm alkali vapor, just the central part of the cell is typically illuminated. The atoms are moving at thermal speeds, causing the polarized atoms to transit through the laser beams and exit the illuminated volume. A collision with the glass wall destroys their polarization~\cite{happer}, and the unpolarized atoms may subsequently re-enter the illuminated region. These transit-time effects reduce the sample polarization.~\cite{jet}
Further, there are 
spin-exchange collisions between spin-polarized Rb atoms which can redistribute the populations in the magnetic sub-states, 
%thus rapidly destroying the spin polarization
causing depolarization.~\cite{happer} 

There are two commonly used approaches~\cite{budker} to suppress the spin-depolarizing effects described above: i) include an inert buffer gas such as Ne, He, or Ar in the Rb vapor cell, or ii) coat the inside of the glass vapor cell with an anti-relaxation coating such as paraffin. 

Coated cells can yield significantly longer spin coherence times than buffer gas cells because the paraffin coating suppresses the Rb spin-depolarization
%preserves the Rb spin polarization 
via ``softer" atom-wall collisions. However, the light-Rb atom interaction dynamics in coated cells is more complicated, yielding a dual-structured EIT lineshape with a narrow central peak sitting atop a broad pedestal (see for example Ref.~\cite{irina} and references cited therein). The broad pedestal is due to Rb atoms interacting with light during a one-time pass through the light beams. The narrow peak, on the other hand, results from contributions by polarized Rb atoms that transit in and out repeatedly through the laser beams, interacting coherently each time with the pump and probe fields. These atoms make many wall collisions before eventually becoming depolarized. The linewidth of the narrow peak is significantly affected by factors such as the coating quality and cell geometry.~\cite{irina}
In the work reported here we do not use coated cells. 

In buffer gas cells, the usual practice for EIT and slow light experiments is to mix noble gas (up to several tens of Torr) with a few microTorr of Rb.~\cite{irina} 
In our experiments we use 10 Torr of Ne. 
Frequent Rb-Ne collisions reduce the Rb atoms' mean free path to values much smaller than the pump beam diameter~\cite{arimondo}, thus confining the atoms within the illuminated volume. Instead of flying through the laser beam at thermal speeds in the absence of a buffer gas, the atoms then slowly diffuse through, thereby increasing the laser-atom interaction time by several orders of magnitude.~\cite{irina} Note that Rb-Ne collisions cause the spin-polarized Rb atom's velocity to change, but with negligible spin relaxation. Such collisions that rapidly redistribute the atomic velocities without re-equilibrating the populations of the atomic levels are termed ``velocity-changing collisions" in the literature.~\cite{budker,arimondo} Rb-Rb and Rb-wall collisions are negligible compared to Rb-Ne collisions. 
Still, decoherence effects remain that are not addressed by the buffer gas (or by wall coatings).
Inhomogeneities in the magnetic field used for Zeeman EIT cause spatial variation of the dark state leading to absorption. In \textcolor{black}{Sec. S2} of the Supplementary Materials we discuss how to suppress these inhomogeneities. Furthermore, an incoherent pumping mechanism, known as radiation trapping, may become significant at high atomic densities. Radiation trapping refers to the reabsorption of spontaneously emitted photons within the illuminated volume, and is expected to become significant inside our probe beam radius  at number densities $N \geq 5 \times 10^{11}$/cm$^3$ (see Sec.~\ref{sec:optsetup}~\ref{subsec:beamsize}).~\cite{welch} Hence, we keep our number density below this value (see Sec.~\ref{sec:chamber} and 
Sec. S2 D).

%%\vspace{1mm}
%\noindent \emph{Power broadening}:
%\newline 
%%and~\ref{eq:highchi1} 
%Evidently, from Eqn.~\ref{eq:eitlinewidth} (along with Eqn.~\ref{eq:chi2_I}) and Eqn~\ref{eq:v_g1} (along with Eqn.~\ref{eq:cond2}), in the high coupling-intensity limit the group velocity $v_g$ of the slowed light pulse depends linearly on coupling intensity $I$, as does the EIT linewidth, i.e., power-broadening occurs. %(see the discussion at the end of Sec.~\ref{sec:theory}\ref{subsec:eit}). 
%On the other hand, lowering the coupling intensity too far (while still staying in the high coupling intensity limit) may reduce the steady-state fraction of atoms in the dark state to the point where slow light effects start to degrade and the slowed light pulse starts to speed up again. Thus we may expect a ``sweet spot" in the coupling intensity where the conditions for slow light are optimized.
%%in the regime where conditions~\ref{eq:cond1} and~\ref{eq:cond2} are satisfied \textcolor{black}{(need to check Condition~\ref{eq:cond2} for our expt)}, 
%%the optimum conditions for slow light are obtained by balancing the coupling intensity on one hand and the EIT linewidth on the other. 
%This is borne out by experimental data in Sec.~\ref{subsec:delay}~\ref{subsec:slowvsI} below. 

%Note that we neglect radiation trapping effects because they are expected to be significant at an atomic density at least an order of magnitude larger than the density $\sim 3 \times 10^{11}$/cm$^3$ used in our experiments~\cite{welch}.  

\subsection{Slow light conditions; estimate of $\gamma_{12}$, $\gamma_{13}$}
%\subsection{Our sample: Estimating decoherence rates $\gamma_{12}$, $\gamma_{13}$}
%\subsection{Conditions for slow light: The numbers} 
\label{subsec:condition}

%Our sample consists of a few $\mu$Torr of isotopically enriched dilute $^{87}$Rb vapor mixed with 10 Torr of Ne buffer gas. 
%Buffer atoms collide frequently with the Rb, keeping illuminated Rb atoms confined within the pump beam
%The frequent Rb-Ne collisions reduce the Rb atoms' mean free path to much smaller than the pump beam diameter
%(see Sec.~\ref{sec:3lvlexpt}B). }
%to render them nearly (but not completely) immobile.}
%suppress broadening of the EIT linewidth due to motion of the Rb atoms, as discussed later in Sec.~\ref{sec:3lvlexpt}\ref{subsec:broaden}. }
%It is well-known that, if spontaneous emission is the dominant dephasing mechanism, the coherence decay rate is half that of the population decay rate~\cite{decoh}, i.e., $\gamma_{13} = \gamma_{23} = \Gamma/2$. 
Let us examine how the conditions for slow light, Eq. (\ref{eq:cond1}) and Eq. (\ref{eq:cond2}), are satisfied in our experiments.
The relation between the Rabi frequency $\Omega_c$ and coupling beam intensity $I$, in the case of the spin-polarized atom in Fig.~\ref{fig:hypzeeEIT}(b), is defined \textcolor{black}{through the well-known expression for the saturation intensity $I_{sat}$}:~\cite{factor2,eberly_budker,steckRb}
\begin{equation}
\frac{{(2 \, |\Omega_c|})^2}{{\Gamma}^2} \equiv \frac{I}{2\,I_{sat}}\,, \,\, \,\,\, \mbox{where} \,\, I_{sat} = \frac{ \pi h c }{ 3{\lambda}^3 } {\Gamma}.
\label{eq:chi2_I}
\end{equation}
Here $\lambda$ is the optical wavelength for the transition, 
%for the $2 \leftrightarrow 3$ transition 
%$I_{sat}$ denotes the saturation intensity, 
and $\Gamma$ is the natural linewidth due to spontaneous emission.~\cite{steckRb} 
$I_{sat}$ is the excitation intensity at which the stimulated emission rate is half the spontaneous emission rate. 
For the $^{87}$Rb D1 transition, $\Gamma = 2\pi \, (5.75$ MHz) and $\lambda = 794.98$ nm ($\omega_p/2\pi = 3.77 \times 10^{14}$ Hz), yielding $I_{sat} =1.5$ mW/cm$^2$.~\cite{steckRb} Our pump intensity $I$ mostly varies between 1.25 and 5.6 mW/cm$^2$
 %{0.15 and 5.6 mW/cm$^2$} 
 (see Sec.~\ref{sec:optsetup}~\ref{subsec:beamsize}), which means $|\Omega_c|/2\pi$ ranges from 1.9 MHz
 to 
 3.9 MHz. 
 
The ground state decoherence rate $\gamma_{12}$ arises from Rb-Rb collisions, Rb collisions with the cell walls, and from Rb atoms eventually diffusing out of the illuminated volume despite the buffer gas. Collisions with the inert buffer gas do not contribute significantly to $\gamma_{12}$. However, since our experiments use Zeeman hyperfine levels, inhomogeneities in the magnetic field contribute to $\gamma_{12}$. Good suppression of stray magnetic fields (\textcolor{black}{Sec. S2, Supplementary Materials}), results in a value $\gamma_{12} \approx 3$ kHz, as reported in the literature.~\cite{walsworth2006prl,shuker} 
%We have carefully characterized these contributions following Ref.~\cite{shuker} but a detailed discussion is beyond the scope of this paper - we refer the reader to Ref.~\cite{shuker}. 
%The precise value of $\gamma_{12}$ is not so important for our experiments because the power-broadened component of $\Gamma_{EIT}$ dominates (see Eqns.~\ref{eq:eitlinewidth} and~\ref{eq:eitlinewidth2}). 
Clearly, $\Omega_c \gg \gamma_{12}$, and since $\Delta_p/2\pi \sim 100$ kHz or less (see Figs.~\ref{fig:EIT} (e,f) and~\ref{fig:eitdata})  our experiments fall in the strong coupling intensity regime, i.e., the condition in Eq. (\ref{eq:cond1}) is well-satisfied.

Next, to check that Eq. (\ref{eq:cond2}) is also well-satisfied in our experiments, we note that the dipole matrix element $|\mu_{13}|$ for the $^{87}$Rb D1 transition in our spin-polarized atom is well-approximated by just the magnetic quantum number-independent value (known as the reduced matrix element) $2.54 \times 10^{-29}$ Cm.~\cite{steckRb} The number density $N$ is $\sim 1.5 - 3.4 \times 10^{11}$/cm$^3$ 
(see Sec.~\ref{sec:chamber}), 
%(from empirical data in Ref.~\cite{steckRb} and the ideal gas law; our vapor temperature ranges from $55^o - 65^o$C). 
yielding 
%$\frac{1}{2\pi}\sqrt{ \frac{N |\mu_{13}|^2} { 2\epsilon_{0} \hbar } \omega_p}  \approx 2 - 4$
$({1}/{2\pi})\sqrt{ {N |\mu_{13}|^2 \omega_p} / ({ 2\epsilon_{0} \hbar }) }  \approx 2 - 4$
GHz, which exceeds the $|\Omega_c|/2\pi$-values above by three orders of magnitude.

%To see this we draw upon the well-known connection between the Rabi frequency $\Omega_c$ and coupling beam intensity $I$~\cite{metcalf,eberly3}: 
%\begin{equation}
%2{|\Omega_c|}^2/{{\Gamma}^2} \equiv I/I_{sat}\,, \,\, \,\,\, \mbox{where} \,\, I_{sat} \equiv \frac{ 2\pi^{2}\hbar c }{ 3{\lambda}^3 } {\Gamma}
%\label{eq:chi2_I}
%\end{equation}
%Here $\lambda$ is the optical wavelength for the $2 \leftrightarrow 3$ transition and the saturation intensity $I_{sat}$ is defined as the excitation intensity at which the stimulated emission rate is half the spontaneous emission rate $\Gamma$. 

Finally, we estimate the excited state damping $\gamma_{13}$ which arises from 
the spontaneous emission rate 
$\Gamma$, the dephasing $\gamma_{coll}$ due to Rb collisions with the buffer gas, and the Doppler broadening $\Gamma_D$. 
Recall from Eq.~\ref{eq:alpha_chi0} and Fig.~\ref{fig:EIT} (a) that $2\gamma_{13}$ is manifested as the width of the absorption peak.  
Our sample consists of a few $\mu$Torr of isotopically enriched dilute $^{87}$Rb vapor mixed with 10 Torr of Ne buffer gas.
To estimate $\gamma_{coll}$ we follow the empirical relation $2\gamma_{coll}/2\pi = 9.84$ MHz/Torr derived in Ref.~\cite{rotondaro}. We see that, for 10 Torr of Ne, $\gamma_{coll}/2\pi \sim 50$ MHz. On the other hand, the Doppler broadening of the absorption profile of Rb vapor, illuminated by a frequency-scanning monochromatic beam, arises due to the thermal motion of the atoms.~\cite{eberly2} In a simple 1D situation, an atom moving with thermal velocity $v_{th}$ toward or away from a laser beam of frequency $\omega_0$ sees a Doppler-shifted frequency $\omega_0 (1 \pm v_{th}/c)$. This means that for a thermal distribution of velocities we may estimate the full width at half maximum (FWHM), $\Gamma_D$, of the Doppler broadened absorption profile to be $\Gamma_D = 2\, \omega_0 \, v_{th}/c = 2k \, v_{th}$. Here 
%we have used $c/\nu_0 = \lambda = 2\pi k$ where $\lambda$ is the wavelength, and 
$k = 2\pi/\lambda$ is the magnitude of the incident field wave vector. \textcolor{black}{If we use the most probable velocity $v_{th} = \sqrt{(2 {\mbox{ln}}2) \, k_B T / m}$, we obtain $v_{th} \approx 200$ m/s at $T = 293\,$K, yielding $\Gamma_D/2 \approx 2\pi (250$ MHz) at $\lambda = 795$ nm.
%comparable to the 815 MHz splitting between the $F_e = 1$ and 2 levels. 
Thus $\gamma_{13}$ is dominated by $\gamma_{coll}$ and $\Gamma_D/2$ (both are significantly larger than $\Gamma$). We use their sum as an estimate for $\gamma_{13}$.}
Note from Eq. (\ref{eq:eitlinewidth2}) and Eq. (\ref{eq:v_g1}), that, 
in the high coupling intensity limit,
both $v_g$ and $\Gamma_{EIT}$ decrease linearly with the intensity.
%(see the discussion at the end of Sec.~\ref{sec:theory}\ref{subsec:eit}). 
However, lowering the intensity too far (although still in the high intensity limit) may reduce the steady-state population in the dark state enough that slow light effects start to degrade, causing the probe pulse to distort and appear to speed up again. Thus we may expect a sweet spot in the pump intensity, where conditions for slow light are optimized (see Sec.~\ref{sec:delay}~\ref{subsec:slowvsI}). 
\section{\textcolor{black}{Experimental} setup}
\label{sec:optsetup}
\begin{figure*}[h]
%%\vspace{-32mm}
\centerline{
\includegraphics[width=18cm]{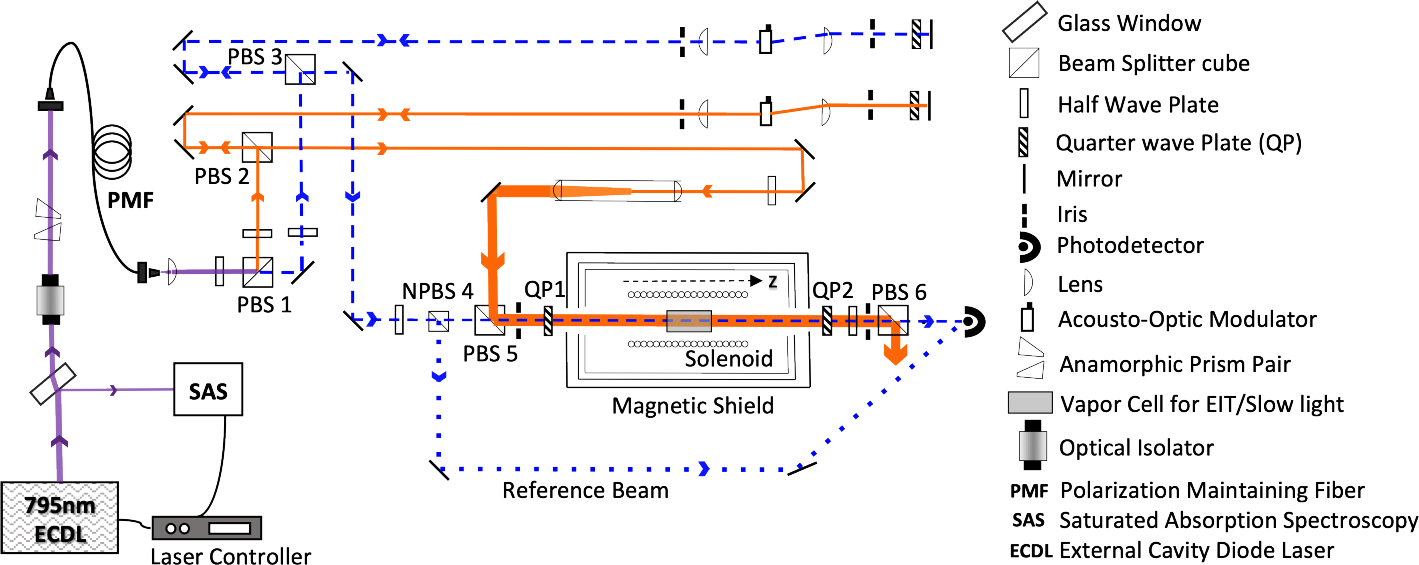}}
%{opticallayout_latest2.pdf}}
%%\vspace{-30mm}
\caption{\textsf{Optical setup. The laser is split at a polarizing beamsplitter (PBS 1) into a strong pump, or coupling, beam (solid lines) and a weak probe beam (dashed lines). Before entering the vapor cell the probe beam is split at a non-polarizing beamsplitter (NPBS 4) to create a reference beam (dotted lines) for the slow light measurement. The $z$-axis is indicated.}} \label{fig:optlayout}
\end{figure*}

We describe how we create pump, probe, and reference beams, tuned to the appropriate alkali transitions for Zeeman EIT shown in Fig.~\ref{fig:hypzeeEIT}(b), and of the required size, intensity, frequency, temporal width, and polarization purity. Our optical layout is depicted in Fig.~\ref{fig:optlayout}. 

%\vspace{-2mm}
\subsection{Laser source}
\label{subsec:opticalayout}

We use a commercial external cavity tunable diode laser (ECDL) to provide 
% 30-35 mW of 
linearly polarized continuous-wave light at $\sim$795 nm. A few percent of the output is split off for use in a saturated absorption spectroscopy (SAS) set-up to enable the precise tuning of the laser to the $F_{g} = 2 \rightarrow F_{e} = 1$ D1 transition, as described below.
%~\cite{bali2012} % (see Ref.~\cite{bali2012}, for example, for a discussion of SAS).
% by the {\textcolor{black}{thin purple line}. 

The main ECDL output is first passed through a Faraday rotation-based optical isolator (OI) which optically isolates the laser from back reflections (e.g., arising from the optics that couple the light into the fiber). The light then passes through an anamorphic prism pair to circularize the elliptical cross-section of the beam,
% 25 mW here just after OI
before coupling into a single-mode polarization-maintaining fiber which preserves the direction of the linear polarization of the input beam (we measured a residual drift of $< 2$\% in the polarization direction at the fiber output). 
%The Faraday rotator optically isolates the laser from back reflections arising, for example, from the optics that couple the light into the fiber. 
The single-mode fiber is typically aligned and operated in most labs at a transmission efficiency of only $30 - 50$\%, but produces a clean Gaussian spatial profile. Gaussian spatial profiles are highly desirable for pump-probe experiments where the pump and probe beams must be well-collimated, and must have good spatial overlap that is amenable to quantitative modeling.
%yielding good Gaussian pump and probe beams.}
%, and isolates the optics downstream from alignments within the ECDL. 
%However, if one wishes to save money on single-mode fiber launchers, or if laser power is an issue (the insertion loss of the fiber is about 50\%), one may forego the fiber and use free-space beams.}

%\vspace{-2mm}
\subsection{Creating the pump and frequency-scannable probe} 
\label{subsec:pumpprobe}
The output from the fiber ($\sim$ 7 mW) is collimated using
 an aspheric lens. 
%(\textcolor{black}{lens in port, so no cage needed?}) 
%and a plano-convex lens ($f = 15$ cm) - 
%The fiber and lens are mounted in a cage for convenience. 
The collimated light is divided by a polarizing beamsplitter (PBS 1) into two orthogonally linearly polarized beams - the strong coupling (or pump) beam (solid line) and weak probe beam (dashed line). 

In pump-probe spectroscopy the probe is swept in frequency around the fixed pump frequency, and the probe transmission spectrum is detected. The sweeping is accomplished with an acousto-optic modulator (AOM), which is a device that uses a sound wave propagating through a crystal to form a diffraction grating for incoming light waves, producing frequency-upshifted ($+1$) and downshifted ($-1$) orders, in addition to the $0$ order at the incident frequency.~\cite{saleh,ajp2011} 
The $\pm 1$ orders emanate at a slight angle to the direction of the incident beam, symmetrically on either side, while the $0$ order continues along the incident beam direction. By carefully aligning the angle of incidence of the beam on the AOM, we achieve about $70 \%$ of the incident power in the first order of choice, with most of the remaining power in the $0$ order, while a small amount leaks into the undesired first order.  The AOM performs a frequency-scan by introducing a fixed offset (80 MHz in our case) and sweeping back and forth about that fixed offset value by some desired amount (chosen by user, anywhere from zero up to a maximum sweep $\pm 20$ MHz for our model). For the purpose of measuring the EIT linewidth in Fig.~\ref{fig:EIT} (c) and (e), a sweep of $\pm$100 kHz suffices (see Fig.~\ref{fig:eitdata}). Because of the fixed frequency-offset of 80 MHz that the AOM introduces, we must insert an identically configured AOM in the path of the pump beam as well, so that both the pump and probe beams are offset in frequency by the same amount. 
To drive the AOMs a dual-output waveform generator which creates twin phase-locked identical 80 MHz, 1 V signals is used. Before feeding to the AOM, each signal is amplified using a standard RF amplifier. 
%The two AOMs are driven by a dual-output waveform generator which creates twin phase-locked identical 80 MHz, 1 V signals that are each amplified using a standard RF amplifier. 
Planoconvex lenses of focal length 30 cm are placed on either side of each AOM, 
%separated by approximately twice their focal length 
with the AOM crystal located at the common focal spot. 

One problem encountered is that the angle of the diffracted orders changes when the output frequency is varied, causing a spatial shift in the beam at the vapor cell located a few feet downstream. This shift is a problem because the probe beam moves off the center of the vapor cell during the course of its frequency scan. To overcome this problem, the desired diffracted order is retroreflected back through the crystal (see top right hand corner of Fig.~\ref{fig:optlayout}), such that the double-shifted order is aligned with the incident beam but in the counter-propagating direction.~\cite{ajp2011} For this double-shifted beam the angular deflection from the second pass cancels the deflection from the first pass. The cancellation is not ideal, due to unavoidable imperfections in alignment, but the spatial shifting of the scanning probe at the site of the vapor cell is highly suppressed in this double-pass AOM configuration. 
The presence of the quarter waveplate just before the retroreflecting mirror in Fig.~\ref{fig:optlayout} serves to orthogonally polarize the double-shifted order with respect to the incident beam, enabling the separation of the double-passed and incident beams at a polarizing beamsplitter (PBS 2 for the pump, PBS 3 for the probe).

%In our experiments The grating is formed by driving the AO crystal with the output from a waveform generator that is connected to an rf-amplifier (\textcolor{black}{see Table in Supplementary Material}). a dual-output waveform generator created twin phase-locked identical 80 MHz, 1 V signals which are each amplified using a standard RF amplifier to drive each AOM with a single-pass efficiency of $\sim 70\%$.  Planoconvex lenses of focal length 30 cm are placed on either side of each AOM, 
%%separated by approximately twice their focal length 
%with the AOM crystal located at the common focal spot. 
%In our experiments, the -1 order is retroreflected by a quarter wave plate and mirror back through the AOM. Thus the -2 order is orthogonally polarized to the incident beam and counter-propagates along the incident beam path. The double-passed and incident beams are separated by a polarizing beamsplitter, PBS 2 for the pump and PBS 3 for the probe in Fig.~\ref{fig:optlayout}.
%The diameter of the two beams is \textcolor{black}{1.13} mm. The probe diameter is left unchanged
% do I need a brief explanation of double-pass mode?

%Details for the fine-tuning of the coupling and probe beams into resonance with their respective transitions are provided in the Supplementary Notes, Sec. S3.

\subsection{Frequency tuning pump and probe to EIT window} 
\label{subsec:tune}
%As mentioned earlier our EIT and slow light experiments are performed on the $m_{F_{g}} = 2, 0$ and $m_{F_{e}} = 1$ Zeeman sub-levels of the D1 transition $5^{2}S_{1/2}, F_{g} = 2 \rightarrow 5^{2}P_{1/2}, F_{e} = 1$ in $^{87}$Rb.
%\subsection{Locating the $^{87}$Rb $F_{g} = 2 \rightarrow F_{e} = 1$ D1 transition}
%\label{subsec:D1}
%%\vspace{-6mm}
The frequency-tuning of the laser for implementing Zeeman EIT is achieved in two steps. First, the laser is tuned to the $5^{2}S_{1/2}, F_{g} = 2 \rightarrow 5^{2}P_{1/2}, F_{e} = 1$ $^{87}$Rb D1 transition, using the the method of 
saturated absorption spectroscopy (SAS). Next, the pump and probe beams are fine-tuned to the $m_{F_{g}} = 2, 0$ and $m_{F_{e}} = 1$ Zeeman sub-levels with the AOMs (see Fig.~\ref{fig:hypzeeEIT}(b)).  

\begin{figure}[h]
\centerline{
\includegraphics[width=6.9cm]{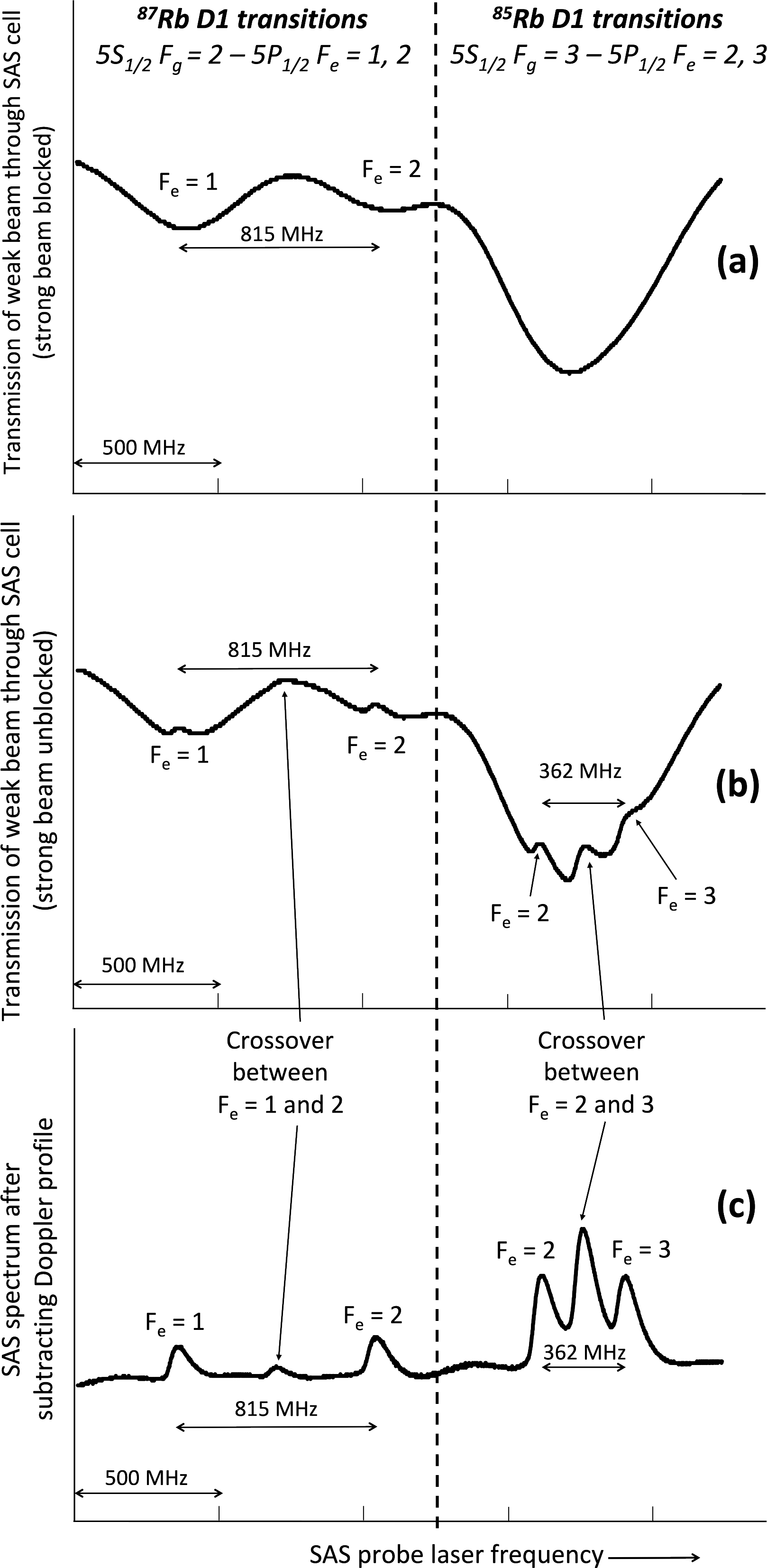}}
%{SB_SAS_Spectra_Improved_Sep16_rev.pdf}}
\caption{\textsf{Tuning the ECDL to the $5^{2}S_{1/2}, F_{g} = 2 \rightarrow 5^{2}P_{1/2}, F_{e} = 1$ $^{87}$Rb D1 transition via saturated absorption spectroscopy. (a) Transmission spectrum for a weak beam propagating through a Rb vapor cell with uncoated walls and no buffer gas (placed inside the box marked SAS in Fig.~\ref{fig:optlayout}). 
%(this cell and the entire SAS setup~\cite{bali2012} is contained 
The Doppler broadened D1 transitions in $^{85}$Rb and $^{87}$Rb are displayed in one continuous scan. The $F_e = 2$ and 3 transitions in $^{85}$Rb are obscured, but the $F_e = 1$ and 2 transitions in $^{87}$Rb are resolved owing to their large separation of 815 MHz. 
%b) The same transmission spectrum, but with a 
(b) A strong counter-propagating beam is introduced, revealing the hyperfine structure, even more clearly in (c) by subtracting away the Doppler component and magnifying the vertical scale.~\cite{bali2012}}} 
\label{fig:spectrum}
%%\vspace{-3mm}
\end{figure}

SAS is performed in a second vapor cell, located inside the box marked SAS in Fig.~\ref{fig:optlayout}, but not explicitly shown. This cell is at room temperature, is filled with natural abundance Rb vapor (72\% $^{85}$Rb, 28\% $^{87}$Rb), has no buffer gas, and the walls are uncoated. The panel in Fig.~\ref{fig:spectrum} shows the transmission spectrum of a weak beam diverted from the ECDL into the SAS cell, where the D1 transitions for the two Rb isotopes are displayed in one continuous scan.
In plot (a), the Doppler-broadened D1 transitions are shown, before SAS is performed. 
Only for the $^{87}$Rb D1 $F_e$ states does the level separation of 815 MHz exceed the 
Doppler broadening $\Gamma_D/2\pi$ for Rb vapor (see Sec.~\ref{sec:3lvlexpt}\ref{subsec:condition}), enabling the two excited states to be resolved. 
The $^{85}$Rb D1 transitions, with their smaller $F_e$ separations, are totally smeared out by Doppler broadening. 
However, when SAS is performed, by introducing in the cell a strong beam that is spatially overlapped with the weak beam but counter-propagating, the absorption is saturated
%(as opposed to the co-propagating pump and probe beams used in the EIT/slow light vapor cell housed inside the magnetic chamber) 
for the velocity-class of atoms that travels in a direction perpendicular to both the strong and weak beams. The hyperfine transitions are revealed as narrow ``holes" that are ``burnt" into the Doppler-broadened absorption spectra  - these holes are manifested as ``bumps" in the transmission spectrum in Fig.~\ref{fig:spectrum}(b). In plot (c) the Doppler component is subtracted away and the vertical scale is magnified. 
%Thus the hyperfine transitions are clearly revealed, now that the Doppler broadening has been mostly suppressed. 
The laser is tuned to the $F_{g} = 2 \rightarrow F_{e} = 1$ $^{87}$Rb D1 peak in plot (c) by reducing the scan to zero, while staying centered on this particular feature, and the goal of SAS is achieved. Note that some atoms that have a velocity component, say, along one of the laser beams, may be down-shifted into resonance with that beam on a particular D1 transition, while being simultaneously up-shifted into resonance with the counter-propagating beam on the higher adjacent D1 transition  - this results in ``crossover" peaks midway between adjacent D1 transitions. See Ref.~\cite{bali2012} for further details on SAS.
%The Doppler-broadened D1 transitions for $^{87}$Rb are resolved in Fig.~\ref{fig:spectrum}(a) because the splitting of 815 MHz between the $F_e = 1$ and 2 states exceeds the Doppler broadening (Sec.~\ref{sec:3lvlexpt}~\ref{subsec:condition}). 

\begin{figure}[h]
\centerline{
\includegraphics[width=7.5cm]{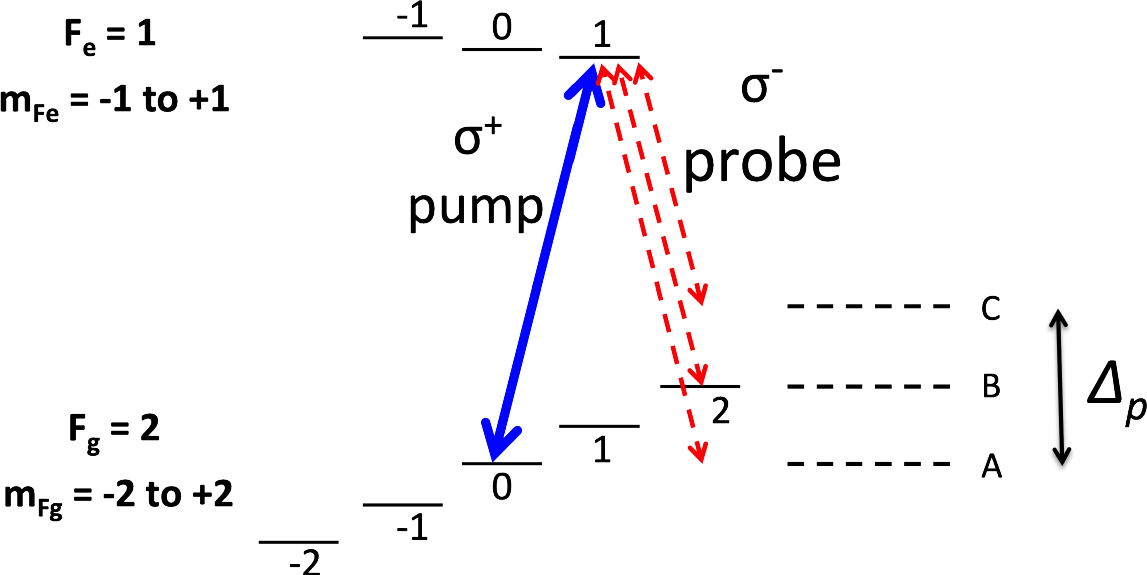}}
%{hypzee_3lvl.pdf}}
\caption{\textsf{Fine-tuning the pump and probe laser frequencies to the $m_{F_{g}} = 0 \rightarrow m_{F_{e}} = 1$ and $m_{F_{g}} = 2 \rightarrow m_{F_{e}} = 1$ sub-levels, respectively. Tuning the ECDL via SAS may put the pump in resonance with the pump transition but leave the probe, which is at the same frequency, detuned from the probe transition by $\Delta_p = 70$ kHz (position A). EIT resonance is achieved by fine-tuning the probe frequency with the probe AOM so that $\Delta_p$ is reduced to zero (position B). 
Scanning the probe frequency symmetrically about B generates the EIT spectra such as shown in Fig.~\ref{fig:eitdata}.}} 
\label{fig:hypzee_3lvl}
%%\vspace{-5mm}
\end{figure}

In order to obtain EIT transmission spectra such as those shown in Fig.~\ref{fig:eitdata}, the frequencies of the pump and probe beams in Fig.~\ref{fig:optlayout} need to be further fine-tuned 
%within the $^{87}$Rb $F_{g} = 2 \rightarrow F_{e} = 1$ D1 transition, 
into resonance with the Zeeman sub-levels $m_{F_{g}} = 0 \rightarrow m_{F_{e}} = 1$ and $m_{F_{g}} = 2 \rightarrow m_{F_{e}} = 1$, respectively (see Fig.~\ref{fig:hypzeeEIT}(b)). This fine-tuning is achieved inside the magnetically shielded vapor cell in Fig.~\ref{fig:optlayout}. This is the Rb-Ne vapor cell in which the slow light experiments are carried out, the one described in Sec.~\ref{sec:3lvlexpt}\ref{subsec:broaden} and~\ref{subsec:condition}.  
The magnetic field $B_z$ for Zeeman splitting is created by the solenoid in Fig.~\ref{fig:optlayout}. Details on the Rb-Ne vapor cell are provided in Sec.~\ref{sec:chamber}, and on the magnetic shielding and solenoid in \textcolor{black}{Sec. S2} of the Supplementary Materials. 
We must take into account the fact that the Zeeman shifts between the magnetic sub-levels of the $F_g = 2$ ground state are 0.7 kHz/mG.~\cite{steckRb} In our experiments, $B_z = 50$ mG, yielding a Zeeman splitting between adjacent ground sub-states of 35 kHz. Because the SAS procedure described above tunes the ECDL frequency before the laser was split into pump and probe, this means that if the pump happens to be in resonance with the $m_{F_{g}} = 0 \rightarrow m_{F_{e}} = 1$ transition, the probe (which is at the same frequency) is detuned from the $m_{F_{g}} = 2 \rightarrow m_{F_{e}} = 1$ probe transition by $\Delta_p = 70$ kHz (position A in Fig.~\ref{fig:hypzee_3lvl}).
Scanning the probe AOM frequency (A $\leftrightarrow$ C) symmetrically about position B in Fig.~\ref{fig:hypzee_3lvl}, where the probe detuning $\Delta_p = 0$ (or more precisely, where the two-photon detuning is zero; see Sec.~\ref{sec:theory}\ref{subsec:window}), yields EIT resonance spectra as displayed in Fig.~\ref{fig:eitdata}. For the slow light experiments, the scan is reduced to zero while staying centered at B. 
%: Points A and B are located 70 kHz apart, which explains why $\Delta_p$ spans $\approx + / - 70$ kHz in Fig. 8. 

It is worth noting that, before the strong (co-propagating) pump is turned on, the weak probe experiences an absorption profile in the Rb-Ne vapor cell that resembles Fig.~\ref{fig:spectrum}(a). In Sec.~\ref{sec:3lvlexpt}\ref{subsec:condition} we estimated the linewidth $\gamma_{13}$, arising from Doppler and collisional broadening in the Rb-Ne cell, for the D1 transitions to be comparable to those shown in Fig.~\ref{fig:spectrum}(a) (see Fig.~\ref{fig:EIT}(a)).
% from Eqn.~\ref{eq:alpha_chi0}). 
Turning on the strong co-propagating pump activates the narrow EIT window predicted by Fig.~\ref{fig:EIT}(b), and observed in Fig.~\ref{fig:eitdata}. 
It is clear from Fig.~\ref{fig:spectrum}(a) that even though the $F_{g} = 2 \rightarrow F_{e} = 1$ $^{87}$Rb D1 transition is advantageous for slow light experiments because this particular transition has the least spectral overlap with neighboring D1 transitions, some residual overlap of the $F_e = 1, 2$ levels is visible. This suggests it is likely there is leakage of atoms from the $F_e = 1$ to the $F_e = 2$ level in the Rb-Ne vapor cell, resulting in a diminished dark state population and a reduced EIT contrast. 

\subsection{Pump and probe beam size and pump intensity}
\label{subsec:beamsize}
The probe beam is Gaussian with a $1/e^2$-radius of 1.13 mm (the distance from the center of the beam where the intensity drops to $1/e^2$ of its  value at the center). To approximate an ideal plane wave (see Sec. 1 in Supplementary Notes) the pump beam, which is also Gaussian, is expanded to a $1/e^2$-radius of 2.26 mm
using a telescope comprising two simple plano-convex lenses mounted in a cage assembly (see Fig.~\ref{fig:optlayout}). 
%(of \textcolor{black}{focal lengths} $f_{1}$,$f_{2} = x,y$ cm, respectively, separated by the sum of their focal lengths). 
%just before entering the vapor cell. 
The pump intensity was mostly varied between {1.25 and 5.6 mW/cm$^2$}, though in a few cases it was as low as {0.15 mW/cm$^2$} (compare to $I_{sat}$, see Sec.~\ref{sec:3lvlexpt}\ref{subsec:condition}).
%For comparison, $I_{sat}$ is 1.5 mW/cm$^2$ for the $^{87}$Rb D1 transition (Eqn.~\ref{eq:chi2_I}). 
%\textcolor{black}{Something about pump and probe beam size and divergence here? pgs. 32 - 34 Ken's thesis}

%\vspace{-1mm}
\subsection{Probe and reference pulses and probe intensity}
\label{subsec:pulses}
In our experiments, the probe beam is in the form of a short (temporal) Gaussian pulse. Slow light demonstrations consist of measurements of the delay of the probe pulse propagating through the sample relative to an identical reference pulse propagating along a similar path-length outside the sample. This reference pulse, depicted by the dotted lines in Fig.~\ref{fig:optlayout}, is split off from the probe pulse by inserting a non polarizing beamsplitter NPBS 4 into the probe path, as shown. The probe and reference pulses are created by amplitude modulation of the probe AOM: \textcolor{black}{The probe RF amplifier output is pulsed on and off by pulsing the input from its waveform generator.} It is straight-forward to measure time-delays between the centers of temporal Gaussian pulses.

Matching the reference pulse path-length to the probe path is a loose requirement because it takes light only about a nanosecond to travel 30 cm in air - this is negligible compared to the delay induced by the slowing down of the light in our experiment (tens of microseconds, see Sec.~\ref{sec:delay}).
% before the probe enters the vapor cell. 
%\begin{figure}[h]
%\centerline{
%\includegraphics[width=8cm]{exptslowlight.pdf}}
%\caption{\textsf{Reference pulse setup for observation of slow light. A beamsplitter is added to the probe path creating a reference pulse which is directed around the vapor cell onto the detector. The reference pulse is blocked when observing the slow light pulse, and vice versa. \textcolor{black}{telescope in probe to shrink beam? Ken pg. 32}}} \label{fig:refpulse}
%\end{figure}
The reference and probe pulses are eventually made incident on the same photodetector, and the delay in their arrival times is measured on an oscilloscope by \textcolor{black}{blocking one pulse while the other is detected}. 
%Because the pulse durations $\tau_p$ are of the order of microseconds, the detector output  is switched to a 10 k$\Omega$ output impedance to dampen signal reflections between the detector and oscilloscope. 
%The modulation signal sent to the amplifier is \textcolor{black}{a sine modulation at 80 MHz with a Gaussian envelope. In our case the modulation peak is set at 800 mV}, which is slightly below the normal operating voltage (1V) because amplitude modulating near 1V begins to saturate the AOM resulting in a truncated Gaussian profile. 
%The AOM response vs. supplied voltage is only linear near 500 mV.\textcolor{black}{ [Then why go w/ 800 mV?]} 

From Sec.~\ref{sec:theory}\ref{subsec:store}, the durations of the Gaussian probe and reference pulses must be long enough that the pulse frequency-bandwidth (estimated as the inverse of the $1/e$-pulse-duration $\tau_p$) fits inside the EIT spectral window $\Gamma_{EIT}$. According to Eq. (\ref{eq:eitlinewidth2}) the EIT linewidth broadens linearly with pump intensity, necessitating probe pulses of progressively longer duration as the pump intensity decreases. 
%However, in our experiments the waveform generator does not provide us with independent control of the amplitude and duration of the temporal Gaussian probe pulse. 
%Due to technical reasons, in order to use the same modulation amplitude as described above, we have to . 
For convenience, we keep the probe pulse duration constant, and select a suitably long duration, for which the frequency bandwidth fits inside the EIT line width for the full range of pump intensities (see Sec.~\ref{sec:protocol} and~\ref{sec:delay}\ref{subsec:window} for details). 
%Of course, the pulse duration should not be so long that measuring a displacement between probe and reference line-centers becomes problematic.

The probe intensity must be kept significantly less than the pump, in order to satisfy the weak probe assumption. However, the probe pulse is too short to register on a typical power-meter. 
Therefore, we illuminated a fast photodiode with continuous-wave light of known intensity, and calibrated the response in volts (as measured on an oscilloscope) per mW. Next, the probe pulse was shone upon the diode and the shape of the voltage response recorded. 
%The probe and reference pulses are prepared as Gaussian temporal waveforms - it is straightforward to measure time-delays between the centers of such pulses. 
In our case the pulses have a $1/e$-width of 170 $\mu$s, and the {peak of the probe Gaussian temporal waveform} is set at {0.3 mW/cm$^2$} (the average intensity of the probe pulse is 0.12 mW/cm$^2$). %Thus, the weak probe assumption is well-satisfied in our measurements, for all except the lowest pump intensity. 
%where we used a pump intensity of 1.25 to 5.6 mW/cm$^2$. In a few cases, however, the pump intensity was as low as 0.15 mW/cm$^2$, comparable to the average probe intensity.   

%\vspace{-2mm}
\subsection{Leakage of pump beam into the detected probe mode}
\label{subsec:polpurity}
%\vspace{-1mm}
The pump and probe beams are recombined at polarizing beamsplitter PBS 5, so that the pump is reflected while the probe is transmitted toward the Rb-Ne vapor cell.
A half-wave plate, placed in each beam before PBS 5, adjusts the relative intensity of the pump and probe beams. Because of the large pump beam size, its half-wave plate (10 mm diameter) is located before the beam expander.
PBS 5 has the property that 99.5\% of $s$-polarization (light polarization normal to plane of incidence) is reflected whereas only 90\% of $p$-polarized light (light polarization parallel to plane of incidence) is transmitted. In our setup, $s$ is vertical polarization (perpendicular to the optics table surface) and $p$ is horizontal (parallel to the table surface). For this reason the pump beam, which must be strong, is chosen to be $s$-polarized, and the probe is $p$-polarized.

The combined beams, which are orthogonal-linearly-polarized, are passed through a quarter-wave plate (QP 1 in Fig.~\ref{fig:optlayout}) and converted to orthogonal-circularly-polarized before entering the vapor cell, as is required for the Zeeman EIT $\Lambda$-scheme. Irises are inserted to assist in day-to-day alignments. 
A second quarter-wave plate (QP 2) placed after the vapor cell converts the $\sigma^+$ and $\sigma^-$ polarizations back to linear polarization so that the pump can be separated from the probe at a polarizing beamsplitter (PBS 6). A half-wave plate placed just after QP 2 is adjusted so that the probe transmits through PBS 6 and is focused onto a photodiode, while the pump is reflected away. The probe transmission spectrum is recorded via a fast photodiode connected to a digital oscilloscope. 
%(see Sec.~\ref{sec:optsetup}\ref{subsec:pulses} above). 
The impedance of the detector is kept low (10 k$\Omega$) in order to reduce electrical reflections in the BNC cable between the detector and oscilloscope, at the cost of reduced overall voltage signal. 

It is obviously important to suppress as much as possible leakage of the strong pump beam into the detector at polarizing beamsplitter PBS 6. Polarizing beamsplitter cubes typically provide an extinction of $10^3$:$1$. For the highest pump powers used in our experiment, we find that a 0.1\% pump leakage at PBS 6 more than doubles the power in the detected probe mode. 
Further, pump leakage may distort the shape of the probe pulse because the pump-profile is not entirely flat. A cleaner probe signal is obtained by the use of a Glan-Thompson polarizer (extinction ratio 10$^5$:1) as PBS 6 for pump-probe separation.  
%~\cite{ken}.  
%when the pump and probe beams are recombined at polarizing beamsplitter PBS 5 just prior to entering the vapor cell (see Fig.~\ref{fig:optlayout}), even a tiny polarization impurity in the pump causes leakage of the pump into the probe beam leading to a significant spurious enhancement of probe power: %Besides, in order to discern genuine EIT effects in the probe transmission, we may wish to see what happens when we block and unblock the pump. 
%It is, therefore, important to suppress as best as possible pump leakage into the probe owing to polarization impurity.
Despite careful attempts to minimize the pump light from leaking into the probe detector, we observed some residual pump leakage. We suspect this is due to the use of low-order wave plates (as opposed to more expensive zero-order wave plates) and to residual errors in the alignment of QP1 and QP2's optical axes.
%aligning QP1's optical axis to 45$^o$ with respect to the input beam polarizations.  
%To minimize this residual pump leakage we insert an iris after PBS 6 and close it down to roughly the size of the probe beam.

\section{Magnetically shielded warm vapor cell 
%with applied magnetic field 
%for Zeeman EIT
} 
\label{sec:chamber}

For our slow light experiments we use a pyrex glass vapor cell, that is a sealed cylinder of 25 mm diameter, containing a small amount of isotopically pure solid $^{87}$Rb metal, along with 10 Torr of Ne buffer gas. Heating the cell is a convenient way to controllably vary the Rb atomic concentration in the vapor phase, yielding a few $\mu$Torr of $^{87}$Rb vapor. A heater assembly is used to vary the cell temperature from 55$^{o}$C to 65$^o$C during the experiment,
%Using empirical data in Ref.~\cite{pressure} for the vapor pressure versus temperature, and converting pressure to number density by applying the ideal gas law, we construct a plot of number density versus vapor temperature (\textcolor{black}{Supplementary Notes, Sec. S2}). Our temperature range mentioned above 
thus varying the Rb atomic number density $N$ participating in EIT between $1.5 \times 10^{11}$ cm$^{-3}$ and 
%$3.8
$3.4 \times 10^{11}$ cm$^{-3}$ \textcolor{black}{(see Supplementary Materials, Sec. S2). } %and of length $L = 2.5$ cm. 
%As stated earlier, the cell is filled with a few $\mu$Torr of isotopically pure $^{87}$Rb vapor and 10 Torr of Ne buffer gas. 
The length of the cell (28 mm) includes 1.6 mm windows (not anti-reflection coated) on either end. Therefore  
the length of the vapor sample $L$ is taken to be 25 mm.
%The demonstration of slow light is carried out in a cell of length 1 inch  which is filled with atomic Ne buffer gas.
%ed two different types of cylindrical pyrex glass vapor cells, each one inch in diameter, filled with isotopically pure $^{87}$Rb vapor and Ne buffer gas. The demonstration of slow light is carried out in a cell of length 1 inch (plus two 1/16-inch windows on either end) which is filled with 10 Torr of atomic Ne buffer gas. 
The cell is placed inside a solenoid as shown in Fig.~\ref{fig:optlayout}. To enable Zeeman EIT, the solenoid applies a small magnetic field $B_z = 50$ mG, co-linear with the laser beam propagation. 
\textcolor{black}{The cell, solenoid, and heater assembly are placed inside a magnetic shield
%a dedicated auxiliary optical pumping technique is used~\cite{shuker}
in order to suppress stray magnetic fields incident on the sample below 0.2 mG, i.e., $< 0.5$\% of $B_z$ (see Supplementary Materials, Sec. S2).}
\section{Setting the probe pulse bandwidth}
\label{sec:protocol}
In  order to observe slow light
%first describe a short experiment conducted in a vapor cell of natural abundance Rb vapor with no buffer gas. The purpose is to measure EIT line width and contrast, and compare with the theory in Sec.~\ref{sec:theory}. Next, we 
%we must first observe and optimize the EIT transparency window. The observed EIT linewidth helps us to 
we need to ensure that the probe (and reference) pulse bandwidth
%. As mentioned earlier (Sec.~\ref{sec:optsetup}\ref{subsec:pulses}), it is important to 
is less than the EIT spectral window $\Gamma_{EIT}$. 
%Next, we measure the delay of the probe pulse that passes through the sample relative to a reference pulse that does not. Finally, we examine the dependence of the probe group velocity $v_g$ on the coupling intensity $I$. 
%\subsection{EIT linewidth measurement and probe pulse width}
%\label{subsec:pulsewidth}
%The probe frequency-scanning is temporarily shut off, and the pump frequency is scanned in order to detect EIT on the $F_g = 2 \rightarrow F_e = 1$ transition. The pump frequency-scan is then turned off and, with the probe frequency-scan back on, the pump frequency is tuned to obtain peak transparency.
The EIT spectrum is obtained by the procedure described in Sec.~\ref{sec:optsetup}\ref{subsec:tune}.
%by scanning the probe frequency over a range larger than the Zeeman splittings of the $F_g = 2$ ground state in Fig.~\ref{fig:hypzee_3lvl}. The probe AOM frequency is then offset to locate the EIT spectrum at the center of the scan. See Sec.~\ref{sec:optsetup}\ref{subsec:tune} for laser-tuning details.
% \textcolor{black}{Further, the pump laser frequency is tuned to the center of the window where peak transparency occurs, by slightly adjusting the ECDL current in order to optimize the EIT spectrum.}

\begin{figure}[h]
\centerline{
%\vspace{-2mm}
\includegraphics[width=8.5cm]{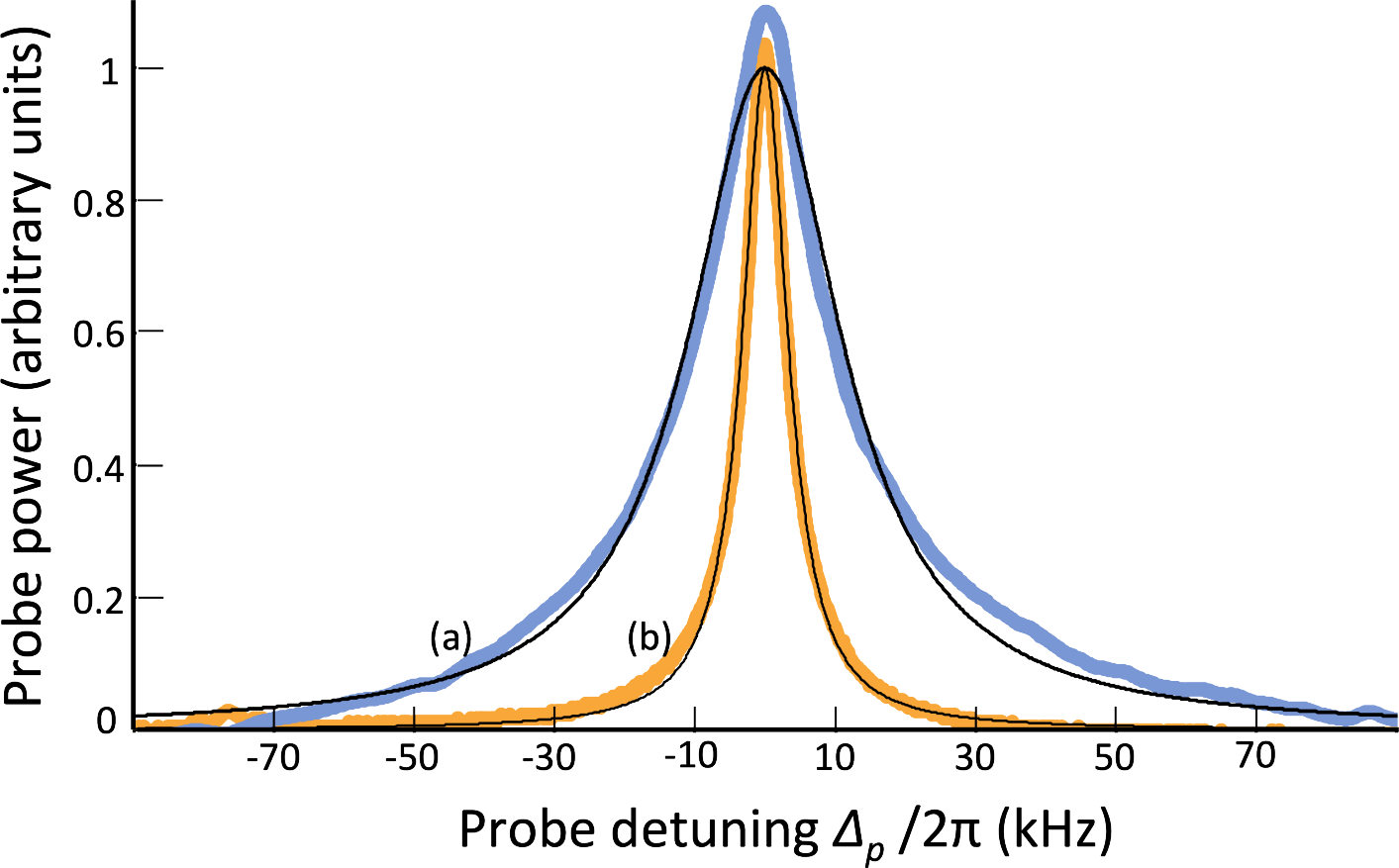}}
%{fig10_EIT.pdf}}
\caption{\textsf{(a) Measured EIT lineshapes (thick lines) in a few $\mu$Torr of isotopically pure $^{87}$Rb vapor with 10 Torr Ne at 65$^o$C, for pump intensity a) $I = 5.5$ mW/cm$^2$ ($\Omega_c/2\pi = 3.9$ MHz), and b) $I = 1.3$ mW/cm$^2$ ($\Omega_c/2\pi = 1.9$ MHz). The FWHM, extracted by Lorentzian fits (thin lines), are 26 kHz and 7.3 kHz, respectively. 
%The observed linewidths are within factor three of the linewidths predicted by Eqn.~\ref{eq:eitlinewidth2}.
%, and the vapor cell temperature is   
 }} \label{fig:eitdata}
\end{figure}

Typical EIT linewidths measured for high and low pump intensities, as a function of the probe detuning $\Delta_p$, are shown in Fig.~\ref{fig:eitdata}.
The maximum contrast measured 
(for the high intensity) 
is only $\sim$25\%, for reasons explained in Sec.~\ref{sec:theory}\ref{subsec:window} and Sec.~\ref{sec:3lvlexpt}\ref{subsec:broaden}. In order to enable a visual linewidth comparison between the lineshapes obtained at high and low intensities, we normalized the lineshapes and displayed them on the same vertical scale. 
%\textcolor{black}{How do we set $\Delta_c = 0$?}
%The graphs are normalized to allow for a comparison of the EIT linewidths at the two different pump intensities. In reality, the maximum contrast we obtain is only about 25\%, much lower than theoretically predicted by Eqn.~\ref{eq:abs} and displayed in Fig.~\ref{fig:EIT}(b). We see that the EIT linewidth, too, is far narrower than theoretically predicted  
%for reasons already discussed in Sec.~\ref{sec:theory}.
%As indicated earlier in Sec.~\ref{sec:theory}\ref{subsec:obe}-\ref{subsec:dark}, we attribute this to: The theory is derived for a closed three-level system but the $^{87}$Rb atom is an ``open" system where population leaks from the dark state to extraneous levels. Further, our simple theory accounts to some extent for collisional and power broadening but does not take into account other decohering mechanisms described in Sec.~\ref{sec:3lvlexpt}\ref{subsec:broaden}. } 
%We may directly compare the theoretically predicted EIT linewidth in Fig.~\ref{fig:EIT}(b) 
%%from Eqn.~\ref{eq:abs} 
%with the experimental linewidth in Fig.~\ref{fig:widthvsint}(b) because a pump intensity of 1.3 mW/cm$^2$ corresponds to $|\Omega_c|/2\pi \approx 4$ MHz (from Eqn.~\ref{eq:chi2_I}): We note that 
By fitting with a Lorentzian, we observe EIT linewidths of 7.3 kHz and 26 kHz for $I = 1.3$ mW/cm$^2$ and $5.5$ mW/cm$^2$, respectively, which correspond to $\Omega_c/2\pi$-values of 1.9 MHz and 3.9 MHz, respectively (see Sec.~\ref{sec:3lvlexpt}\ref{subsec:condition}). Data are spectrally narrower at the center of the Lorentzian owing to coherent contributions from atoms that diffuse in and out of the laser beam multiple times without decohering (this is known as Ramsey-narrowing).~\cite{irina,walsworth2006prl}

We chose a $1/e$-width of 170 $\mu$s for our Gaussian probe and reference pulses (see Sec.~\ref{sec:optsetup}\ref{subsec:pulses}). The corresponding bandwidth of $\leq 1$ kHz fits inside the EIT window which, from Eq. (\ref{eq:eitlinewidth2}), is at least $2\gamma_{12}/2\pi \approx 6$ kHz wide even for the smallest pump intensities.

\textcolor{black}{Generating a EIT lineshape measurement as in Fig.~\ref{fig:eitdata} typically takes 40 ms for the higher pump intensities, and 160 ms for the lower intensities (we set the AOM for a $\pm 100$ kHz frequency-sweep in 10 ms; each spectrum in Fig.~\ref{fig:eitdata} is an average of either 4 or 16 such sweeps). Our laser system's passive stabilization yields a frequency drift of $< 10$ MHz/hour, resulting in a drift of less than 0.5 kHz while generating these EIT data-profiles. 
%This drift is much less than the measured EIT linewidths, and even smaller than the Zeeman shifts between adjacent ground magnetic sub-states, and is therefore acceptable. 
%There is no need to frequency-lock the laser with an electronic servo loop. 
} 

%Finally, note in Fig.~\ref{fig:widthvsint} that the EIT data are spectrally narrower at the transparency center than the Lorentzian fits.
% %causing the data-curves to extend beyond the fits in the vertical direction. 
% This is ascribed to Ramsey-narrowing induced by atoms that diffuse out of the laser beam and return without decohering~\cite{walsworth2006prl}.

\section{Results and discussion}
\label{sec:delay}
\begin{figure}[h]
\centerline{
\includegraphics[width=7.6cm]{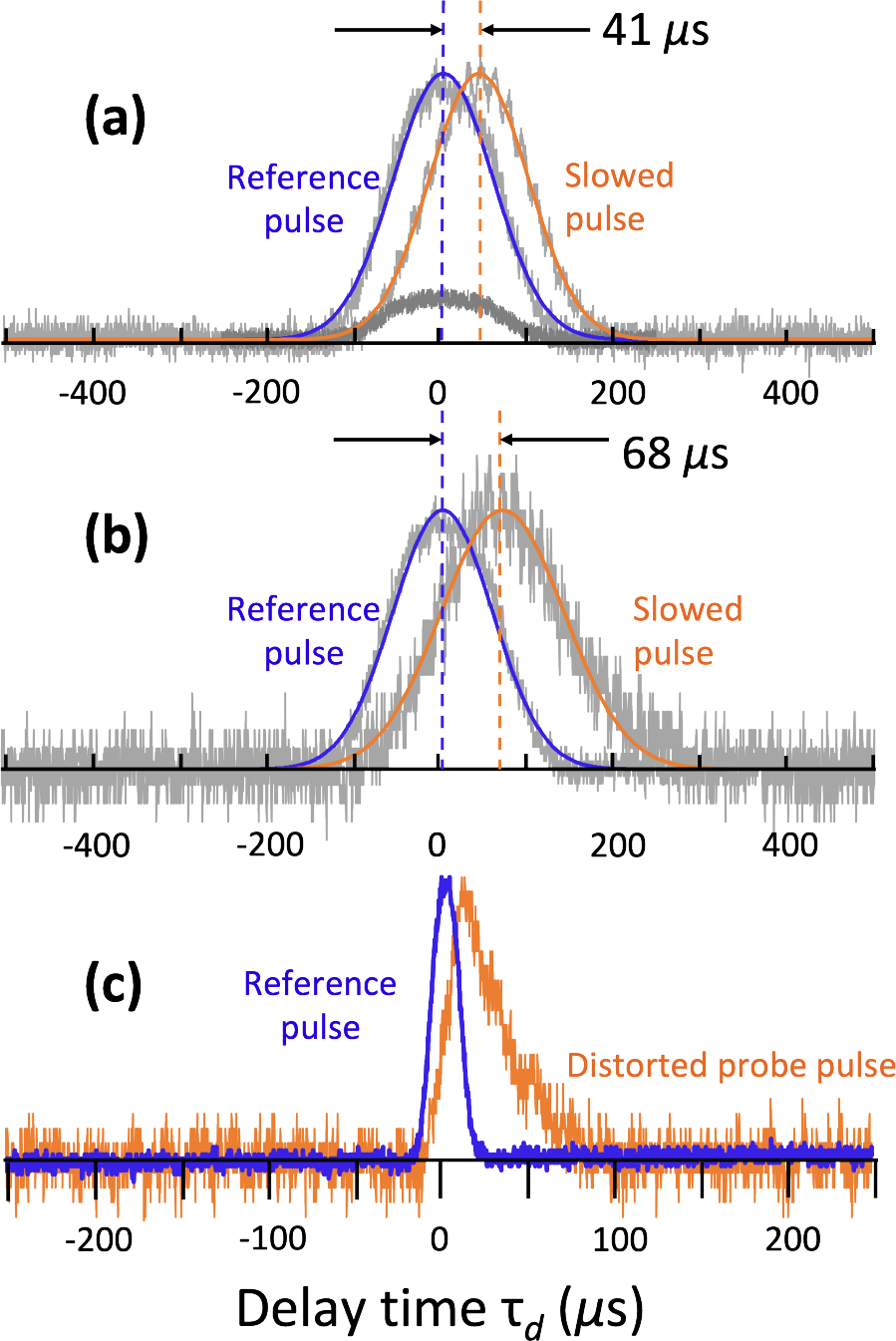}}
%{slow_rev.pdf}}
\caption{\textsf{Measurement of time-delay $\tau_d$ between reference and probe pulses traveling through air and the EIT medium, respectively. a) $\tau_d = 41 \, \mu$s yields $v_g = 610$ m/s. Pump intensity {$I = 3$ mW/cm$^2$}. The squat data-curve is the transmitted probe pulse when the pump is blocked, in which case EIT ceases so that large absorption occurs with no slowing. b) $\tau_d = 68 \, \mu$s yields $v_g = 368$ m/s, our slowest observed velocity. Pump intensity {$I = 1.2$ mW/cm$^2$}. c) Significant distortion in the transmitted probe pulse occurs if we select a pulse duration so narrow that the probe bandwidth does not fit inside the EIT transparency window. All measurements are in $^{87}$Rb vapor with 10 Torr Ne at 65$^o$C. 
%The peaks of the reference pulse do not exactly align with zero owing to a slight delay from the photodetector, which does not impact the measurement of the relative delay between the probe and reference pulses.  
 }} 
\label{fig:slow}
%\vspace{-2mm}
\end{figure}

In order to create short pulses, the probe frequency scan is now turned off, the probe AOM offset voltage is tuned to the EIT peak, and the amplitude modulation is turned on, 
as described in Sec.~\ref{sec:optsetup}\ref{subsec:pulses}. Once both slow and reference pulses are detected on an oscilloscope,
%~(see Fig.~\ref{fig:optlayout}), 
the temporal waveform for each is fit to a gaussian curve, and the relative delay time $\tau_d$ is extracted.
% - this is just the time taken by the slowed pulse to propagate a certain distance $L$ through the vapor minus the time taken by the reference pulse to travel the same distance through air. 
%We calculated the group velocity $v_g$ using 
%\begin{equation}
%\delta = \frac{L}{v_g} - \frac{L}{c} \Rightarrow v_g = \frac{cL}{c\delta + L} 
%\xrightarrow{c\delta >> L} \frac{L}{\delta}
%\label{eq:vgexpt}
%\end{equation}
%where $L = 1$ inch from Sec.~\ref{sec:chamber}\ref{subsec:cell} above. The inequality in the equation is satisfied for delays on the order of several nanoseconds and beyond - in our case $\delta$ is in the microsecond regime. Note that the reference pulse travels an additional path length (see Fig.~\ref{fig:optlayout}) compared to the probe pulse, but the extra time incurred is only $\sim 1$ ns. 
Typical measurements of $\tau_d$ are presented in Fig.~\ref{fig:slow}, where the data (fuzzy lines) are shown along with Gaussian fits (smooth lines). 

In Figs.~\ref{fig:slow}(a) and (b) the reference pulse arrives at the detector earlier than the probe by $\tau_d = 41$ $\mu$s and 68 $\mu$s, respectively, which corresponds to a slowed probe group velocity $v_g$ of 610 m/s and 368 m/s (using $v_g = L / \tau_d)$. 
%(see Sec.~\ref{sec:theory}\ref{subsec:slow}).
%(approximation is well satisfied for $\tau$ in $\mu$s regime). 
%In Fig.~\ref{fig:slow}(b) we show the slowest group velocity of 361 m/s, corresponding to a delay of 68 $\mu$s, that we achieved with our setup. 
In each case, the peaks of the fits for reference and probe are scaled to the same value for easier comparison. The reference and probe pulse intensities are similar at the non-polarizing beamsplitter NPBS4 in Fig.~\ref{fig:optlayout}, but the probe pulse suffers some absorption in the vapor (recall that our EIT contrast never exceeds 25\%) which explains the increased noise on the slowed pulses. The slower group velocity was achieved by lowering the pump intensity which further reduces the overall signal-to-noise ratio. 

\subsection{Observed $\Gamma_{EIT}$, $\tau_d$, and $v_g$ vs. theoretical prediction}
\label{subsec:compare}
To predict the EIT linewidth $\Gamma_{EIT}$ we insert in Eq. (\ref{eq:eitlinewidth2}) parameter-values that are relevant to our experiment, as discussed in Sec.~\ref{sec:3lvlexpt}\ref{subsec:condition}: $\gamma_{12}/2\pi \approx 3$ kHz, $\gamma_{13}/2\pi \approx 300$ MHz. The optical depth is estimated using $N = 3.4 \times 10^{11}$ cm$^{-3}$ and $L = 2.5$ cm (see Sec.~\ref{sec:chamber}), yielding $OD \approx 25$.
For control intensities $I$ of 1.3 and 5.5 mW/cm$^2$, corresponding to $\Omega_c/2\pi = 1.9$ and 3.9 MHz (see Sec.~\ref{sec:protocol}), we find that Eq. (\ref{eq:eitlinewidth2}) predicts $\Gamma_{EIT}/2\pi =$ 10.8 and 26.3 kHz, respectively, remarkably close to the observed values in Fig.~\ref{fig:eitdata}. 

To predict the delay time $\tau_d$ and the slowed probe group velocity $v_g$, we refer to Eq. (\ref{eq:v_g1}) and Eq. (\ref{eq:delay}), which use just the power-broadened component of $\Gamma_{EIT}$. For intensities $I = 3$ mW/cm$^2$ and 1.2 mW/cm$^2$ as in Fig.~\ref{fig:slow}, the calculated delays $\tau_d$ are 72 $\mu$s (yielding a predicted $v_g$ of 347 m/s) and 181 $\mu$s (predicted $v_g =$ 138 m/s), respectively. The predicted $v_g$-values are a factor two to three slower than the observed $v_g$-values in Fig.~\ref{fig:slow} (a) and (b). To make more accurate predictions, we must move away from an idealized three-level atom model and include the full hyperfine structure
(Fig.~\ref{fig:Rb_energylevels}), 
%(see Sec.~\ref{sec:3lvlexpt}), 
which is beyond the scope of this article.~\cite{lukin}

\subsection{Role of EIT transparency window}
\label{subsec:window_expt}
Fig.~\ref{fig:slow}(a) highlights the importance of the EIT process. The squat data-waveform is the transmitted probe pulse when the pump beam is blocked causing the EIT window to cease to exist: Large absorption and no slowing is observed. The transmitted power of the slow pulse for the unblocked pump is $\sim 20-25\%$ of the incident probe power, while for the blocked pump is less than 5\%. 

To further highlight the important role played by EIT, we show in Fig.~\ref{fig:slow}(c) what happens if we select a pulse duration for which the probe bandwidth does not fit inside the EIT transparency window.
The input Gaussian pulse in this case, namely $\sim 20 \mu$s (which yields a {$1/e$-pulse bandwidth of $\sim 10$ kHz}), is {nearly an order of magnitude shorter than the 170 $\mu$s pulse employed in Fig.~\ref{fig:slow}(a) and (b)}. Thus, in the case of Fig.~\ref{fig:slow}(c) the pulse bandwidth is comparable to the EIT window which ranges from a few kHz to $\leq$ 30 kHz for our experiments. The frequency components of this pulse that do not fit within the EIT window are not slowed,  and are instead significantly absorbed. This leads to a significant temporal stretching and distortion of the pulse, as shown in Fig.~\ref{fig:slow}(c) (we scaled the vertical size of the reference pulse to about the same as the transmitted probe). 

%For our longest $\tau_d$ of 68 $\mu$s, the delay-bandwidth product from Eqn.~\ref{eq:product} is 0.4, which is two orders of magnitude less than values achieved using other techniques~\cite{irina}.    

In fact, some probe pulse broadening is visible even in the case of the slowest group velocity measured in Fig.~\ref{fig:slow}(b). In this case, higher frequency components of the 170 $\mu$s Gaussian probe pulse ($1/e$-frequency bandwidth $\sim 1$ kHz) extrude past the 7.3 kHz Lorentzian EIT transparency window, and are strongly absorbed. This causes an effective narrowing of the probe bandwidth, which leads to a broadening of the pulse duration.   

\subsection{Slow group velocity vs. pump intensity}
\label{subsec:slowvsI}
\begin{figure}[h]
\centerline{
\includegraphics[width=8.5cm]{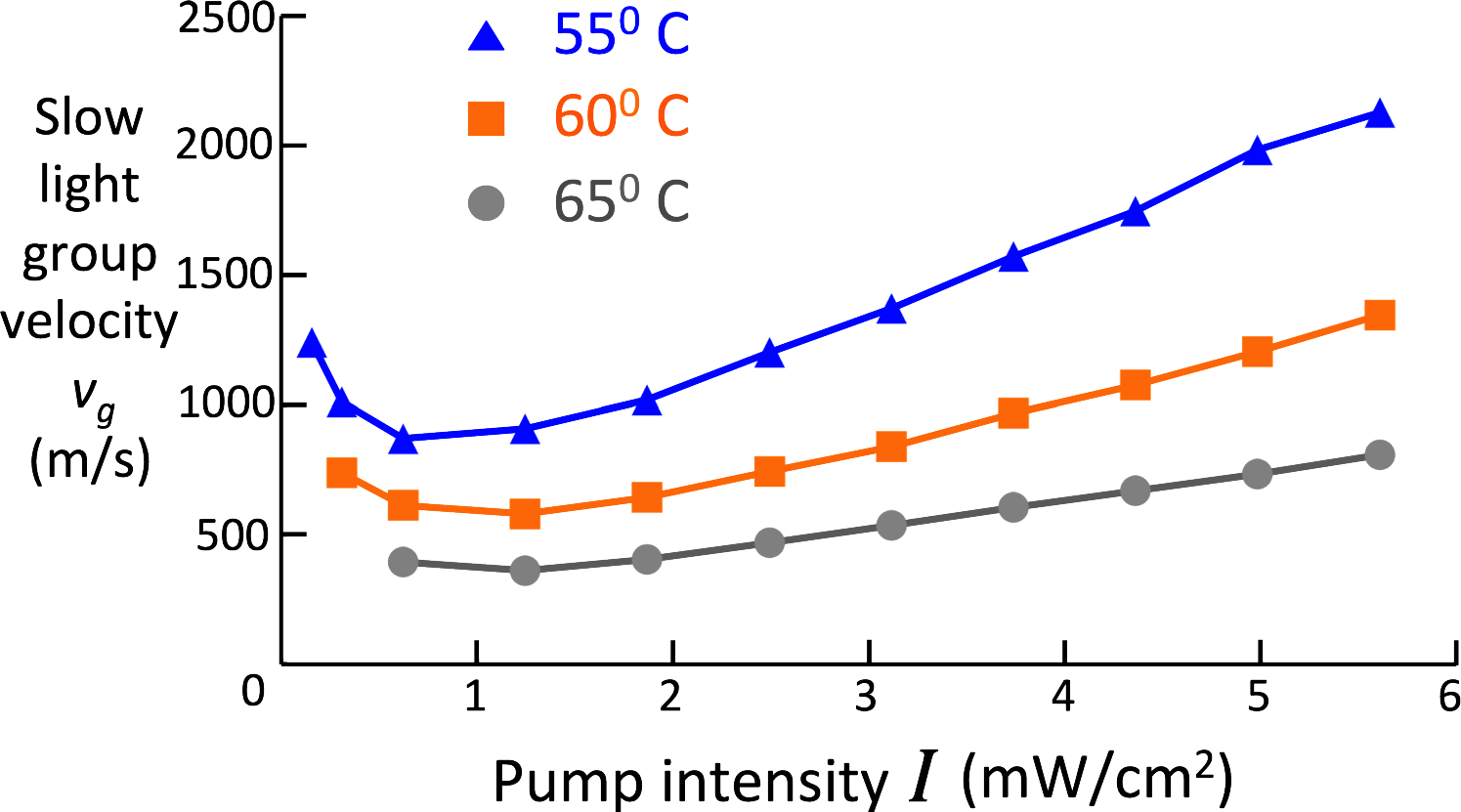}}
%{sweetspot3.pdf}}
%\vspace{-2mm}
\caption{\textsf{ Slow light at different pump intensities $I$ and temperatures $T$ (lines drawn to guide the eye). 
%Increasing the temperature $T$ raises the atomic density $N$ (see Fig. S2 in supplementary notes), resulting in lower group velocities $v_g$ in accordance with Eqn.~\ref{eq:v_g1}. 
%At fixed $T$ (and $N$), $v_g$ decreases as $I$ decreases in agreement with Eqn.~\ref{eq:v_g1}, though creeps up at very low pump intensities where the weak-probe assumption breaks down. 
All measurements are in isotopically pure $^{87}$Rb vapor with 10 Torr Ne.
% and the expectation that slowing effects decrease as the EIT linewidth broadens. 
%For really low intensities $v_g$ creeps back up, yielding a ``sweet spot" in intensity where slow light effects are optimized. 
%In the inset we discard the six data-points for the lowest intensities ($< 1$ mW/cm$^2$) and plot $v_g$ versus the ratio $I/N$ : The three $v_g$-curves are approximately linear with nearly equal slopes in accordance with Eqn.~\ref{eq:v_g1}, but the lines do not all coincide as predicted by the simple three-level theory.
}} 
\label{fig:slowvgI}
%\vspace{-4mm}
\end{figure}

In Fig.~\ref{fig:slowvgI} we plot the experimentally observed slow light group velocity for several pump intensities at three different vapor temperatures. 
The observed temperature-dependence of $v_g$ is in accordance with what we expect from Eq. (\ref{eq:v_g1}): 
%The atom number density 
$N$ increases with $T$ (see \textcolor{black}{Sec. S2 D} in supplementary materials), causing $v_g$ to decrease. 
%Increasing $N$ further by raising $T$ may cause undesirable radiation trapping effects to start arising~\cite{welch}. 

At a fixed temperature we surmise from Fig.~\ref{fig:slowvgI} that two competing slow light effects occur when we vary the pump intensity.
%Based on our simple three-level theory in Sec.~\ref{sec:theory}\ref{subsec:eit}-\ref{subsec:vg} above, 
%we  which originate from the fact that as the intensity increases the EIT contrast increases (Eqn.~\ref{eq:contrast}), but so does the linewidth (Eqn.~\ref{eq:eitlinewidth}). 
The linear increase of $v_g$ with increasing pump intensity $I$ is expected from Eq. (\ref{eq:v_g1}). Recall that this originates from the power-broadening of the EIT transparency window as described by Eq. (\ref{eq:eitlinewidth2}), thereby reducing the ``tightness of the pinch" of the ``wrinkle" in the real refractive index $n_r$ at $\Delta_p = 0$ in Figs.~\ref{fig:EIT} (d,f). 
%reducing the pump intensity $I$ reduces the steady-state fraction of atoms being pumped to the dark state thereby tending to \emph{increase} the group velocity $v_g$ - this is consistent with Eqn.~\ref{eq:contrast} which describes a decrease in EIT contrast as the intensity decreases. On the other hand we expect from Eqn.~\ref{eq:v_g1} (in conjunction with Eqns.~\ref{eq:chi2_I} and~\ref{eq:cond2}) that reducing $I$ \emph{decreases} $v_g$.    
%
%At large pump intensity (but still satisfying Eqn.~\ref{eq:cond2}), the linear intensity-dependence of $v_g$ in Eqn.~\ref{eq:v_g1} originates from the power broadening of 
%%As indicated in Sec.~\ref{sec:theory}\ref{subsec:it}, 
%the EIT linewidth (see Sec.~\ref{sec:3lvlexpt}\ref{subsec:broaden} above) which depends linearly on intensity as well (see Eqn.~\ref{eq:eitlinewidth}). 
%The narrowing of the EIT transparency window with decreasing intensity effectively increases the magnitude of the positive gradient $dn_r/d\omega_p$ in Eqn.~\ref{eq:grvel} (in other words, tends to ``further pinch the wrinkle" in the , thereby causing $v_g$ to decrease. Fig.~\ref{fig:slowvgI} shows that for all but the lowest intensities, this EIT linewidth-narrowing effect dominates over the EIT contrast reduction.
{However, it is obvious that one cannot keep reducing the pump intensity to achieve lower and lower group velocities as there is no slow light in the absence of a pump. 
% the power-broadening term in Eqn.~\ref{eq:eitlinewidth} becomes negligible and the EIT linewidth bottoms out at $2\gamma_{12}$. 

At really low pump intensities, \textcolor{black}{the weak probe assumption starts to break down and, as mentioned in Sec.~\ref{sec:3lvlexpt}\ref{subsec:condition} already, the population of atoms being pumped to the dark state may decline to the point that the 
%group velocity starts to creep back up.
probe pulse appears to start speeding up again.}    
%Thus there occurs a ``sweet spot" in the coupling intensity at which the conditions for slow light are optimized. 
%The slowest $v_g$-value achieved is further reduced when we increase the atomic number density $N$ by 
%increasing the sample temperature, in accordance with Eqn.~\ref{eq:v_g1}. 
%The lowest speed we achieved is 361 m/s at 65$^o$C. 

%To further test Eqn.~\ref{eq:v_g1} we plot $v_g$ versus the ratio of the pump intensity $I$ to number density $N$ (Fig.~\ref{fig:slowvgI}, inset), after removing the six data-points obtained at the weakest intensities ($< 1$ mW/cm$^2$). The slopes for the $v_g$ data-curves versus $I/N$ at the three different temperatures are approximately equal in accordance with Eqn.~\ref{eq:v_g1}, though an unexplained systematic constant offset in the $v_g$ values at the different temperatures remains. 
%
%A modest rise in temperature, by 10 to 20$^o$C, would significantly further reduce $v_g$, possibly less than 100 m/s. The reason we did not proceed to higher temperatures is that the white plastic acetal flanges on our solenoid frame began to soften, as mentioned earlier in Sec.~\ref{sec:chamber}\ref{subsec:solenoid}. This choice of material is easily avoided in future designs. 

\textcolor{black}{
%The slowing shown here is more than adequate: 
Group velocities of several thousand m/s suffice for most cutting-edge experiments on quantum memory and image storage in warm vapor.~\cite{irina,ma} At higher temperatures, the density increases, but so do spin depolarization mechanisms such as Rb-Rb spin exchange collisions and radiation trapping, as indicated in Sec.~\ref{sec:3lvlexpt}\ref{subsec:broaden}. In vapor cells that use paraffin-based anti-spin relaxation coatings instead of buffer gas, temperatures exceeding 80$^o$C may cause coating breakdown. Coatings such as OTS (octadecyltrichlorosilane) permit higher temperatures.~\cite{ran,ots}}
% up to 200$^o$C~\cite{ran,ots}.}
% but break down from prolonged contact with alkali vapor, necessitating collection of the alkali into a cold finger when the cell is not in use~\cite{ran,ots}.}

%
\section{Conclusion}
\label{sec:conclusion}
We have presented detailed theoretical and experimental undergraduate-friendly instructions on how to produce light pulses propagating through warm alkali vapor with speeds as low as few hundred m/s. We elucidated the role played by EIT in producing slow 
%\textcolor{black}{and stored} 
light. 
%Equipment prices and vendor information are \textcolor{black}{ in supplementary materials.}

The experimental setup described here is remarkably versatile. It can be used for investigations into slow and stored light, including detailed measurements of pulse delay $\tau_d$ as the pulse width $\tau_p$ is varied, and further measurement of the subtle Ramsey narrowing seen at the EIT line-center in Fig.~\ref{fig:eitdata} as the pump and probe beam sizes are varied.~\cite{walsworth2006prl} By slightly varying the angle between the pump and probe beams, one can study EIT linewidth-narrowing due to the spatial localization of alkali atoms from frequent velocity-changing collisions with the buffer gas.~\cite{dicke} Furthermore, one may study how atomic diffusion degrades storage times for slow light pulses~\cite{image} of various transverse profiles, e.g., Laguerre-Gaussian~\cite{vortex} and Bessel beams~\cite{bessel}, which possess topological phase features that permit significantly more robust storage in comparison to the usual Gaussian beams. The setup may be adapted for innovative magnetometry with potential applications in magnetic induction tomography and the detection of concealed objects.~\cite{renzoni1,renzoni2}

\vspace{-2mm}
\section{Acknowledgements}
%KD gratefully acknowledges
We acknowledge support
% from the Army Research Office. 
%This material is based upon work supported by, or in part by, 
by the U. S. Army Research Office under grant number W911NF2110120.
We appreciate insightful discussions with Drs. Irina Novikova, David Phillips, Ofer Firstenberg, and Ran Finkelstein, and detailed invaluable feedback from three anonymous referees. We thank 
Hong Cai, Kaleb Campbell, Richard Jackson, Dillon deMedeiros, Bradley Worth, Amanda Day, Somaya Madkhaly, Yuhong (Iris) Zhang, Peter Harnish, and Jason Barkeloo 
%[names withheld]
for help during the initial setup. 
We thank the Miami University Instrumentation Laboratory for crucial help in machining and electronics. The authors have no conflicts to disclose.
%We are 
%%also 
%indebted to the Dean of the College of Arts and Science 
%at Miami University
%for providing generous seed funding to our advanced optics and lasers teaching laboratory for undergraduate seniors and first-year Masters' students. 

\vspace{1mm}
$^{\dag}$ These two authors contributed equally.
\newline 
$^{*}$ Corresponding author: balis@miamioh.edu

%\vspace{-2mm}


\begin{thebibliography}{10}
\bibitem{factor2} Our definition of 
$\Omega_c, \Omega_p$ 
%Rabi frequency 
follows Refs.~\cite{irina,eberly,lukin}. Some authors omit the 2 in the denominator (e.g.,~\cite{steckRb,budker}).
\bibitem{identical1} Eq. (\ref{eq:delay}) is identical to Eq. (6) in Ref.~\cite{irina}, Eq. (38) in \cite{lukin2}.
\bibitem{identical2} Eq. (\ref{eq:product}) is identical to Eq. (39) in Ref.~\cite{lukin2}.
\end{thebibliography}
\end{document}

%\twocolumn[ 

\title{SUPPLEMENTARY NOTES\\
Producing slow 
%and stored 
light in warm alkali vapor using electromagnetically induced transparency}

\author{Kenneth DeRose$^{\dag}$, Kefeng Jiang$^{\dag}$, Jianqiao Li, Linzhao Zhuo, Scott Wenner, and Samir Bali*}
\address{
Department of Physics, Miami University, Oxford, Ohio 45056-1866,
USA \\
}

\begin{abstract}
In order to complement the undergraduate-friendly detailed experimental description in the main article, here, in Sec.~\ref{sec:theory}, we provide background details for the basic physics behind electromagnetically induced transparency (EIT). 
%In the sections below we follow closely the excellent treatment in Ref.~\cite{eberly}, but include non-zero pump laser detuning, and a section on dark and bright states.
%, and for completeness,
%In the spirit of providing a complete stand-alone reference, 
%we show in the following sections how to calculate $n_{r}$ and $n_{i}$ in order to reveal the basic physics behind EIT and slow light, 
We endeavor to keep the discussion amenable to advanced juniors and seniors. Furthermore, in Sec.~\ref{sec:chamber} we provide useful technical details on magnetic shielding of the alkali vapor cell in order to suppress stray magnetic fields in the laser-atom interaction region below 0.2 mG, the implementation of a heater assembly 
%to raise the alkali vapor temperature 
in order to increase the atomic density participating in EIT, and the construction of a solenoid to apply a small magnetic field to the vapor cell in order to enable Zeeman EIT. 
%We include a number density versus temperature plot in Sec.~\ref{sec:nvsT}. 
A price and vendor list for parts is provided in Table 1.
%We conclude the theory section by addressing a couple popular misconceptions about slow light and the behavior of the refractive index near resonance.
\end{abstract}

%] %activate for two-column activation
\renewcommand\thesection{S\arabic{section}}
\section{Theoretical Background}
\label{sec:theory}
The theoretical discussion of EIT and slow light centers around calculating the complex refractive index $n = n_{r} + in_{i}$ of the medium. 
%This is because EIT in atomic media is a resonant phenomenon. 
%Near-resonance absorption is included in the theory as an imaginary contribution to the refractive index. 
The imaginary part $n_{i}$ yields the absorption spectrum, where EIT in atomic media is revealed as a narrow transparency window near resonance. The real part $n_{r}$ yields the group velocity of light propagating through the medium. Slow light is revealed as a consequence of the steep frequency-variation of $n_{r}$ within the EIT transparency window. 

We begin by 
%defining a three-level system, the simplest atomic energy-level configuration in which to elucidate the basic concepts of EIT and slow light. Next, we 
writing down the Hamiltonian and the Schr\"odinger equation for the illuminated three-level atomic model in Fig. 1 (reproduced in Fig.~\ref{fig:fig1} a), and derive
% when illuminated by near-resonant pump and probe fields, leading to a derivation of the so-called ``c-dot equations" - a set of coupled differential equations for the probability amplitude of the atom being in each of the three atomic levels. However, experiments measure probabilities and coherences, not wave functions, which motivates conversion of the c-dot equations into 
the widely used optical Bloch equations. Next, we use these equations to calculate the complex refractive index $n$ of a dilute gas (or vapor), and reveal the required conditions for EIT and slow light. We follow closely the treatment in Ref.~\cite{eberly} - however, we define the laser detunings $\Delta_p$ and $\Delta_c$ opposite in sign to Ref.~\cite{eberly}.
%\subsection{Three-level system for EIT and slow light} 
%\label{subsec:3lvl}
%Following the notation in Ref.~\cite{eberly} we depict a fictitious three-level atomic model in Fig.~\ref{fig:fig1}(a), in which transitions are allowed between states 1 and 3 and between states 2 and 3 but not between states 1 and 2. 
%This is the so-called lambda-system because the lower-levels can be close-lying resulting in the three levels forming a $\Lambda$-shaped configuration.
% - is the simplest possible atomic system for understanding the basic concepts of EIT.  
\renewcommand\thefigure{S\arabic{figure}}
\begin{figure}[h]
\centerline{
\includegraphics[width=8.3cm]{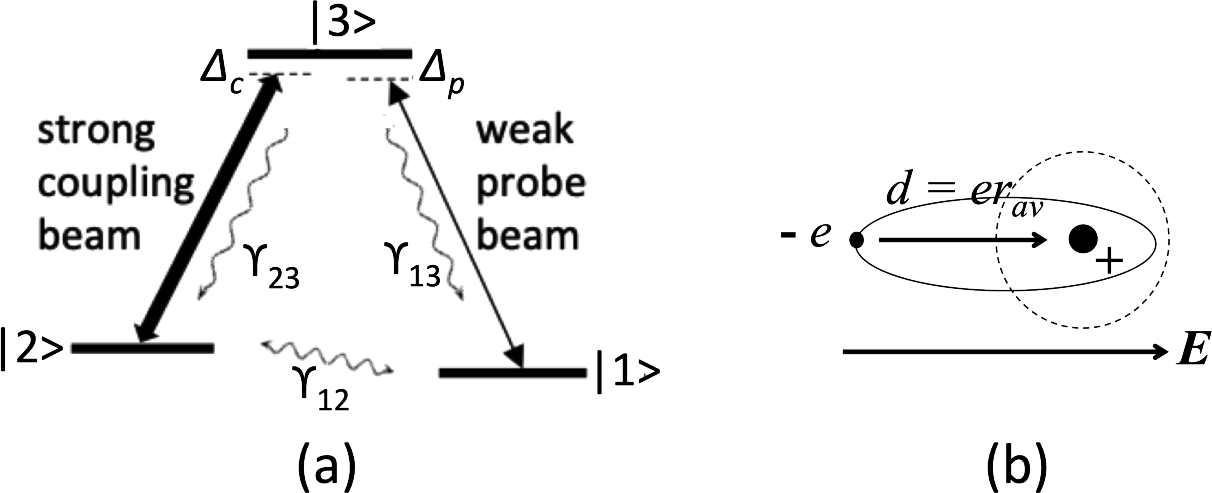}}
%{fig1.pdf}}
\caption{\textsf{a) Three-level atomic model from Fig. 1. 
%A strong coupling (or pump) beam of frequency $\omega_{c}$ and a weak probe beam of frequency $\omega_{p}$, are both tuned to near-resonance with $|3\rangle$ from levels $|2\rangle$ and $|1\rangle$ respectively, with detunings $\Delta_{c}$ and $\Delta_{p}$. Decoherence rates between levels 3 and 1, 3 and 2, 2 and 1, are denoted by $\gamma_{13}, \gamma_{23}$, $\gamma_{12}$, respectively (see Eqns.~\ref{eq:obedamp} and preceding text). 
b) Depiction of an atom showing a lone valence electron orbiting the nucleus (big dot). In the absence of an external field, the mean dipole moment, averaged over an electron revolution, is zero. When illuminated by an electric field $\vec{E}$, the electron cloud shifts in the opposite direction (elliptical orbit) creating an induced dipole moment $\vec{d}$ of magnitude $er_{av}$, where $r_{av} \, (\propto E)$ represents the average displacement between the positive and negative regions of the atom. The polarizability $\alpha_{p}$ is defined by $\vec{d} = \alpha_{p}\vec{E}$.} }\label{fig:fig1}
\end{figure}
%The standard EIT experiment consists of a strong coupling (or pump) beam of frequency $\omega_{c}$ and a weak probe beam of frequency $\omega_{p}$, both tuned to near-resonance with $|3\rangle$ from levels $|2\rangle$ and $|1\rangle$ respectively, with detunings $\Delta_{c}$ and $\Delta_{p}$~\cite{eberly2}. We assume that the pump beam only addresses the coupling transition $|2\rangle \rightarrow |3\rangle$ and the probe beam only addresses the $|1\rangle \rightarrow |3\rangle$ transition. The probe frequency is scanned around the (fixed) pump frequency, and the probe transmission spectrum is measured.

%\vspace{-3mm}
\subsection{Hamiltonian for light-atom interaction} 
\label{subsec:ham}
If we denote the wavefunction for the atom interacting with an incident light field as $\Psi (\vec{r}, t)$, the time evolution of $\Psi$ is described by the Schr$\ddot{{\mbox o}}$dinger equation 
%(any variable below with a caret above denotes an operator)
\renewcommand\theequation{S\arabic{equation}}
\begin{equation}
i \hbar \frac{\partial \Psi (\vec{r}, t)}{\partial t} = \hat{H} \Psi (\vec{r}, t) = (\hat{H_{0}} + \hat{V}) \Psi (\vec{r}, t).
\label{eq:schro}
\end{equation}
Here, $\hat{H}$ is the Hamiltonian for the system, i.e., the energy operator of the system, acting upon the wavefunction $\Psi$. Since we are focusing on just the atom and its interaction with the incident field, we may ignore the contribution to the Hamiltonian from the incident field alone (also known as the free field). In that case it is convenient to express the Hamiltonian operator $\hat{H}$ as a sum of two terms,
%\begin{equation}
%\hat{H} = \hat{H_{0}} + \hat{V},
%\label{eq:schro2}
%\end{equation}
where $\hat{H}_{0}$ is the bare Hamiltonian which describes the atom in the absence of any external field, and $\hat{V}$ is the interaction Hamiltonian which describes the interaction between the atom and the external field. 

Typically the outermost electrons of the atom are the least tightly bound and respond most readily to the incident field. Indeed, this is an important reason why alkali atoms are favorites with physicists - relatively low optical energies are needed to resonantly excite the lone valence electron. Further, the interaction of the single electron with the incident field is simple to model. The incident electromagnetic field's electric and magnetic vectors, $\vec{E}(\vec{r}, t)$ and $\vec{B}(\vec{r}, t)$ respectively, interact with the valence electron (negative charge $e$) through the Lorentz force. 
%$\vec{F} = e (\vec{E} + \vec{v} \times \vec{B})$, 
%which in Gaussian units is $\vec{F} = $. 
The electron speed is non-relativistic, 
so the effect of the $\vec{B}$-field may be ignored.~\cite{gaussian}
The incident $\vec{E}$-field induces an electric dipole in the atom.

Thus, $\hat{V}$ is given by the potential energy stored in this induced dipole, 
%\begin{equation}
$\hat{V} = - \hat{\vec{d}} \boldsymbol{\cdot} \hat{\vec{E}}(\vec{r}, t)$ 
%\rightarrow - \hat{\vec{d}}.\hat{\vec{E}}(t),
%\label{eq:Hint}
%\end{equation}
where $\vec{d}$ is the induced dipole moment (see Fig.~\ref{fig:fig1}(b)) and $\vec{r}$ is the location of the atomic center of mass. We ignore the variation of $\vec{E}$ over the spatial extent of the atom ($\sim$ nm) because the optical wavelength ($\sim$ hundreds of nm) is much larger.
% (this is known as the ``long wavelength approximation").  
%\textcolor{blue}{homogeneity of $\vec{E}$ across sample?}

%\subsection{Light shift calculation for two-level atoms} 

For the simple three-level atom shown in Fig.~\ref{fig:fig1}(a) we denote its states as $|i\rangle$ $(i = 1, 2, 3)$, with bare energies $\xi_{i}$.
Clearly, the states $|i\rangle$ are eigenstates of the bare Hamiltonian $\hat{H}_{0}$, i.e.,
%\begin{equation}
$\hat{H}_{0}|i\rangle  = \xi_{i}|i\rangle$
%\label{eq:eigen}
%\end{equation}
and may be defined to form an orthonormal basis: $\langle i | j \rangle$ is 1 if $i = j$ and 0 otherwise.
This means that any arbitrary wavefunction for the three-level atom, such as $\Psi(\vec{r}, t)$ in Eq. (\ref{eq:schro}), may be expressed as a linear superposition of the states 
$|i\rangle$ with time-dependent weighting coefficients $a_i$:
\begin{equation}
\Psi(\vec{r}, t) = \displaystyle\sum_{i=1}^{3} a_{i}(t) |i\rangle, \, \, \mbox{where} \, \, \displaystyle\sum_{i=1}^{3} |a_{i}|^{2} = 1.
\label{eq:linsup}
\end{equation}
%Here the $a$-coefficients carry the time-dependence while the spatial dependence is carried in the $|i\rangle$'s. 
Substituting this 
%Eqn.~\ref{eq:linsup} 
into Eq. (\ref{eq:schro}), and  introducing $\hbar \omega_{ij} \equiv \xi_{i} - \xi_{j}$ for $i \neq j$, we project the resultant equation onto each of the $\langle i|$-states to obtain
\begin{eqnarray}
i \hbar \dot{a}_{1} & = & V_{13}a_{3} \nonumber\\
i \hbar \dot{a}_{2} & = & \hbar \omega_{21} a_{2} + V_{23}a_{3} \nonumber\\
i \hbar \dot{a}_{3} & = & \hbar \omega_{31} a_{3} + V_{32}a_{2} + V_{31}a_{1},
\label{eq:a1}
\end{eqnarray}
where 
%used Eqn.~\ref{eq:eigen} and we 
$\xi_{1}$ is set equal to 0. We assume that the $1\leftrightarrow2$ transition is dipole-forbidden, i.e., not coupled by a dipole transition. In Eq. (\ref{eq:a1})  the interaction terms $V_{ij}$ are: $V_{ij} = \langle i | -\hat{\vec d}| j \rangle \boldsymbol{\cdot} \vec{E}(t) = -{\vec d}_{ij} \boldsymbol{\cdot} \vec{E}$. Note that the ${\vec d}_{ii}$ terms are zero because the spatial states $|i \rangle$ have well-defined parity and 
$\hat{\vec{d}}$ is an odd spatial function. 
%($d = e\bar{r}$, see Fig.~\ref{fig:fig1}). 
Also note that ${\vec{d}}_{ij}^{*} = {\vec{d}}_{ji}$ (see Eq. (9.3.8) in Ref.~\cite{eberly}).
%because \textcolor{black}{$\hat{\vec{d}}$ is real.}
%V_{23} & = &-\langle \phi_{2}| \hat{\vec d}|\phi_{3} \rangle \left( \frac{1}{2} \vec{\epsilon_{c}} E_{c} e^{-i\omega_{p}t} + c.c.\right)
%defined
%\begin{equation}
%$V_{ij} = \langle i | \hat{V} | j \rangle = V^*_{ji}$.
%\label{eq:V}
%\end{equation}

Assuming the incident field $\hat{\vec{E}}(t)$ %in Eqn.~\ref{eq:Hint} 
to be a sum of the coupling and probe fields each of which is a monochromatic plane-wave of frequency $\omega_{c}$ and $\omega_{p}$, amplitude $E_{c}$ and $E_{p}$, and polarization unit vectors $\vec{\epsilon}_{c}$ and $\vec{\epsilon}_{p}$, we write: 
\begin{equation}
\hat{\vec{E}} (t) 
%\rightarrow \vec{\epsilon} E_{0} \mbox{cos}\omega_{L}t 
= \frac{1}{2} \vec\epsilon_{c} E_{c} e^{-i\omega_{c}t} + \frac{1}{2} \vec\epsilon_{p} E_{p} e^{-i\omega_{p}t} + c.c.,
\label{eq:mono}
\end{equation}
where $c.c.$ denotes complex conjugate, and we have assumed the atom to be located at the origin for convenience.~\cite{extendedsample} 
%and using 
%\begin{equation}
%V_{ij} =  \langle \phi_{i}|-\vec{d}|\phi_{j}\rangle.\vec{E}(t) = V^*_{ji}
%\label{V}
%\end{equation}
Substituting Eq. (\ref{eq:mono}) in Eq. (\ref{eq:a1}) we find the optical Bloch equations: 
\begin{eqnarray}
\dot{a}_{1} & = & \frac{i}{2\hbar} \vec{d}_{13} \boldsymbol{\cdot}\left( \vec\epsilon_{p} E_{p} e^{-i\omega_{p}t} + \vec\epsilon_{p}^{\,\,*} E_{p}^{*} e^{i\omega_{p}t} \right) a_{3} \nonumber\\
\dot{a}_{2} & = & -i\omega_{21}a_{2} + \frac{i}{2\hbar} \vec{d}_{23} \boldsymbol{\cdot} \left( \vec\epsilon_{c} E_{c} e^{-i\omega_{c}t} + \vec\epsilon_{c}^{\,\,*} E_{c}^{*} e^{i\omega_{c}t} \right) a_{3} \nonumber\\
\dot{a}_{3} & = & -i\omega_{31}a_{3} + \frac{i}{2\hbar} \vec{d}_{32}\boldsymbol{\cdot} \left( \vec\epsilon_{c} E_{c} e^{-i\omega_{c}t} + \vec\epsilon_{c}^{\,\,*} E_{c}^{*} e^{i\omega_{c}t} \right) a_{2} 
\nonumber\\
&  & + \frac{i}{2\hbar} \vec{d}_{31} \boldsymbol{\cdot} \left( \vec\epsilon_{p} E_{p} e^{-i\omega_{p}t} + \vec\epsilon_{p}^{\,\,*} E_{p}^{*} e^{i\omega_{p}t} \right) a_{1}\,\,.
\label{eq:a2}
\end{eqnarray}

In order to solve the differential equations in Eq. (\ref{eq:a2}), we take a cue from the case of zero incident fields (i.e., $E_{p} = E_{c} = 0$), for which the solutions are trivially obtained as
%and then guess at a similar form for the solutions for nonzero incident fields. 
%\begin{eqnarray}
$a_{1}(t) = a_{1}(0), \, \, a_{2}(t) = a_{2} (0) \mbox{e}^{-i\omega_{21}t}$, 
%where $\omega_{21} = \omega_{31} - \omega_{32}$, 
and
$a_{3}(t) = a_{3} (0) \mbox{e}^{-i\omega_{31}t}$. Next, we guess at the nonzero incident field solutions by simply replacing the constants $a_{i}(0)$ in the zero-incident field solutions with the slowly-varying coefficients $c_{i}(t)$ (which are as yet unknown, and must be solved for as shown below), and replacing $\omega_{31}$ and $\omega_{32}$ by $\omega_{p}$ and $\omega_{c}$ respectively (because we expect the driving frequencies $\omega_{c}$, $\omega_{p}$ to dominate after long times). This yields
\begin{eqnarray}
a_{1}(t) & = & c_{1}(t) \nonumber\\
a_{2}(t) & = & c_{2} (t) \mbox{e}^{-i(\omega_{p}-\omega_{c})t}  \nonumber\\
a_{3}(t) & = & c_{3} (t) \mbox{e}^{-i\omega_{p}t} \,\,.
\label{eq:c}
\end{eqnarray}
We now replace the $a_{i}$'s in Eq. (\ref{eq:a2}) with the $c_{i}$'s from Eq. (\ref{eq:c}), and 
invoke the rotating wave approximation (RWA) which lets us ignore terms oscillating at twice the optical frequency because they average to zero (measurement times exceed a few optical cycles). 
%(this is because measurements typically last at least a few optical cycles and such terms would average to zero over time). 
We obtain:
\begin{eqnarray}
\dot{c}_{1} & = & i\Omega_{p}^{*} c_{3} \nonumber \\
\dot{c}_{2} & = & i(\Delta_{p}-\Delta_{c})c_{2} + i\Omega_{c}^{*}c_{3} \nonumber \\
\dot{c}_{3} & = & i\Delta_{p}c_{3} + i\Omega_{c} c_{2} + i\Omega_{p} c_{1} \,\,.
\label{eq:c1}
\end{eqnarray}
where we have defined the probe and coupling Rabi frequencies as $\Omega_{p} \equiv \vec{d}_{31} \boldsymbol{\cdot} \vec{\epsilon}_{p} E_{p} / 2\hbar$ and $\Omega_{c} \equiv \vec{d}_{32} \boldsymbol{\cdot} \vec{\epsilon}_{c} E_{c} / 2\hbar$, respectively. The Rabi frequency is the rate at which the state-occupation probability of a resonantly excited two-level atom oscillates between the excited and ground states. In writing Eq. (\ref{eq:c1}) we have introduced the pump and probe detunings $\Delta_{c} \equiv \omega_{c} - \omega_{32}$ and $\Delta_{p} \equiv \omega_{p} - \omega_{31}$. Further, note that in going from Eq. (\ref{eq:a2}) to Eq. (\ref{eq:c1}) we have used $\vec{d}_{13} \boldsymbol{\cdot} \vec{\epsilon}_{p}^{\,\,*} E_{p}^{*} / 2\hbar = {\vec{d}_{31}}^{\,\,*} \boldsymbol{\cdot} \vec{\epsilon}_{p}^{\,\,*} E_{p}^{*} / 2\hbar = \Omega_{p}^{*}$, and a similar argument to derive $\Omega_{c}^{*}$. 
%We point this out because students should exercise caution that, for instance, $\vec{d}_{13}.\vec{\epsilon}_{p} E_{p} / 2\hbar \neq \Omega_{p}$ or $\Omega_{p}^{*}$.
%Note that ${\Omega}_{31}^{(p)*} = \vec{d}_{13}.\vec{\epsilon}_{p}^{*} E_{p}^{*} / 2\hbar. 

\subsection{Dark and bright states in EIT}
\label{subsec:dark}
%In order to better understand the results in Sec.~\ref{sec:theory}~\ref{subsec:eit} above, 
An important question is: What are the new eigenstates 
%and eigenenergies 
for the Hamiltonian in Eq. (\ref{eq:schro}) describing the three-level atomic system in Fig.~\ref{fig:fig1}(a) under illumination by the coupling and probe fields? 
To answer this question, we re-cast Eq. (\ref{eq:c1}) in a form that enables us to write down the Hamiltonian under the rotating wave approximation which we denote by ${\hat H}_{RWA}$. Eq. (\ref{eq:c1}) now yields:
\begin{equation}
i\hbar \left( \begin{array}{c} \dot{c}_1 \\ \dot{c}_2 \\ \dot{c}_3 \end{array} \right) = -\hbar \left( {\begin{array}{ccc} 0 & 0 & {\Omega_p}^{*}\\ 0 & \Delta_p - \Delta_c & {\Omega_c}^{*}\\ {\Omega_p} & {\Omega_c} & \Delta_p \end{array}} \right) \left( \begin{array}{c} {c}_1 \\ {c}_2 \\ {c}_3 \end{array} \right),
\label{eq:c2}
\end{equation}
\noindent We rewrite the Schrodinger equation in Eq. (\ref{eq:schro}) as $i\hbar |\dot{\Psi}_{RWA}\rangle = {\hat H}_{RWA} |{\Psi}_{RWA}\rangle$ where $|{\Psi}_{RWA}\rangle = \displaystyle\sum_{i=1}^{3} c_{i}(t) |i\rangle$, and compare with Eq. (\ref{eq:c2}) to find:
\begin{equation}
{\hat H}_{RWA} =  -\hbar \left( {\begin{array}{ccc} 0 & 0 & {\Omega_p}^{*}\\ 0 &\Delta_p - \Delta_c & {\Omega_c}^{*}\\ {\Omega_p} & {\Omega_c} & \Delta_p \end{array}} \right).
\label{eq:Hdark}
\end{equation}
Setting the pump and probe detunings equal to zero for simplicity, we deduce three eigenenergies:
%(for $\Delta_p = \Delta_c = 0$):
\begin{equation}
\left( \lambda_0, \lambda_\pm \right) = \left( 0, \, \, \, \mp\hbar\Omega \right) \,\,\,\, \mbox{for} \,\,\, \Delta_p = \Delta_c = 0,
%\left( \Delta \pm \sqrt{ {\Delta}^2 + {\Omega}^{2} } \right) \right\}
\label{eq:evalue}
\end{equation}
where $\Omega$ is a generalized Rabi frequency defined as
$\Omega \equiv \sqrt{|\Omega_c|^2 + |\Omega_p|^2}$. The three orthonormal eigenstates corresponding to these eigenvalues are: 
%(for $\Delta_p = \Delta_c = 0$):
\begin{eqnarray}
|0\rangle & = & \frac{1}{\Omega} \left( \Omega_c |1\rangle - \Omega_p |2\rangle \right) \nonumber\\
|\pm\rangle & = & \frac{1}{\sqrt 2} \left( \frac{ {\Omega_p}^{*} |1\rangle + {\Omega_c}^{*} |2\rangle}{\Omega} \mp |3\rangle \right).
\label{eq:estate}
\end{eqnarray}

Eq. (\ref{eq:estate}) reveals a few key insights. An atom that happens to be in eigenstate $|0\rangle$ has no overlap with state $|3\rangle$, hence cannot excite into state $|3\rangle$. For this reason we refer to $|0\rangle$ as a \emph{dark state}. In EIT experiments, the probe is much weaker than the coupling field 
($\Omega_p << \Omega_c$), yielding the result $|0\rangle \approx |1\rangle$. Further, if we re-cast the two non-dark states $|\pm\rangle$ in Eq. (\ref{eq:estate}) in terms of two new \emph{non-stationary} states $|+'\rangle$ and $|-'\rangle$:
%\egin{equation}
$|\pm'\rangle = \frac{1}{\sqrt2} \left( |-\rangle \pm |+\rangle \right)$ 
%\label{eq:pm'}
%\end{equation}
we see that, in this weak probe approximation, $|+'\rangle \approx |2\rangle$ and $|-'\rangle \approx |3\rangle$. Here, $|+'\rangle$ and $|-'\rangle$ are not stationary states because the strong coupling field oscillates the atoms back and forth between $|2\rangle$
%$|+'\rangle \approx |2\rangle$ and $|-'\rangle \approx |3\rangle$
and $ |3\rangle$. 
We therefore refer to the $|+'\rangle \approx |2\rangle$-state, which is strongly coupled to $|3\rangle$, as the \emph{bright state}. Thus, in the case of on-resonant pump and probe excitation and a weak probe, the states $|1\rangle$ and $|2\rangle$ in the three-level system of Fig.~\ref{fig:fig1}(a) form a dark state and bright state, respectively. 

%Note that atoms in the bright state soon get optically pumped into the dark state where they stay. However, as indicated in the main article, our simple ``closed" three-level model for a single atom
 %described in Sec.~\ref{sec:3lvlexpt} below 
%does not adequately replicate the dense samples of moving atoms with multiple levels that are used in our experiments, and this transfer to the dark state is reduced, leading to significantly reduced EIT contrast.
% and broadening of the EIT linewidth. 
 
%(see Sec.~\ref{sec:3lvlexpt} below). 
%There are many energy-levels close by to which the population can leak, besides several additional EIT line-broadening mechanisms described in Sec.~\ref{sec:3lvlexpt}~\ref{subsec:broaden} below. 
%We cannot stop with the``$c$-dot equations" in Eqns.~\ref{eq:c1} because 
The formation of a dark state - this new eigenstate into which resonantly illuminated three-level atoms are pumped - lies at the heart of EIT:
%The atoms are bathed by on-resonance light, yet are uncoupled to it, i.e., the atoms are transparent. 
These atoms are transparent to the resonant illumination.

\subsection{Density-matrix: Populations, coherences, and decay} 
\label{subsec:obe} 
%We cannot stop with the``$c$-dot equations" in Eqns.~\ref{eq:c1} because 
The previous section shows the formation of a dark state which leads to a ``transparency window" called EIT. 
But what is the spectral width of this transparency window? Can we estimate the amplitude of the window, i.e., the EIT contrast? To address these questions we need to include decay processes into our discussion. The density matrix formalism allows for simple inclusion of these processes, phenomenologically. 

%We introduce $\rho_{ii} \equiv c_{i}c_{i}^{*}$ the \emph{probability} of the atom being in state $|i\rangle$, and $\rho_{ij} \equiv c_{i}c_{j}^{*}$ the \emph{coherence} between states $|i\rangle$ and $|j\rangle$. We straightforwardly re-write the ``$c$-dot equations" Eqns.~\ref{eq:c1} as
% %``$\rho$-dot" equations which are 
% differential equations for $\rho_{ij}$, by using $\dot{\rho}_{ij} = c_{i}\dot{c}^{*}_{j} + \, \dot{c}_{i}c^{*}_{j}$. These are the optical Bloch equations.
% %As indicated in Sec.~\ref{sec:intro} above, 
%The $\rho_{ij}$'s form the elements of a probability density matrix for light-atom interaction denoted by $\hat{\rho}$.
%, which is encapsulated in the induced dipole operator 
%for the potential energy stored in the induced dipole 
%$\hat{V}$ in Eqn.~\ref{eq:Hint}. 

For an illuminated three-level atom, we introduce the probability density matrix $\hat{\rho}$ which is a 3$\times$3 matrix with diagonal elements $\rho_{ii} \equiv c_{i}c_{i}^{*}$ representing the \emph{probability} of the atom being in state $|i\rangle$ (i.e, the population in state $|i\rangle$ for a sample of many atoms) and off-diagonal elements $\rho_{ij} \equiv c_{i}c_{j}^{*} \, (i \neq j)$ representing the \emph{coherence} (or strength or amplitude) of the induced dipole operator $\hat{V}$ between states $|i\rangle$ and $|j\rangle$, or, in other words, the complex amplitude of the electron displacement $\vec{r}_{av}$ in Fig.~\ref{fig:fig1}(b).
%~\cite{coh}. 
The density matrix, rather than the dark state, has a tangible connection to the lab because experiments measure probabilities and coherences, not eigenstates.
%The use of the word ``coherence" here, associated with the product of probability amplitudes for the atom to be in either of two different quantum states can be confusing to students - it may help to consider an analogy to the coherence of an electromagnetic wave which is determined by measuring the product of probability amplitudes for the detected photon to have taken either path in a double-slit experiment (or alternatively, the product of electric field amplitudes for each of the two interfering light fields).  

As discussed in the previous section, a strong pump and weak probe causes level $|1\rangle$ in Fig.~\ref{fig:fig1}(a) to take on the role of the dark state, meaning that an atom, once it falls into $|1\rangle$, stays there, unable to interact with the light any more. On the other hand, an atom in level $|2\rangle$, the bright state, once optically excited to $|3\rangle$, can either fall into $|1\rangle$ where it then stays, or fall into $|2\rangle$ where it is re-excited to $|3\rangle$ and the process repeats. Over many absorption-emission cycles, this causes the atomic population to be progressively transferred, or ``optically pumped",  
%almost all population to pump 
into level $|1\rangle$, meaning that $\rho_{11} \approx 1$ and $\rho_{22} \approx \rho_{33} \approx 0$ (also, $\rho_{32} \approx 0$). In this weak probe approximation we find from 
%the c-dot equations in 
Eq. (\ref{eq:c1}):
%{\textcolor{black}{GIVE ME REF FOR STRONG PROBE}}
\begin{eqnarray}
\dot{\rho}_{13} \, \, \, = \, \, \, c_{1}\dot{c_{3}^{*}} + \dot{c_{1}}c_{3}^{*} & = & -i\Delta_{p}\rho_{13} - i\Omega_{c}^{*}\rho_{12} - i\Omega_{p}^{*} \nonumber\\
\dot{\rho}_{12} \, \, \, = \, \, \, c_{1}\dot{c_{2}^{*}} + \dot{c_{1}}c_{2}^{*}  & = & -i(\Delta_{p} - \Delta_{c})\rho_{12} - i\Omega_{c}\rho_{13}  \nonumber\\
\dot{\rho}_{23} & \approx & 0.
\label{eq:obe}
\end{eqnarray}

We have so far ignored decay and relaxation/line-broadening mechanisms.
%these are briefly discussed in the main article. 
%Sec.~\ref{sec:3lvlexpt}\ref{subsec:broaden} below. 
Elastic collisions, for example, cause an exponential decay of the phase of the induced dipole moments $\rho_{12}$ and $\rho_{13}$, while inelastic collisions cause an exponential decay of their magnitudes. For warm samples ($T = 55^{0}$C - $65^{0}$C in our experiments), Doppler broadening is another significant line-broadening mechanism. 
\textcolor{black}{We denote all decay processes from $3 \rightarrow 1$ by the optical decoherence rate $\gamma_{13}$. The dominant contributions to $\gamma_{13}$ in our experiment come from Doppler broadening and Rb collisions with the Neon buffer gas (see Sec. 3C). }We lump all elastic and inelastic damping processes causing relaxation of the induced dipole moment for the \emph{non}-allowed transition 1 to 2 into $\gamma_{12}$, the ground state decoherence rate. 
%(see Fig.~\ref{fig:fig1}(a)). 
%The same elastic and inelastic damping processes, but now with spontaneous emission added in, cause 
\textcolor{black}{Recall that $\gamma_{12} << \gamma_{13}$ in our experiments.}
%(or $|2\rangle$) 
%for the \emph{allowed} optical transition 1 to 3.}
%the spontaneous emission rate $\Gamma$, in which case it is well-known that the coherence decay rate is half that of the population decay rate~\cite{decoh}, i.e., $\gamma_{13} = \gamma_{23} = \Gamma/2$. 
%This is readily deduced from the definition of $\rho_{ij}$ above - it is obvious, for example, that $|\rho_{13}| = (\rho_{11}\rho_{33})^{1/2}$. For the weak probe $\rho_{11} \approx 1$ and the population in the excited state $\rho_{33}$ decays as $e^{-\Gamma t}$, implying that $\rho_{13}$ decays as $e^{-\frac{\Gamma}{2} t}$.}
%As detailed in the main article, our experiments are performed on hyperfine levels of the D1 transition 
%%$5^{2}S_{1/2}, F_{g} = 2 \rightarrow 5^{2}P_{1/2}, F_{e} = 1$ 
%in $^{87}$Rb vapor, which we model by a three-level system and use~\cite{steckRb}: $\Gamma/2\pi 
% = 5.75$ MHz, $\gamma_{12}/2\pi = 3$ kHz. The coupling Rabi frequency $|\Omega_c|/2\pi$ 
%%is comparable to $\gamma_{13}$ 
%ranges from about 1 to 8 MHz. 

 %The Zeeman ground states $m_{F_{g}} = 2, 0$ and the excited state $m_{F_{e}} = 1$, as depicted in Fig. 3(b) of the main article, approximate the three levels $|1\rangle, |2\rangle, |3\rangle$, respectively, in Fig.~\ref{fig:fig1}(a).
%In this case, $\gamma_{13}/2\pi 
%= \frac{1}{2}(5.75$ MHz), and $\gamma_{12}/2\pi = 3$ kHz. Also, in our experiments, as mentioned in the main article \textcolor{black}{section?}, 
Including these exponential dephasing terms in Eq. (\ref{eq:obe}) we obtain
\begin{eqnarray}
\dot{\rho}_{13} & = & -i\left(\Delta_{p} - i\gamma_{13}\right)\rho_{13} - i\Omega_{c}^{*}\rho_{12} - i\Omega_{p}^{*} \nonumber\\
\dot{\rho}_{12} & = & -i\left(\Delta_{p} - \Delta_{c} - i\gamma_{12}\right)\rho_{12} - i\Omega_{c}\rho_{13}.
\label{eq:obedamp}
\end{eqnarray}

We are interested in solutions to Eq. (\ref{eq:obedamp}) for times $t >> 1/\gamma_{12}, 1/\gamma_{13}$. In this long-time limit, 
%at times long after 
the atomic variables have ``relaxed" to their steady-state values and are no longer functions of time. Setting $\dot{\rho}_{12}$ and $\dot{\rho}_{13}$ equal to zero, we obtain the ``steady-state solutions":
\begin{equation}
\rho_{13} = \frac{\Omega_{p}^{*} \, (\Delta_{p} - \Delta_{c} - i\gamma_{12})}{|\Omega_{c}|^{2} - (\Delta_{p} - \Delta_{c} - i\gamma_{12})(\Delta_{p} - i\gamma_{13})}.
\label{eq:rho13ss}
\end{equation}
Eq. (\ref{eq:obedamp}) and Eq. (\ref{eq:rho13ss}) are derived in Ref.~\cite{eberly}, except that our equations include nonzero pump detuning $\Delta_{c}$. 

\subsection{Probe-induced dipole moment, polarizability, refractive index, and absorption}
\label{subsec:n} 

We may evaluate the expectation value of the induced dipole moment operator $\hat{\vec{d}}$ in the state $\Psi$, in terms of the density matrix coherences above. 
%The connection of the off-diagonal terms $\rho_{ij} \, (i \neq j)$ with the induced dipole moment between states $|i\rangle$ and $|j\rangle$ is clearly brought out by directly evaluating the expectation value $\langle \hat{\vec{d}} \rangle$ of the induced dipole moment operator 
%$\hat{\vec{d}}$ in the state $\Psi$. 
From Eq. (\ref{eq:linsup})
%Using Eqn.~\ref{eq:c}
%the equation for $\Psi$ just before Eqn.~\ref{eq:a1}, and ,
% and recalling that $\langle i | \hat{\vec{d}} | j \rangle = \vec{d}_{ij}$ 
we obtain $\langle \hat{\vec{d}} \rangle = \displaystyle\sum_{i\neq j}^{3} a_{i}^{*}a_{j} \vec{d}_{ij} $ (recall from Sec.~\ref{sec:theory}\ref{subsec:ham} that the $\vec{d}_{ii}$ terms are zero).
%, i.e., the spatial states $|\phi_{i}\rangle$ have well-defined parity and $\hat{\vec{d}}$ is an odd spatial function. 
Using Eq. (\ref{eq:c}) we find for the dipole moment induced by the coupling and probe fields:
\begin{equation}
\langle \hat{\vec{d}} \rangle =  \rho_{12}\,e^{i(\omega_{p}-\omega_{c})t} \vec{d}_{21} + \rho_{13}\,e^{i\omega_{p}t} \vec{d}_{31} 
%\nonumber \\
%& & + \, \, \rho_{23}\,e^{i\omega_{c}t} \vec{d}_{32} 
+ c.c.
\label{eq:conn}
\end{equation}

If our goal is to calculate the polarizability, refractive index, and absorption as detected by the \emph{probe} beam propagating through the coherently pumped three-level atomic sample, we should focus in Eq. (\ref{eq:conn}) on the probe-induced term in which $\rho_{13}$ appears as the weighting coefficient of the induced dipole moment $\vec{d}_{31}$ oscillating at the probe frequency. 
%between levels $|1\rangle$ and $|3\rangle$.
Extracting this term from Eq. (\ref{eq:conn}) we write for the \emph{probe-induced dipole moment}, denoted by $\langle \hat{\vec{d}} \rangle_{{probe}}$:
% In order to calculate the \emph{probe absorption} we extract this term from Eqn.~\ref{eq:conn} and write: 
%we may write the following equation for the induced dipole moment oscillating at the probe frequency, 
%which we denote by $\langle \hat{\vec{d}} \rangle_{{probe}}$:
\begin{equation}
\langle \hat{\vec{d}} \rangle_{{probe}} =  \rho_{13}\,e^{i\omega_{p}t}\vec{d}_{31} + c.c.
\label{eq:dprobe1}
\end{equation}
%from which it is clear our next step is to calculate $\rho_{13}$.

%The optical Bloch equations are deduced from the c-dot equations by using $\dot{\rho}_{ij} = c_{i}\dot{c}^{*}_{j} + \, \dot{c}_{i}c^{*}_{j}$. In the usual situation the pump is strong and the probe weak causing almost all population to pump into level $|1\rangle$ in Fig.~\ref{fig:fig1}(a), meaning that $\rho_{11} \approx 1$ and $\rho_{22} \approx \rho_{33} \approx 0$ (also, $\rho_{32} \approx 0$). In this \emph{weak probe approximation} we find 
%{\textcolor{black}{GIVE ME REF FOR STRONG PROBE}}
%\begin{eqnarray}
%\dot{\rho}_{13} & = & i\Delta_{p}\rho_{13} - i\Omega_{c}^{*}\rho_{12} - i\Omega_{p}^{*} \nonumber\\
%\dot{\rho}_{12} & = & i(\Delta_{p} - \Delta_{c})\rho_{12} - i\Omega_{c}\rho_{13} \nonumber\\
%\dot{\rho}_{23} & \approx & 0
%\label{eq:obe}
%\end{eqnarray}

The induced dipole moment is also defined in terms of the polarizability, as in the caption to Fig.~\ref{fig:fig1}. From Eq. (\ref{eq:mono}), we write for the \emph{probe-induced polarizability} $\alpha_p$:
\begin{equation}
\langle \hat{\vec{d}} \rangle_{{probe}} =  \alpha_{p} \left(  \frac{1}{2} \vec\epsilon_{p} E_{p} e^{-i\omega_{p}t} + c.c.\right).
\label{eq:dprobe2}
\end{equation}
Equating terms oscillating at the same frequency in Eq. (\ref{eq:dprobe1}) and Eq. (\ref{eq:dprobe2}), 
%we find
%\begin{equation}
%\frac{1}{2} \alpha_{p}\vec{\epsilon}_{p}E_{p} = \rho_{31}\vec{d}_{13}
%\label{eq:alphap1}
%\end{equation}
and taking the dot product with $\vec{\epsilon}_{p}^{\,\,*}$ on both sides (note: $\vec{\epsilon}_{p}\boldsymbol{\cdot} \vec{\epsilon}_{p}^{\,\,*} = 1$), we find the polarizability $\alpha_p$ in terms of the density matrix element $\rho_{13}$:
\begin{equation}
\frac{1}{2} \alpha_{p}E_{p} = \rho_{31}\vec{d}_{13} \boldsymbol{\cdot} \vec{\epsilon}_{p}^{\,\,*}.
\label{eq:alphap2}
\end{equation}

We substitute the complex conjugate of Eq. (\ref{eq:rho13ss}) in Eq. (\ref{eq:alphap2}) to obtain for the probe-induced polarizability 
\begin{equation}
\alpha_{p} = \frac{|\mu_{13}|^{2}}{\hbar} \frac{\Delta_{p} - \Delta_{c} + i\gamma_{12}}{|\Omega_{c}|^{2} - (\Delta_{p} - \Delta_{c} + i\gamma_{12})(\Delta_{p} + i\gamma_{13})},
\label{eq:alphap3}
\end{equation}
where we used the definition of $\Omega_{p}$ as stated immediately after Eq. (\ref{eq:c1}), and we introduced the symbol $\mu_{13} = \vec{d}_{13} \boldsymbol{\cdot} \vec{\epsilon}_{p}^{\,\,*}$ to denote the projection of the transition dipole moment on the direction of the field polarization. 
%(note that $|\mu_{13}|^{2}$ is directly related to the Clebsch-Gordan coefficient for the probe transition). 

The relation between the macroscopic refractive index and the polarizability is well-known from undergraduate electrodynamics.~\cite{eberly2} For a linear non-magnetic medium with no free currents and dielectric constant $\epsilon$ we obtain for the \emph{refractive index} $n$ \emph{detected by the probe}:
%~\cite{maxwell}: 
\begin{equation}
n = \sqrt \epsilon = \sqrt{1+N\alpha_{p}/\epsilon_{0}} \xrightarrow[n\sim1]{\text{gas}} 1 + N\alpha_{p} / 2\epsilon_{0}, 
\label{eq:n}
\end{equation}
where $N$ is the atomic density, or more precisely, the number of atomic dipoles per unit volume, $\epsilon_0$ is the dielectric permittivity of vacuum, 
and we have used the fact that $n \sim 1$ for a gas or vapor. 
Clearly, from Eq. (\ref{eq:alphap3}) and Eq. (\ref{eq:n}), $n$ is complex, i.e., we may write $n = n_r + i \, n_i$.

%Recall from Eqn.~\ref{eq:dprobe1} that $\rho_{13}$ is related to the steady-state magnitude of the induced dipole moment at the probe frequency. Hence $\rho_{13}$ has a direct connection to the polarizability of the medium (defined as the induced dipole moment per atom, which we denote by $\alpha_{p}$).
%In the lab, however, we measure probe absorption. 
%%(denoted by $\alpha$). 
%Therefore, we need to connect $\rho_{13}$ with probe absorption.
%% $\alpha$. and the imaginary refractive index $n_{i}$. 
%For the benefit of undergraduates who may not have had a course in optics, we briefly describe in the next sub-section how we arrive at a well-known relation between the absorption 
%%$\alpha$ 
%and the polarizability $\alpha_{p}$. 
The \emph{probe absorption}, denoted by $\alpha$, and the imaginary refractive index $n_i$ are straightforwardly related. Consider the simplest case of a plane wave $\vec{E}(z,t) = \vec{E}_{0} \mbox{e}^{i(kz-\omega t)}$ propagating in the $z$-direction through a medium, with the well-known wave relation $k = n\omega/c$.
% between the spatial frequency $k$ and temporal frequency $\omega$. 
 This wave attenuates according to Beer's Law which states: $I(z) = I_{0} \, \mbox{exp}(-\alpha z)$, where $I_{0}$ is the intensity at $z=0$, and $\alpha$ is the attenuation coefficient.  The incident intensity attenuates exponentially due to both scattering and absorption - the latter dominates at near-resonance incident frequencies, in which case we may simply refer to $\alpha$ as the absorption coefficient. The exponent $\alpha z$ is known as the \emph{optical depth}. 
% given by the well-known alternate expression $N\sigma z$ where $\sigma$ is the absorption cross-section. 
% The value for $\sigma$ is largest on-resonance where the cross-section becomes on the order of the square of the optical wavelength: $\sigma = \frac{3\lambda^2}{2\pi}$~\cite{metcalf}. 
The on-resonance optical depth seen by the probe as it traverses an atom sample of length $L$ (with the strong coupling beam turned off) is denoted by $OD$ in the main article.~\cite{metcalf,lukin,od} 
%for a sample of length $L$ is $N L$, which is nothing but the factor $\eta k L$ from Ref.~\cite{lukin} (that we introduced in the main article in text preceding Eqn. 11).
 
Substituting the complex expression for $n$ into the plane wave amplitude above, we find $\vec{E}(z,t) = \vec{E}_{0} \mbox{e}^{-n_{i}\omega z/c}\mbox{e}^{i(n_{r}\omega z/c-\omega t)}$, from which we derive $I(z) = I_{0} \, \mbox{exp}(-2n_{i}\omega z/c)$. Comparing this relation with Beer's Law we arrive at the well-known relation between the absorption coefficient and the imaginary refractive index: 
\begin{equation}
\alpha = 2 n_{i}\omega / c.
\label{eq:alpha}
\end{equation}

%obtain $n = 1+N\alpha_{p}/ 2\epsilon_{0}$, which yields the real refractive index $n_{r}$ and imaginary refractive index $n_{i}$) in terms of the real and imaginary parts of the polarizability ($\alpha_{p,r}$ and $\alpha_{p,i}$, respectively): 
%\begin{eqnarray}
%n_{r} & = & 1 + N\alpha_{p,r} / 2\epsilon_{0} \\
%n_{i} & = & N\alpha_{p,i} / 2\epsilon_{0}
%\label{eq:nrni}
%\end{eqnarray}

%Now, how to calculate the polarizability $\alpha_{p}$? The answer lies in an examination of Eqns.~\ref{eq:dprobe1} and~\ref{eq:rho13ss} and the realization that the definition of the polarizability stated in the caption to Fig.~\ref{fig:fig1} allows us to use Eqn.~\ref{eq:mono} to write:
%\begin{equation}
%\langle \hat{\vec{d}} \rangle_{{probe}} =  \alpha_{p} \left(  \frac{1}{2} \vec\epsilon_{p} E_{p} e^{-i\omega_{p}t} + c.c.\right).
%\label{eq:dprobe2}
%\end{equation}

Substituting Eq. (\ref{eq:alphap3}) in Eq. (\ref{eq:n}), and rationalizing the denominator, we may equate the real and imaginary parts separately in Eq. (\ref{eq:n}). Using Eq. (\ref{eq:alpha}), we obtain for the absorption coefficient $\alpha$:
\begin{eqnarray}
\alpha & = & \frac{N\omega_{p}}{\epsilon_{0}c} \frac{|\mu_{13}|^{2}}{\hbar}  \label{eq:abs} \\
& \times& \frac{\gamma_{13}\Delta'^{2} + \gamma_{12}(\gamma_{12}\gamma_{13} + |\Omega_{c}|^{2})}{{\left[ \Delta_{p}\Delta' - \gamma_{12}\gamma_{13} - |\Omega_{c}|^{2} \right]^{2} + \left[ \gamma_{12}\Delta_{p} + \gamma_{13}\Delta' \right]^{2}}}, \nonumber
\end{eqnarray}
%(\Delta'^{2} + \gamma_{12}^{2})(\Delta_{p}^{2} + \gamma_{13}^{2})  - 2 |\Omega_{c}^{2}| (\Delta'\Delta_{p} - \gamma_{12}\gamma_{13}) + |\Omega_{c}|^{4}} 
and for the real part of the refractive index $n_{r}$
\begin{eqnarray}
n_{r} & = & 1- \frac{N}{2\epsilon_{0}} \frac{|\mu_{13}|^{2}}{\hbar}  \label{eq:n_r} \\
& \times & \frac{\Delta'(\Delta'\Delta_{p} - |\Omega_{c}|^2) + \gamma_{12}^{2}\Delta_{p}}{{\left[ \Delta_{p}\Delta' - \gamma_{12}\gamma_{13} - |\Omega_{c}|^{2} \right]^{2} + \left[ \gamma_{12}\Delta_{p} + \gamma_{13}\Delta' \right]^{2}}}, \nonumber 
\end{eqnarray}
where $\Delta' \equiv$ relative pump-probe detuning $\Delta_{p}-\Delta_{c}$ (also referred to as the Raman, or two-photon, detuning).
Eq. (3) and Eq. (4) in the main article assume $\Delta_c = 0$ to obtain simplified versions of \textcolor{black}{Eq. (S22) and Eq. (S23)}. 

\section{Magnetic shielding, heating the sample, $B_z$
%with applied magnetic field 
%for Zeeman EIT
} 
\label{sec:chamber}
This section describes the magnetic shielding used to suppress stray magnetic fields thereby creating a ``zero gauss chamber", in which the laser-atom interaction occurs. We describe the solenoid used to apply a small $B$-field to the alkali sample in a direction co-linear with the laser beam, in order to enable Zeeman EIT. We present our method used to degauss the magnetic shields and measure the residual stray magnetic fields after degaussing. Finally, we describe the implementation of a heater system to raise the alkali vapor temperature, in order to increase the atomic density participating in EIT.   
%how we .
%are suppressed below 0.2 mG.
%The zero gauss chamber and all the components inside it are shown in Figs.~\ref{fig:0gauss}-\ref{fig:9}. 
%n Sec. 3 of the Supplementary Notes we describe how to suppress below 0.2 mG any stray magnetic fields in the laser-atom interaction region.

\subsection{Zero gauss chamber:} 
\label{subsec:shield}
\begin{figure}[h]
\centerline{
\includegraphics[width=9cm]{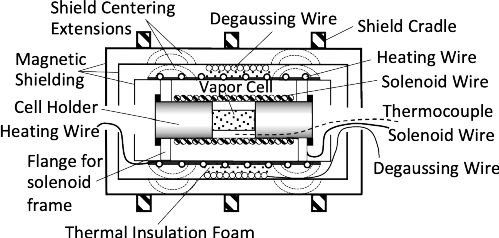}}
%{0gausschamber.pdf}}
\caption{\textsf{Cross-section of the zero-gauss chamber.}
%, suppress stray magnetic fields, and increase the participating atomic density.
%Cross-sectional view of the components inside the zero gauss chamber used for applying a small axial magnetic field to enable Zeeman EIT, suppressing stray magnetic fields that broaden EIT linewidth, and heating the vapor cell to increase the participating atomic density. 
%See text for further explanation of the components. 
} \label{fig:0gauss}
\end{figure}

%\subsection{Magnetically shielded, heated alkali vapor cell:}
%\label{subsec:cell}
The zero gauss chamber and all the components inside it are shown in 
%Figs.~\ref{fig:0gauss}-\ref{fig:9}. 
%As shown in 
Fig.~\ref{fig:0gauss}. The chamber itself is formed by three layers of mu-metal sheets, each of thickness 0.64 mm, in the form of open concentric cylinders, with close-fitting end caps on the outermost cylinder. Mu-metal alloy has high magnetic permeability and is specially engineered to divert incident magnetic field lines to ride along the material walls rather than penetrate through. A hole is cut in the end caps to allow optical access, as well as allow access for wires used for various functions described in the following sub-sections. 
%heating the cell and monitoring its temperature, supplying current to the solenoid, and degaussing the shields as described below
We purchased commercial shielding (\textcolor{black}{see Table 1}), but magnetic shields can be built in-house to cut costs. {\cite{rsi2012}} The manufacturer specifies an attenuation by at least a factor of 1575 of external magnetic fields through our three mu-metal sheets. For example, if the Earth's magnetic field is 0.5 G, the residual magnetic field inside the enclosed cylindrical volume should be 0.3 mG or less. In our case, we find that the residual magnetic field is typically suppressed below 0.2 mG, as described below. 

%The outermost cylinder is 17.25" long, with a diameter of 8.25". The innermost cylinder's length is 15" and inside diameter is 6".  A mu-metal end-cap on either end of the triple-shield forms a closed chamber which suppresses by a factor of 1500 any external magnetic fields leaking through. Two holes, about an inch in diameter, are drilled into the end-caps to allow the pump and probe laser beams through (and also, as shown, wiring leads for the solenoid, the heater wire, the thermocouple, and the degaussing current).  
%The inner layers of the shield have rounded extensions attached on the outside as shown, 
%as can be seen in Fig.~\ref{fig:0gauss}.
%which concentrically locate each shield at the radial center of the larger shield it is placed in. The extensions did not protrude outward enough, causing inner layers to be shifted and rotated. Inserting a few strips of thermal tape to fill the $\sim 3$ mm space between
%each extension and the outer shield is the simplest fix for this problem. 
\begin{figure}[h]
\centerline{
\includegraphics[width=8.5cm]{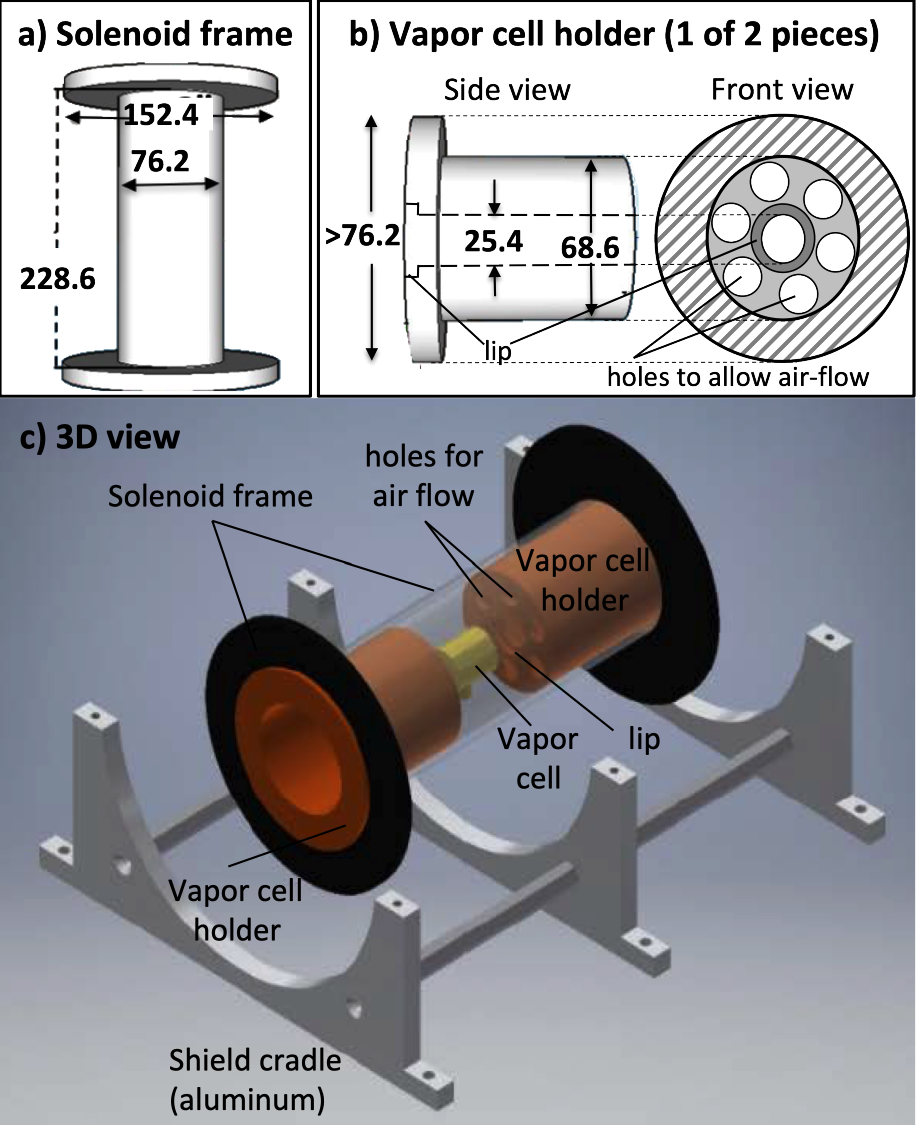}}
%{fig7_final.pdf}}
\caption{\textsf{a) The dimensions (in mm) for the plastic solenoid frame on which the wire is later wrapped. b) Side view of the plastic holder that goes on each end of the vapor cell, and a Front view (i.e., axial view of the holder as seen looking outward from the vapor cell) depicting six through-holes drilled through the entire length of each holder, to allow air-flow. c) A 3D view of the components inside the zero gauss chamber, with the shields and solenoid wiring removed to display the vapor cell and holders. The holders are drawn slightly back to reveal the lip in which the ends of the vapor cell securely fit. The solenoid frame's end-flanges (shown in black) are a close slide-fit inside the innermost mu-metal cylinder. 
%The inset show the complete aluminum cradle for housing the zero gauss chamber and locking it to the table.
}} \label{fig:3dview}
\vspace{-3mm}
\end{figure}

\subsection{Solenoid for creation of $B_z$:}
\label{subsec:solenoid}
For slow light investigations based on Zeeman EIT, we rely upon a well-defined quantization axis (the $z$-axis), which in our case is collinear with the weak external constant magnetic field $B_z$ and the propagation direction of the pump and probe laser beams. $B_z$ creates Zeeman sub-levels, for each of which the magnetic dipole moment precesses around $B_z$ with an angular momentum component $m\hbar$ along the $z$-axis.   
For this choice of quantization axis, if the electric field vector of the incident light field is perpendicular to the $z$-axis, as is the case for our pump and probe beams, the induced (i.e., absorptive and stimulated) transitions among the Zeeman sub-levels obey the selection rule $\Delta m = \pm 1$.~\cite{along} In other words, the absorption depends upon the projections of the precessing magnetic dipoles along the $z$-axis. Spontaneous emission from the excited Zeeman states is not associated with any special quantization axis and is constrained only by the selection rule $\Delta m = 0, \pm1$. 

The Zeeman magnetic sub-levels are separated via an axial magnetic field $B_z \sim 50$ mG which is created by the solenoid wiring depicted in Fig.~\ref{fig:0gauss}. Any magnetic field inhomogeneity would cause spatial variation of the dark state, leading to absorption. 
%Ideally, the magnetic field needs to be completely uniform over the entire illuminated volume of the vapor cell. 
%Rb atoms near the walls of the cell may experience a smaller magnetic field than at the center. 
The uniformity of $B_z$ depends on the length-to-radius
ratio of the solenoid. 
We referred to magnetic field calculations for finite solenoids of different length-to-radius ratios,~\cite{nasa} and concluded that a solenoid radius $r$ of 38.1 mm and length $L_s$ of 228.6 mm (ratio of 6:1) yields a uniform magnetic field with a tolerance of
1\% within a cylindrical volume 25.4 mm in diameter and 76.2 mm long located at the center of the solenoid. The plastic frame for the solenoid (in our case, a polycarbonate tube with end-flanges made from acetal  - also known as Delrin), and its placement relative to the vapor cell and other components, is depicted in Fig.~\ref{fig:3dview}.
%The frame for the solenoid should be made from a sturdy plastic such as polycarbonate 
%%or a less expensive plastic (e.g., acetal) 
%that does not soften at the temperatures used in our experiment. To center the solenoid inside the innermost mu-metal cylinder, the solenoid frame is fitted with white plastic flanges made of a less expensive plastic called acetal which are just under 6" in diameter. However, we find that acetal begins to soften at temperatures above 65$^o$C, hence should not be used if an experiment requires higher vapor temperatures. Once softened, the flanges do not acquire their original shape upon cooling back down, compromising their good fit inside the innermost mu-metal cylinder.  
%We decided to use a radius $r$ of 1.5" and length $L_s$ of 9".
Copper wire (gauge 21 or 22 is appropriate) is wound around the frame in two layers, with 220 turns/layer. The current $I$ needed is calculated using $\mu_{0}{N_t}I = B_z{\sqrt{{L_s}^2 + 4r^2}}$, where $N_{t}$ is the number of turns, $\mu_0$ is the magnetic permeability of vacuum, and $B_z$ is the magnitude of the desired axial magnetic field.~\cite{nasa} \textcolor{black}{We need a current $\leq$ 1 mA to generate the desired $B_z$.} 
%to generate the small axial $B$-field needed 
% - we built an elementary circuit using a LM317HV current regulator capable of producing stable adjustable currents on the order of a $\mu$A with a potentiometer~\cite{ken}. 

\subsection{Suppression of stray magnetic fields}
\label{subsec:lowB}
It is clear from the previous subsection that any small departure from collinearity between the laser propagation direction and the applied longitudinal magnetic field results in a small component of the pump and probe beams causing a persistent undesired $\Delta m = 0$ cycling of some atoms between the $m_{F_g} =1$ and $m_{F_e}=1$ levels (Fig. 4(b) of the main article). 
%Similar effects from the weak probe may be neglected). 
This weakens our approximation to a three-level lambda scheme, with the extraneous levels causing a decrease in EIT contrast and a significant reduction of slow light effects.

Here, we describe how to suppress below 0.2 mG (less than 0.5\% of the Zeeman field $B_z$) any stray magnetic fields in the entire laser-atom interaction volume - a cylinder of diameter equal to that of the probe beam and of length equal to that of the vapor cell. First, we degauss the mu-metal shields. Next, we use an optical method to accurately measure the low-level residual magnetic field in the laser-atom interaction region after the degaussing.~\cite{shuker} As a reference, typical high-end commercial gauss meters 
%(e.g., Hall flux meters such as Sypris Model \# XXX) 
reliably measure magnetic fields down to 10-20 mG, but not below. 

\vspace{-2mm}
\subsubsection{Degaussing the magnetic shields}
\label{subsubsec:degauss1}
The mu-metal sheets lose shielding effectiveness over a period of a few days, primarily owing to the mechanical stress created by removing and replacing the end-caps on the triple-shield when changing the optical setup inside the magnetic chamber. This is attributed to atomic dipoles becoming progressively magnetically locked together through their own interactions rather than through external fields. We therefore have to degauss the shields on days we take any data (and definitely each time we open the magnetic chamber, e.g., to use a different vapor cell).  

To degauss the shields, 
%(\textcolor{black}{we degauss only the innermost cylinder, right?}) 
we subject the mu-metal to a
large AC magnetic field of which the magnitude is slowly lowered to zero. The large AC field will
overcome the locked dipoles, forcing their alignment with
%tear the dipoles toward 
the AC field.
% \textcolor{black}{As the field is reduced, the dipoles begin aligning themselves to combat external fields - from Ken, pg 47...unclear}. 
In our experiments, we use
a cable wound around the layer of thermal foam ($\sim 80$ turns over a length of $\sim 20$ cm) surrounding the innermost mu metal shield to supply the degaussing current (see Fig.~\ref{fig:0gauss}).
%There are $\sim 80$ turns over a length of $\sim 20$ cm. 

The appropriate rate of decreasing the current is determined empirically. If not decreased optimally, residual fields continue to circulate in the innermost shield which can be detected, as discussed in the next sub-section.
%Sec.~\ref{sec:lowB}\ref{subsec:degauss2} below. 
By repeatedly modifying and iterating the degaussing process to optimally suppress the residual field 
%was suppressed below 0.5 mG, at least two orders of magnitude below the axial $B$-field used to create Zeeman EIT, 
we determined an adequate degaussing procedure:  
We passed a peak current of 5 A through the cable which is slowly lowered over 1-2 minutes using a Variac; the current is reduced twice
as slowly when under 200 mA. 

\subsubsection{Optical measurement of sub-mG residual magnetic fields}
\label{subsubsec:degauss2}
%\vspace{-1mm}
The residual magnetic field after degaussing may be measured using the well-known fact that the magnetic moment of an atom precesses around an applied magnetic field $\vec{B}$ (this is known as Larmor precession) at the Larmor frequency $\nu_L$, given by
%\begin{equation}
$h\nu_{L} =  {g_f}\mu_{B}|\vec{B}|$.~\cite{eberly}
%\label{eq:larmor}
%\end{equation}  
The constants $g_f$ and $\mu_B$ are the Lande g-factor and Bohr magneton, respectively. 
%$\mu_B = e\hbar/2 m_e = 9.274\times 10^{-24}$ JT$^{-1}$ where $e$ and $m_e$ are the electron charge and mass, and for the $F_g = 2 \rightarrow F_e = 1$ transition, $g_f = 1/2$. 
%(Sec. 3 of the main article).

If we illuminate the $^{87}$Rb vapor with a single strong circularly-polarized beam, say $\sigma^{+}$, propagating along the $z$-direction, tuned to the $F_g = 2 \rightarrow F_e = 1$ transition, and apply a small magnetic field $B_z$ (i.e., the $z$-direction is the quantization axis), then, as described in detail in Sec. 3A of the main article, this beam 
%intensity $I$ is strong enough ($I > I_{sat}$), and illuminates the atoms for long enough (many absorption-emission cycles), that 
optically pumps the atoms into the $F_g = 2, m_{F_{g}} = + 1, +2$ magnetic sub-levels after illumination over a time-scale of milliseconds. 
%(see Fig. 4(b) in the main article). 
Turning the strong laser beam off at this point allows the ``spin polarization" of the sample to begin decaying via the mechanisms discussed in Sec. 3B of the main article.
%causing the atoms to move toward states with lower $m_{F_{g}}$-values. 
If the relaxing atoms are now illuminated by a very weak $\sigma^{+}$ probe beam, so weak that its effect on the ongoing atomic relaxation is negligible, then the probe absorption continually increases in time (and the transmitted power increases) as the atoms relax toward ground sub-states with lower $m$-values. This is because the transition strength, and hence the absorption, for a particular $m_{F_{g}} \rightarrow m_{F_{g+1}}$ transition, increases as atoms move toward lower $m_{F_{g}}$-values 
(see Fig. 4(b) in the main article). 
The absorption settles to a steady-state value after a few tens of milliseconds, once the population distribution among the $F_g = 2$ magnetic sub-levels attains steady-state equilibrium.
%Note that the transition strengths 
%%are proportional to the Clebsch-Gordan (C-G) coefficients which 
%are different for each specific $\Delta m$-transition. \textcolor{black}{Thus sample absorption changes appreciably as the spin polarization decays from strongly polarized 
%%(high absorption, since C-G coefficients are highest for transitions from sub levels with $|m_{F_g}| = 1, 2$) 
%to unpolarized. }
%%(low absorption, since the population spreads to sub levels with low $|m_{F_g}|$-values which have low C-G coefficients).   

\begin{figure}[h]
\centerline{
\includegraphics[width=8.5cm]{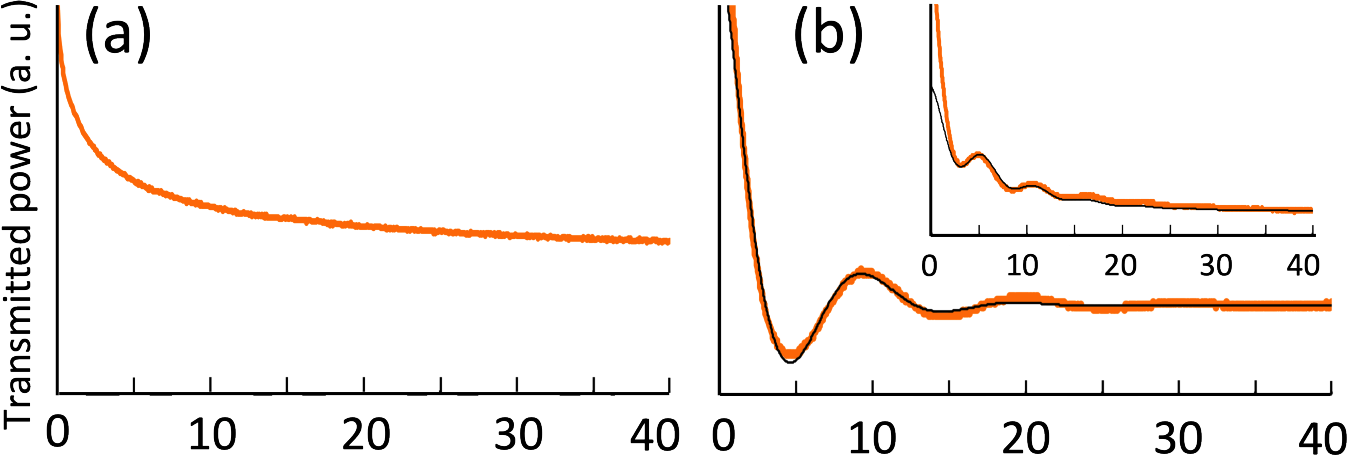}}
%{Larmor_final.pdf}}
\caption{\textsf{a) Monitoring the sample depolarization with a weak $\sigma^{+}$-probe, in the presence of a small solenoid current that applies an axial magnetic field of 2.4 mG. 
%The absorption increases in time as the atoms migrate toward ground sub-states with lower $m_{F_{g}}$-values and higher transition strengths. The absorption levels off once the population distribution in the $F_g = 2$ sub-levels reaches steady-state. 
b) If the solenoid current is set to zero, the depolarization (and hence the absorption, see text) oscillates at the Larmor precession frequency $\nu_L$ which is determined by fitting the transmission with a decaying oscillatory function.
%Using the equation connecting Larmor frequency to magnetic field (see text), 
We find that $\nu_L$ here corresponds to a stray field of 139 $\mu$G, which is a typical value obtained after degaussing. If the degaussing procedure is not optimized, the measurement yields a significantly higher $\nu_L$-value (see inset) and therefore a higher residual magnetic field, 244 $\mu$G in the case shown here. }} \label{fig:larmor}
\end{figure}

Fig.~\ref{fig:larmor}(a) shows the result of a measurement of the probe absorption as a function of time, as described above. 
%\vspace{-3mm}
We use $B_z = 2.4$ mG, and illuminate the atoms with a single circularly polarized beam for which the intensity is modulated between two states, by sending a square-wave-modulation to the AOM (see Fig. 5 in main article): A high-intensity state (the pump phase), in which the beam is intense ({4.3 mW/cm$^2$} in our case) for a period of 500 ms, followed by a low-intensity state (the probe phase) where the beam intensity is kept at {0.044 mW/cm$^2$}. The low-intensity state is at 500 ms long, though only the first 40 ms need to be recorded (since, by that time, the population is distributed evenly among the ground state magnetic sub-levels and the absorption settles to a steady-state value, as can be seen in Fig.~\ref{fig:larmor}(a)). The switching-time between the two intensity-states is 15 $\mu$s, which is much smaller than the time-scale over which the depolarization takes place.

However, if a stray non-axial magnetic field $\vec{B}$ is present, the probe absorption shows oscillations, as shown in Fig.~\ref{fig:larmor}(b). If we set the solenoid current to zero during the measurement above so that the only contribution to a magnetic field in the $z$-direction comes from the z-component of $\vec{B}$, the atomic magnetic dipoles precess around $\vec{B}$ at the Larmor frequency $\nu_L$ during the depolarization process, i.e., the projections on the $z$-axis of the magnetic dipoles precessing around $\vec{B}$ oscillate at this frequency, as do the energy separations between the Zeeman sub-levels. The laser detunings for the transitions, and hence the populations, change in accordance causing observable oscillations in the sample absorption at $\nu_L$. By measuring the oscillation frequency, we can determine the magnitude of the weak stray $B$-field from the expression for $\nu_L$ above. The plot in Fig.~\ref{fig:larmor}(b) corresponds to a stray magnetic field of 139 $\mu$G.

The inset in Fig.~\ref{fig:larmor}(b) shows that this stray $B$-field measurement technique can be used to optimize the mu-metal shield degaussing procedure. In the example shown, persistent currents in the innermost shield lead to detectable Larmor oscillations in the absorption, corresponding to a residual non-axial magnetic field nearly twice as strong as the typical case depicted in the main plot in Fig.~\ref{fig:larmor}(b).

The measurements in Fig.~\ref{fig:larmor} were performed in a vapor cell of isotropically pure $^{87}$Rb with 30 Torr of Ne buffer gas (sample length $L = 51$ mm) warmed to 
%To perform this measurement with best signal-to-noise ratio we create as many spin-polarized atoms as possible by 
%using the vapor cell with higher buffer gas pressure (30 Torr) mentioned in Sec. 5B of the main article, 
%warming to 
$65^o$C (number density $\sim 3.8 \times 10^{11}$/cm$^3$). At the time this measurement was performed we happened to have this particular vapor cell in the magnetically shielded chamber. We do not expect the results in Fig.~\ref{fig:larmor}
%for residual magnetic field determination 
to be significantly different if we had instead used the 10 Torr vapor cell on which all the data in the main article was collected. 

\subsection{Heated alkali vapor cell and cell-holder:}
\label{subsec:heat}
The most direct method to increase the density $N$ of the atoms participating in EIT is to raise the temperature of the alkali vapor cell. Following the Novikova group at College of William and Mary~\cite{wm} we directly wound an electrically insulated {non-magnetic} heating wire (see Fig.~\ref{fig:0gauss}) around the outer surface of the innermost mu-metal shield, for heating the entire volume of space enclosed by the innermost shield. \textcolor{black}{The heating wire was
coated in a thermal paste to improve thermal contact with the shield.} The wire is held firmly in place with thermal tape and a layer of thermally insulating foam. 

\begin{figure}[h]
\centerline{
\includegraphics[width=6cm]{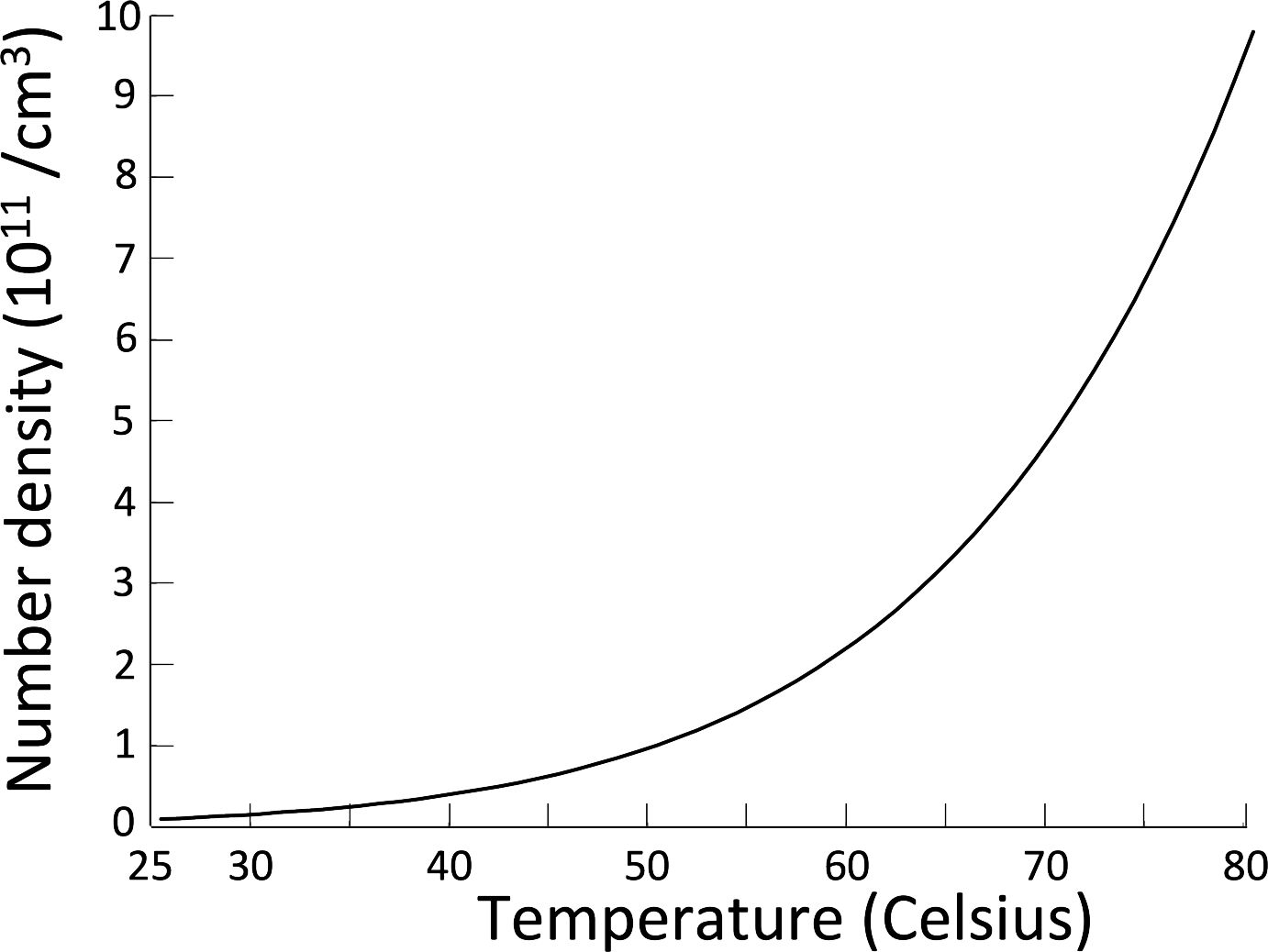}}
%{density.pdf}}
\caption{\textsf{Number density $N$ versus vapor temperature $T$ for Rb.
%In our experiments we vary the temperature between 55$^o$C and 65$^o$C, corresponding to a number density variation of $1.5 \times 10^{11}$ cm$^{-3}$ to 
%$3.8
%$3.4 \times 10^{11}$ cm$^{-3}$.
}} 
\label{fig:density}
\end{figure}

The vapor cell is held in a pair of plastic holders (in our case, acetal). The vapor cell fits securely inside a lip on each holder, as shown in Fig.~\ref{fig:3dview}(b) and (c). A rubber O-ring is placed in the lip on each end of the cell to prevent the cell from jostling around. The holders are a close slide-fit inside the solenoid frame. 
%\textcolor{black}{- the frame is shown in the top left inset}.
The vapor cell should be heated uniformly, so that ``cold spots" due to thermal gradients do not occur on the cell. Holes are drilled through the body of the cell holders, as shown,
%Figs.~\ref{fig:9}(c, d), 
to allow heated air to circulate, which significantly reduces the time needed for heating and increases the uniformity of the heating. 
%Fig.~\ref{fig:9}(b), which is placed concentrically inside the innermost cylindrical mu-metal shield. 
To accurately control the cell temperature, 
\textcolor{black}{the heating wire current is controlled by reference to a \textcolor{black}{thermocouple} placed near the cell. During data-taking, we turn the heater off to prevent any extraneous magnetic fields arising from the heater current.}
An empirical equation for the pressure $p$ of Rb vapor in thermodynamic equilibrium with its solid phase is:~\cite{pressure}
\begin{eqnarray}
\mbox{log}_{10}p & = & -94.04826 \, \, - \, \, \frac{1961.258}{T} \, \, - \, \, 0.03771687 \times T, \nonumber \\
& & + \, \, 42.57526 \times \mbox{log}_{10}T
\label{eq:pressure}
\end{eqnarray}
where $T$ is in Kelvin, and $p$ is the vapor pressure in Torr. Assuming the ideal gas law holds, we write:
\begin{equation}
N = \frac{133.3 \times p}{k_B T}.
\label{eq:ideal}
\end{equation}
Here the factor 133.3 converts from Torr to Pa and $k_B$ is the Boltzmann constant. Fig.~\ref{fig:density} shows a plot of the number density versus vapor temperature, which is useful while designing EIT and slow light experiments.

\section{Price and vendor list for key components}
\label{table}
Table 1 provides a price and vendor list for all the key components of the experiment. 
\begin{table*}[h!]
\caption{\label{tab2}\textsf{Key components used for the slow light experiment. }} 
%\vspace{2mm}
\centering
\begin{tabular}{p{6cm}p{6cm} p{2cm}}
\hline \hline
\hfil Item & \hfil Model  & \hfil Cost\\
\hline
\hfil ECDL, current/temperature control & \hfil Vitawave (Russia) ECDL7940R & \hfil \hspace{-2mm}\$ 8000~\cite{arnold1998} \\
%\hline
%\hfil SAS vapor cell (72\% $^{85}$Rb, 28\% $^{87}$Rb) & \hfil Precision Glassblowing TG-ABRB & \hfil \$ 350\\
\hfil Anamorphic prism pair & \hfil Thorlabs PS871-B & \hfil \$ 164\\
%\hline
\hfil Faraday Rotator & \hfil Conoptics 712B & \hfil \$ 1985~\cite{conoptics}\\
%\hline
\hfil PM single mode fiber & \hfil Thorlabs P3-780PM-FC-2 & \hfil \$ 197\\
\hfil Fiberport for each end of fiber (2) & \hfil Thorlabs PAF2A-11B & \hfil \$ 587 ea\\
%\hfil 3-axis stage & \hfil Klinger & \hfil \$ xx \\
%\hfil \textcolor{black}{Aspheric lens; is this the one in setup?} & \hfil \textcolor{black}{Thorlabs C140TMD-B} & \hfil \$ 73 \\
\hfil Cage assembly for telescope & \hfil Thorlabs CP33 (plate); SR4 (rods) & \hfil \$ 211 \\
\hline
\hfil AOMs (2) & \hfil Gooch and Housego 3080-122 & \hfil \$ 660 ea\\
\hfil Dual channel waveform generator & \hfil Rigol DG4102 & \hfil \$ 899 \\
\hfil \textcolor{black}{RF amplifier (Gain 30 dB)} & \hfil \textcolor{black}{RF-Bay MPA-22-30}	& \hfil \$ 330 \\
%\hfil Collimating lens pair & \hfil Edmund Optics & \hfil \$60?\\
\hline
\hfil Glan-Thompson polarizer %& \hfil Thorlabs GTH10M-B & \hfil \$ 652 ea \\
& \hfil Newlight Photonics Inc. GSC 0108 & \hfil \$ 549 \\
%\hline
\hfil Polarizing beamsplitters (3) & \hfil Thorlabs PBS 252 & \hfil \$ 253 ea \\
%\hline
\hfil Nonpolarizing beamsplitter & \hfil Thorlabs BS 013 & \hfil \$ 225 \\
%\hline
\hfil Half-wave plates (8) & \hfil Singapore Optics WP-05UM-LH-795 &\hfil  \$ 95 ea \\
%\hline
\hfil Quarter-wave plates (4) & \hfil Singapore Optics WP-05UM-LQ-795 & \hfil \$ 95 ea \\
%\hline
\hfil Photodiode & \hfil New Focus Inc. 1621 & \hfil \$ 594\\
\hline
\hfil Triple mu-metal shield \& degauss wire & \hfil Magnetic Shield Corporation ZG-206 & \hfil \$ 2130~\cite{rsi2012}\\
%\hline
\hfil Heater wire (15 ft) & \hfil ARi Industries 2HN063B-13 & \hfil \$ 6.67/ft\\
\hfil Temperature controller & \hfil Omega CN743 & \hfil \$ 115\\
\hfil Thermocouple & \hfil Digikey 317-1305-ND & \hfil \$ 3.18\\
\hfil Vapor cell for EIT and slow light & \hfil Precision Glassblowing TG-ABRB-I87 & \hfil \$ 650\\
\hfil & \hfil (pure $^{87}$Rb, 10 Torr Ne) & \hfil \\
%\hline
%\hfil \textcolor{black}{ANYTHING ELSE?} & \hfil ?? & \hfil \$ ??\\\
%\hfil \textcolor{black}{ANYTHING ELSE?}  & \hfil ?? & \hfil \$ ??\\
\hline \hline
\end{tabular}
%\vspace{-4mm}
\end{table*}
Expensive items such as the external cavity tunable diode laser system (ECDL), and the mu-metal magnetic shield with degaussing capability, can be built in-house to cut costs.~\cite{arnold1998,rsi2012}.. In addition to some basic optics, the SAS setup~\cite{bali2012} (see Sec. 4C of the main article) requires a simple uncoated vapor cell (natural abundance Rb, no buffer gas) which can be purchased from Precision Glassblowing (model TG-ABRB) for \$350.

\newpage
$^{\dag}$ These two authors contributed equally.
\newline 
$^{*}$ Corresponding author: balis@miamioh.edu

%\vspace{-3mm}
%\renewcommand\thebibliography{S\arabic{bibitem}}
%\renewcommand*{\citenumfont}[1]{S#1}
%\renewcommand*{\bibnumfmt}[1]{[S#1]}
%\newpage